\documentclass[11pt,ams]{article}%
\usepackage{geometry}
\usepackage{amsmath}
\usepackage{amsfonts}
\usepackage{amssymb}
\usepackage{graphicx}%

\newcommand{\s}{\sigma}

\renewcommand{\d}{\delta}
\renewcommand{\S}{\Sigma}

\begin{document}

\date{}
\title{\textbf{Study of the properties of the Gribov region in $SU(N)$ Euclidean Yang-Mills theories
in the maximal Abelian gauge}}
\author{\textbf{M.\thinspace A.\thinspace L.\thinspace
Capri$^{a}$\thanks{malcapri@cbpf.br}}\thinspace,
\textbf{A.\thinspace J.\thinspace
G\'{o}mez$^{b}$\thanks{ajgomez@uerj.br}}\thinspace,
\textbf{M.\thinspace S.\thinspace
Guimaraes$^{b}$\thanks{marceloguima@gmail.com}}\thinspace,
\\\textbf{V.\thinspace E.\thinspace R.\thinspace Lemes$^{b}$%
\thanks{vitor@dft.if.uerj.br}}\thinspace,
\textbf{S.\,P.\,Sorella}$^{b}$\thanks{sorella@uerj.br}\\[0.5cm]
\textit{\ }$^{a}$\textit{\small CBPF $-$ Centro Brasileiro de
Pesquisas F\'{\i}sicas,}\\
\textit{\small Rua Xavier Sigaud 150, 22290-180 Urca, Rio de
Janeiro, Brasil}$
$\\[2mm]
\textit{$^{b}${\small {UERJ $-$ Universidade do Estado do Rio de
Janeiro}}}\\\textit{{\small {Instituto de F\'{\i}sica $-$
Departamento de F\'{\i}sica Te\'{o}rica}}}\\\textit{{\small {Rua
S{\~a}o Francisco Xavier 524, 20550-013 Maracan{\~a}, Rio de
Janeiro, Brasil}}}$$} \maketitle

\begin{abstract}
\noindent In this paper we address the issue of the Gribov copies in
$SU(N), N>2,$ Euclidean Yang-Mills theories quantized in the maximal
Abelian gauge. A few properties of the  Gribov region
in this gauge are established.  Similarly to the case of $SU(2)$, the Gribov region turns out to be convex, bounded
along the off-diagonals directions in field space, and unbounded along the diagonal ones. The implementation of the restriction to the Gribov region in  the functional integral is discussed through the introduction of the horizon function, whose construction will be outlined in detail.  The influence of this restriction on the behavior of the gluon and ghost propagators of the theory is also investigated together with a set of dimension two condensates.

\end{abstract}

\baselineskip=13pt

\section*{Introduction}
In his seminal work  \cite{Gribov:1977wm}, Gribov pointed out that the quantization of Yang-Mills theories through the Faddeev-Popov method is plagued by the existence of the Gribov copies. Even if one imposes a subsidiary gauge fixing condition in an attempt to remove the gauge redundancy, there still exist field configurations belonging to the same gauge orbit  which fulfill such condition, {\it i.e.} there will be equivalent
field configurations, or copies, in the gauge fixed theory.  As a consequence, the functional measure in the Feynman path integral becomes ill-defined.  \\\\ In order to circumvent this problem, Gribov suggested  \cite{Gribov:1977wm} that the  domain of integration in the Feynman path integral should be
restricted to a certain region, known as the  Gribov region, which was supposed to contain  inequivalent field configurations only. Nowadays, it is known that the Gribov region is not completely free from  Gribov copies, {\it i.e.} additional copies still exist within the Gribov region  \cite{vanBaal:1991zw,Dell'Antonio:1991xt}.  In order to avoid the presence of these additional copies, a further restriction to a smaller region, known as the fundamental modular region, should be
implemented  \cite{vanBaal:1991zw,Dell'Antonio:1991xt}. Though, a procedure of effectively achieving the restriction to the fundamental modular region is still beyond our present capabilities. Therefore, in the following, we shall limit ourselves to investigate the Gribov region. \\\\In \cite{Gribov:1977wm}, the Landau gauge was employed in order to illustrate several aspects related to the existence of the Gribov copies. In subsequent works \cite{Dell'Antonio:1991xt,Zwanziger:1982na,Dell'Antonio:1989jn,Zwanziger:1988jt},
properties of the Gribov region in this gauge were
established. In particular, in \cite{Zwanziger:1989mf,Zwanziger:1992qr},
Zwanziger was able to show that the restriction to the Gribov region could be achieved by
adding to the Faddeev-Popov action an additional nonperturbative term, called the horizon function. Remarkably, the resulting action, known as the Gribov-Zwanziger action, can be cast in local form and proven to be multiplicatively renormalizable \cite{Zwanziger:1989mf,Zwanziger:1992qr,Maggiore:1993wq,Dudal:2005na,Dudal:2010fq}.   More recently, in  \cite{Dudal:2007cw,Dudal:2008sp,Dudal:2008xd,Dudal:2008rm}, dynamical effects related to the condensation of local dimension two operators have been accommodated in the Gribov-Zwanziger action in a way compatible with locality and renormalizability, giving rise to the so called refined Gribov-Zwanziger model. As far as the  behavior of the gluon and ghost propagators is concerned, the refined model yields a positivity violating gluon propagator which is suppressed in the infrared and which does not vanish at zero momentum, while giving a ghost propagator which is no longer enhanced in the infrared, behaving essentially as $\frac{1}{k^2}$ for $k \approx 0$. Such behavior, referred as to the decoupling or massive solution, has also been obtained within the context of the Schwinger-Dyson equations     \cite{Aguilar:2004sw,Aguilar:2008xm,Boucaud:2008ky,Fischer:2008uz}. So far, the behavior of the gluon and ghost propagators in the infrared region obtained from the refined model seems to be in agreement with recent lattice data in the Landau gauge \cite{Cucchieri:2007rg,Cucchieri:2008fc,Cucchieri:2008mv, Cucchieri:2009zt,Bogolubsky:2009dc,Bogolubsky:2009qb}. Though, we have to underline that  no unanimous consensus on this matter has yet been reached, see e.g. \cite{Fischer:2008uz,Maas:2009se,Maas:2009ph} for other possible positions. At present, the issue of the infrared behavior of the gluon and ghost propagators in the Landau gauge is object of a rather interesting and intensive debate.   \\\\The effects stemming from the existence of the Gribov copies turn out to be relevant in the nonperturbative infrared region, and might play an important role for color confinement in $QCD$. Moreover,  this is a central issue towards a correct quantization of Yang-Mills theories. As pointed out in \cite{Singer:1978dk}, the existence of the Gribov copies is in fact a general feature of the gauge fixing procedure.
Besides the Landau gauge, the issue of the
Gribov copies has been addressed in other gauges such as: the Coulomb gauge \cite{Zwanziger:2007zz} and
the linear covariant gauge \cite{Sobreiro:2005vn}, which has the Landau
gauge as a particular case. Concerning the maximal Abelian gauge, only the
particular case of  $SU(2)$ has been discussed so far \cite{Bruckmann:2000xd,Capri:2005tj}. \\\\In the present work
we shall consider the case of Euclidean $SU(N)$, $N>2$,  Yang-Mills theories in the
maximal Abelian gauge. Let us also remind that this gauge is suitable for the study of the so called Abelian dominance hypothesis \cite{Ezawa:1982bf,Suzuki:1989gp,Hioki:1991ai}, which is one of the main ingredient of the dual superconductivity mechanism for color confinement \cite{Nambu:1975ba,Mandelstam:1974vf,'tHooft:1982ns}, according to which Yang-Mills theories in the low energy region should be described by an effective Abelian theory in the presence of monopoles. A dual Meissner effect arising as a consequence of
the condensation of these magnetic charges might give rise to the formation of flux tubes which confine the quarks. The maximal Abelian gauge displays also the important property of possessing a lattice formulation \cite{Amemiya:1998jz, Bornyakov:2003ee,Mendes:2006kc}, while being a renormalizable
gauge in the continuum \cite{Min:1985bx,Fazio:2001rm,Dudal:2004rx,Kondo:2001tm}.\\\\As already mentioned, several works have been devoted to investigate
the effects of the Gribov copies in the maximal Abelian gauge in the case of $SU(2)$. A study of the influence of the Gribov copies on the two-point gluon and ghost correlation functions was
reported in \cite{Capri:2005tj}. These propagators have been also investigated within the Schwinger-Dyson framework, see \cite{Huber:2009wh}.  In \cite{Capri:2008vk},
general properties of the Gribov region in this gauge were established. In
\cite{Capri:2006cz}, the horizon function was obtained under the
requirements of localizability and renormalizability. Finally, the
inclusion of a set of local dimension two operators in the
presence of the horizon function was presented in
\cite{Capri:2008ak}, giving rise to a model which is analogous to the refined
Gribov-Zwanziger of the Landau gauge. It is worth to point out that the resulting behavior of the tree level
gluon and ghost propagators obtained in \cite{Capri:2008ak} are in good  agreement with the most recent lattice numerical simulations \cite{Mendes:2006kc} done in  the case of $SU(2)$.\\\\In the present work, we attempt to generalize
the entire program outlined in the series of papers
\cite{Capri:2005tj,Capri:2008vk,Capri:2006cz,Capri:2008ak} to $SU(N)$
Yang-Mills theories in the maximal Abelian gauge. The general
case of $SU(N)$, $N>2$,  is interesting for, at least, two reasons.
Firstly, as the symmetry group of $QCD$ is $SU(3)$, the case of $N = 3$
is of particular interest. Secondly, as we shall see, for $N > 2$, the Faddeev-Popov operator
of the maximal Abelian gauge possesses an additional term which is absent in the case of $SU(2)$. This term
will induce modifications on the behavior of the correlation functions.
For $N=2$, only the diagonal
components of the gluon propagator are affected by the restriction
to the Gribov region. However, for $N>2$, the off-diagonal
components are affected too, through a term
proportional to $(N-2)$. This feature might raise interest from the
lattice community. A numerical study of the gluon propagator in the case of SU(3) is, to our knowledge,
still lacking.  \\\\The paper is organized as follows. In
section~1, the Faddeev-Popov quantization in the maximal Abelian gauge is shortly reviewed.
In section~2, we introduce the Gribov region $\Omega$ and we establish some of its properties. This region turns out to be
convex, unbounded in all diagonal directions in field space, and bounded along the
off-diagonal directions. Section~3 is devoted to
the proof of a useful statement which ensures that for any
field configuration belonging to the Gribov region $\Omega$ and located near its
boundary, there is a Gribov copy, close to the
boundary of $\Omega$, however located outside of $\Omega$. This statement,
originally proven by Gribov in the case of the Landau gauge \cite{Gribov:1977wm}, provides us
a support for restricting the domain of integration in the functional integral to the region $\Omega$. The effective implementation of such
a restriction will be the main subject of section~4. In section~5, in order to go
beyond Gribov's quadratic approximation, we introduce the horizon function. Although in the case of $SU(N)$ several possible candidates for the horizon function can be written down, only one term turns out to be selected by the requirements of being localizable,  of reproducing the horizon function already introduced in the case of $SU(2)$, and of coinciding with the  expression obtained from Gribov's no pole condition \cite{Gribov:1977wm}.
In section~6, a local action containing a suitable set of
dimension two operators chosen in such a way as to preserve
the symmetry content of the theory is presented.
The resulting gluon and ghost propagators will be worked out. Finally, in section~7, we
collect our conclusion and discuss some
perspectives.

\section{The maximal Abelian gauge}

\subsection{Some useful definitions}
In order to introduce the maximal Abelian gauge fixing condition, one makes use of the following decomposition for the gauge field $A_{\mu}(x)$:
\begin{equation}
A_{\mu}=A^{A}_{\mu}T^{A}=A^{a}_{\mu}T^{a}+A^{i}_{\mu}T^{i}\,,
\end{equation}
where $T^A$ are the $(N^2-1)$ generators of $SU(N)$, while  $T^a$ and $T^i$ stand, respectively, for the off-diagonal and diagonal generators. There are $N(N-1)$ off-diagonal
generators and $(N-1)$ diagonal ones. The
diagonal generators give rise to the Abelian subgroup
$U(1)^{N-1}$, which is known as the Cartan subgroup. Here, we have adopted
three sets of indices, namely,
\begin{eqnarray}
A,B,C,D,E,\dots &\in&\{1,\dots,(N^{2}-1)\}\,,\nonumber\\
a,b,c,d,e,\dots&\in&\{1,\dots,N(N-1)\}\,,\nonumber\\
i,j,k,l,\dots&\in&\{1,\dots,(N-1)\}\,,
\end{eqnarray}
where the capital Latin indices $\{A,B,C,D,\dots\}$ are the usual
$SU(N)$ group indices. The off-diagonal indices are represented by
$\{a,b,c,d,e,\dots\}$ and run from $1$ to $N(N-1)$, while the
diagonal indices are given by $\{i,j,k,l,\dots\}$ running from $1$
to $(N-1)$. The commutation relations between the generators $T^A, T^a, T^i$ are given by
\begin{eqnarray}
\bigl[T^{A},T^{B}\bigr]&=&if^{ABC}\,T^{C}\,,\nonumber\\
\bigl[T^{a},T^{b}\bigr]&=&if^{abc}\,T^{c}+if^{abi}\,T^{i}\,,\nonumber\\
\bigl[T^{a},T^{i}\bigr]&=&-if^{abi}\,T^{b}\,,\nonumber\\
\bigl[T^{i},T^{j}\bigr]&=&0\,,\label{Lie}
\end{eqnarray}
where $f^{ABC}$ denote the structure constants of $SU(N)$.
From the last equation of \eqref{Lie} we conclude that
$f^{aij}=f^{ijk}=0$. The remaining nonvanishing structure
constants $f^{abc}$ and $f^{abi}$ are totally antisymmetric by the
exchange of indices and obey the following
identities\footnote{These identities follow from the Jacobi
identity:$$f^{ABC}f^{CDE}+f^{ADC}f^{CEB}+f^{AEC}f^{CBD}=0\,.$$}:
\begin{eqnarray}
0&=&f^{abi}f^{bcj}+f^{abj}f^{bic}\,,\nonumber\\
0&=&f^{abc}f^{cdi}+f^{adc}f^{cib}+f^{aic}f^{cbd}\,,\nonumber\\
0&=&f^{abc}f^{cde}+f^{abi}f^{ide}+f^{adc}f^{ceb}+f^{adi}f^{ieb}
+f^{aec}f^{cbd}+f^{aei}f^{ibd}\,.
\end{eqnarray}
The off-diagonal and diagonal components of the gauge field
transform under an infinitesimal gauge transformation as
\begin{eqnarray}
\d
A^{a}_{\mu}&=&-(D^{ab}_{\mu}\omega^{b}+gf^{abc}A^{b}_{\mu}\omega^{c} +gf^{abi}A^{b}_{\mu}\omega^{i})\,,\\
\d
A^{i}_{\mu}&=&-(\partial_{\mu}\omega^{i}+gf^{abi}A^{a}_{\mu}\omega^{b})\,.
\label{gaugetransf}
\end{eqnarray}
Here, $D^{ab}_\mu$ is the covariant derivative with respect to the
diagonal components,
\begin{equation}
D^{ab}_\mu=\d^{ab}\partial_{\mu}-gf^{abi}A^{i}_{\mu}\,,
\end{equation}
and $(\omega^{a},\omega^{i})$ are the gauge parameters. The Yang-Mills action
can also be written in terms of these diagonal and off-diagonal
components as
\begin{equation}
S_{\mathrm{YM}}=\frac{1}{4}\int d^{4}x\,(F^{a}_{\mu\nu}F^{a}_{\mu\nu}
+F^{i}_{\mu\nu}F^{i}_{\mu\nu})\,,\label{YM}
\end{equation}
with
\begin{eqnarray}
F^{a}_{\mu\nu}&=&D^{ab}_{\mu}A^{b}_{\nu}-D^{ab}_{\nu}A^{b}_{\mu}+gf^{abc}A^{b}_{\mu}A^{c}_{\nu}\,,\nonumber\\
F^{i}_{\mu\nu}&=&\partial_{\mu}A^{i}_{\nu}-\partial_{\nu}A^{i}_{\mu}+gf^{abi}A^{a}_{\mu}A^{b}_{\nu}\,.
\end{eqnarray}

\subsection{Describing the maximal Abelian gauge condition}
Since we have split the gauge field $A_{\mu}(x)$ into two
components, a diagonal (or Abelian) one represented by
$A^{i}_{\mu}(x)$ and an off-diagonal one given by $A^{a}_{\mu}(x)$, we
can now choose to fix the gauge invariance of expression \eqref{YM} by imposing a gauge condition on each component of the gauge field. For the off-diagonal components $A^{a}_{\mu}(x)$ one imposes the condition
\begin{equation}
D^{ab}_{\mu}A^{b}_{\mu} = \partial_{\mu}A^{a}_{\mu}-gf^{abi}A^{i}_{\mu}A^{b}_{\mu}= 0 \,, \label{MAG}
\end{equation}
while for the diagonal ones one requires
\begin{equation}
\partial_{\mu}A^{i}_{\mu}=0\,. \label{abmag}
\end{equation}
Note the nonlinearity in the gauge fixing of the off-diagonal components, eq.\eqref{MAG}.  This particular
choice of gauge fixing is supported by the interesting property that it follows from the requirement  that  the auxiliary functional
\begin{equation}
\mathcal{F}[A]=\int d^{4}x\,A^{a}_{\mu}A^{a}_{\mu}\,,\label{auxfunctional}
\end{equation}
is stationary with respect to the local gauge
transformations \eqref{gaugetransf}.  The nonlinear condition  \eqref{MAG} still allows for a remaining local $U(1)^{N-1}$ invariance which is removed by imposing a gauge condition on the diagonal components, for which a Landau type condition is usually chosen, as expressed by eq.\eqref{abmag}. Conditions \eqref{MAG} and \eqref{abmag} are referred as to the maximal Abelian gauge.\\\\For further use, let us also display here conditions \eqref{MAG} and \eqref{abmag} in momentum space.  By performing
a Fourier transformation, one straightforwardly obtains
\begin{eqnarray}
k_{\mu}A^{a}_{\mu}(k)&=&igf^{abi}\int\frac{d^{4}p}{(2\pi)^{4}}\,A^{i}_{\mu}(k-p)A^{b}_{\mu}(p)\,,\nonumber\\
k_{\mu}A^{i}_{\mu}(k)&=&0\,.\label{MAGmomentum}
\end{eqnarray}

\subsection{The Faddeev-Popov quantization}
The gauge fixing term, which naturally arises from the Faddeev-Popov quantization method, assumes the following form in the case of the maximal Abelian gauge:
\begin{eqnarray}
S_{\mathrm{MAG}}&=&s\int{d^{4}x}\,\left(\bar{c}^{a}\,D^{ab}_{\mu}A^{b}_{\mu}
+\bar{c}^{i}\,\partial_{\mu}A^{i}_{\mu}\right)\nonumber\\
&=&\int{d^{4}x}\,\Bigl[ib^{a}\,D^{ab}_{\mu}A^{b}_{\mu}
-\bar{c}^{a}\mathcal{M}^{ab}c^{b}
-gf^{abc}(D^{ad}_{\mu}A^{d}_{\mu})\bar{c}^{b}c^{c}
-gf^{abi}(D^{ac}_{\mu}A^{c}_{\mu})\bar{c}^{b}c^{i}\nonumber\\
&&+ib^{i}\,\partial_{\mu}A^{i}_{\mu}
+\bar{c}^{i}\,\partial_{\mu}(\partial_{\mu}c^{i}+gf^{abi}A^{a}_{\mu}c^{b})\Bigr]\,.
\end{eqnarray}
In this expression, the auxiliary fields $(b^{a},b^{i})$  are the
Lagrange multipliers enforcing the maximal Abelian gauge fixing
conditions \eqref{MAG}, \eqref{abmag}; $(c^{a},c^{i})$ are the Faddeev-Popov
ghost fields, while $(\bar{c}^{a},\bar{c}^{i})$ are the anti-ghost
fields; $\mathcal{M}^{ab}$ stands  for the Faddeev-Popov
operator,  given by
\begin{equation}
\mathcal{M}^{ab}=-D^{ac}_{\mu}D^{cb}_{\mu}-gf^{acd}A^{c}_{\mu}D^{db}_{\mu}
-g^{2}f^{aci}f^{bdi}A^{c}_{\mu}A^{d}_{\mu}\,;\label{FPop}
\end{equation}
and finally, $s$ is the nilpotent BRST operator acting on the
fields as
\begin{equation}
\begin{tabular}{cclccl}
$sA^{a}_{\mu}$&$\!\!\!\!=\!\!\!\!$&$-(D^{ab}_{\mu}c^{b}
+gf^{abc}A^{b}_{\mu}c^{c}
+gf^{abi}A^{b}_{\mu}c^{i})\,,\hspace{6pt}$
&$sA^{i}_{\mu}$&$\!\!\!\!=\!\!\!\!$&$-(\partial_{\mu}c^{i}
+gf^{abi}A^{a}_{\mu}c^{b})\,,$\vspace{5pt}\\

$sc^{a}$&$\!\!\!\!=\!\!\!\!$&$gf^{abi}c^{b}c^{i}
+\frac{g}{2}f^{abc}c^{b}c^{c}\,,$
&$sc^{i}$&$\!\!\!\!=\!\!\!\!$&$\frac{g}{2}f^{abi}c^{a}c^{b}\,,$\vspace{5pt}\\

$s\bar{c}^{a}$&$\!\!\!\!=\!\!\!\!$&$ib^{a}\,,$
&$s\bar{c}^{i}$&$\!\!\!\!=\!\!\!\!$&$ib^{i}\,,$\vspace{5pt}\\

$sb^{a}$&$\!\!\!\!=\!\!\!\!$&$0\,,$&$sb^{i}$&$\!\!\!\!=\!\!\!\!$&$0\,.$
\end{tabular}
\label{brst_fields}
\end{equation}
The nonlinearity of the gauge fixing implies that an additional
self-interacting ghost term is required in order to ensure
renormalizability \cite{Min:1985bx,Fazio:2001rm,Dudal:2004rx,Kondo:2001tm}. This additional term is defined as follows
\begin{eqnarray}
S_{\alpha}&=&-\frac{\alpha}{2}s\int
d^{4}\!x\,\biggl(i\bar{c}^{a}b^{a}
-gf^{abi}\bar{c}^{a}\bar{c}^{b}c^{i}
-\frac{g}{2}f^{abc}c^{a}\bar{c}^{b}\bar{c}^{c}\biggr)\nonumber\\
&=&\frac{\alpha}{2}\int d^{4}\!x\,\biggl(b^{a}b^{a}
+2igf^{abi}b^{a}\bar{c}^{b}c^{i} +igf^{abc}b^{a}\bar{c}^{b}c^{c}
+\frac{g^{2}}{2}f^{abi}f^{cdi}\bar{c}^{a}\bar{c}^{b}c^{c}c^{d}\nonumber\\
&&+\frac{g^{2}}{2}f^{abc}f^{adi}\bar{c}^{b}\bar{c}^{c}c^{d}c^{i}
+\frac{g^{2}}{4}f^{abc}f^{ade}\bar{c}^{b}\bar{c}^{c}c^{d}c^{e}\biggr)\,.\label{alpha}
\end{eqnarray}
Notice that this term is proportional to a gauge parameter
$\alpha$ and thus the original gauge fixing is recovered when
$\alpha$ is set to zero. However, the limit $\alpha\to 0$ has to
be taken after the removal of the ultraviolet divergences. In fact, some of the
terms proportional to $\alpha$ would reappear due to radiative corrections, even if
$\alpha = 0$, \cite{Min:1985bx,Fazio:2001rm,Dudal:2004rx,Kondo:2001tm}. Moreover, the
action
\begin{equation}
S=S_{\mathrm{YM}}+S_{\mathrm{MAG}}+S_{\alpha}\label{YM+MAG+alpha}
\end{equation}
is multiplicatively renormalizable to all orders of perturbation
theory \cite{Min:1985bx,Fazio:2001rm,Dudal:2004rx,Kondo:2001tm}. \\\\To conclude this section, let us make a few remarks on the partition function of the theory, namely
\begin{equation}
\mathcal{Z}=\int
\mathcal{D}A^{a}\mathcal{D}A^{i}
\mathcal{D}b^{a}\mathcal{D}b^{i}
\mathcal{D}c^{a}\mathcal{D}\bar{c}^{a}
\mathcal{D}c^{i}\mathcal{D}\bar{c}^{i}\,e^{-S[A,b,c,\bar{c}]}\,,
\end{equation}
where $S$ is given by \eqref{YM+MAG+alpha}. Now, taking the limit
$\alpha\to0$ and integrating over the Lagrange multipliers
$(b^{a},b^{i})$, one gets
\begin{equation}
\mathcal{Z}=\int
\mathcal{D}A^{a}\mathcal{D}A^{i}
\mathcal{D}c^{a}\mathcal{D}\bar{c}^{a}
\mathcal{D}c^{i}\mathcal{D}\bar{c}^{i}\,\d(D^{ab}_{\mu}A^{b}_{\mu})
\d(\partial_{\mu}A^{i}_{\mu})\,e^{-S_{\mathrm{YM}}
+\int{d^{4}x}\,[\bar{c}^{a}\mathcal{M}^{ab}c^{b}
-\bar{c}^{i}\,\partial_{\mu}(\partial_{\mu}c^{i}+gf^{abi}A^{a}_{\mu}c^{b})]}\,.
\end{equation}
To deal with the diagonal ghosts $(c^{i},\bar{c}^{i})$ we perform
the following change of variables
\begin{equation}
c^{i}\to\xi^{i}=c^{i}+gf^{abi}\frac{\partial_{\mu}}{\partial^{2}}A^{a}_{\mu}c^{b}\,,\qquad
\bar{c}^{i}\to\bar\xi^{i}=\bar{c}^{i}\,,
\end{equation}
with all other fields unchanged. Being linear in
$(c^{i},\bar{c}^{i})$ this change of variables leads to a
Jacobian which is field independent. Thus, we can verify that
\begin{equation}
\bar{c}^{i}\,\partial_{\mu}(\partial_{\mu}c^{i}+gf^{abi}A^{a}_{\mu}c^{b})
\to \bar\xi^{i}\partial^{2}\xi^{i}\,.
\end{equation}
Therefore, for the partition function we have
\begin{eqnarray}
\mathcal{Z}&=&\left(\int
\mathcal{D}\xi^{i}\mathcal{D}\bar\xi^{i}\,
e^{-\int{d^{4}x}\,\bar\xi^{i}\partial^{2}\xi^{i}}\right)
\int\mathcal{D}A^{a}\mathcal{D}A^{i}
\mathcal{D}c^{a}\mathcal{D}\bar{c}^{a}
\,\d(D^{ab}_{\mu}A^{b}_{\mu})
\d(\partial_{\mu}A^{i}_{\mu})\,e^{-S_{\mathrm{YM}}
+\int{d^{4}x}\,\bar{c}^{a}\mathcal{M}^{ab}c^{b}}\nonumber\\
&=&\mathcal{N}\int\mathcal{D}A^{a}\mathcal{D}A^{i}
\mathcal{D}c^{a}\mathcal{D}\bar{c}^{a}
\,\d(D^{ab}_{\mu}A^{b}_{\mu})
\d(\partial_{\mu}A^{i}_{\mu})\,e^{-S_{\mathrm{YM}}
+\int{d^{4}x}\,\bar{c}^{a}\mathcal{M}^{ab}c^{b}}\,,
\end{eqnarray}
with $\mathcal{N}$ being an irrelevant constant factor. Finally,
integrating over the off-diagonal ghosts $(c^{a},\bar{c}^{a})$, it
follows that
\begin{equation}
\mathcal{Z}=\mathcal{N}\int\mathcal{D}A^{a}\mathcal{D}A^{i}\,
\d(D^{ab}_{\mu}A^{b}_{\mu})
\d(\partial_{\mu}A^{i}_{\mu})
\det(\mathcal{M}^{ab})
\,e^{-S_{\mathrm{YM}}}\,.\label{Z}
\end{equation}
As we shall see later, the expression \eqref{Z} will be taken as
the starting point for the implementation of the restriction of
the domain of integration to the Gribov region $\Omega$.

\section{The Gribov region $\Omega$ and some of its properties}

\subsection{On the existence of Gribov copies in the maximal Abelian gauge}
Before entering into the analysis of the Gribov region $\Omega$, let us spend a few words on the existence of copies in the maximal Abelian gauge. In order to deal with this issue, let us
first take a field configuration $(A^{a}_{\mu},A^{i}_{\mu})$
fulfilling the gauge conditions \eqref{MAG}, \eqref{abmag}. Then, we can ask if
there exists a gauge transformed configuration,  $(\widetilde{A}^{a}_{\mu},\widetilde{A}^{i}_{\mu})$, which fulfills
the same gauge conditions as $(A^{a}_{\mu},A^{i}_{\mu})$, $i.\,e.$
\begin{equation}
D^{ab}_{\mu}(\widetilde{A})\widetilde{A}^{b}_{\mu}=0\,\qquad\partial_{\mu}\widetilde{A}^{i}_{\mu}=0\,.
\end{equation}
In the case of infinitesimal gauge transformations, this question can be answered in the affirmative  if the following
conditions can be verified for some value of the gauge parameters
$(\omega^{a},\omega^{i})$:
\begin{equation}
\mathcal{M}^{ab}(A)\omega^{b}=0\,,\qquad -\partial_{\mu}(\partial_{\mu}\omega^{i}
+gf^{abi}A^{a}_{\mu}\omega^{b}) =0\,,\label{copy_existence}
\end{equation}
where $\mathcal{M}^{ab}$ is the Faddeev-Popov operator given by
\eqref{FPop}. Notice that the diagonal parameter $\omega^{i}(x)$ is completely determined
in terms of the off-diagonal quantities $\omega^{a}(x)$ and $A^{a}_{\mu}(x)$. In fact, from eqs.\eqref{copy_existence}, it follows
\begin{equation}
\omega^{i} = -gf^{abi} \frac{1}{\partial^2}  \partial_{\mu} \left( A^{a}_{\mu}\omega^{b} \right) \;. \label{abc}
\end{equation}
Thus, we can say that  the
condition for the existence of equivalent field configurations, or
Gribov copies, relies on the existence of zero-modes of the Faddeev-Popov operator $\mathcal{M}^{ab}(A)$. Explicit examples of such zero modes can be found in \cite{Bruckmann:2000xd}.

\subsection{Determining the Gribov region $\Omega$ in the maximal Abelian gauge}
The so-called Gribov region $\Omega$ in the maximal Abelian gauge can be defined in a way similar to the
case of the Landau gauge \cite{Zwanziger:1989mf,Zwanziger:1992qr}. More precisely, the region $\Omega$ consists of all
field configurations which are relative minima of the auxiliary functional
\eqref{auxfunctional}, which is the same as to require that
$\mathcal{M}^{ab}$ has to be positive, namely
\begin{equation}
\d^{2}\mathcal{F}[A]=2\int d^{4}x\,\omega^{a}\mathcal{M}^{ab}(A)\omega^{b}>0
\qquad\Rightarrow\qquad\mathcal{M}^{ab}(A)>0\,.
\end{equation}
Thus, we can write
\begin{equation}
\Omega=\{A^{a}_{\mu}, A^{i}_{\mu}\,|\,D^{ab}_{\mu}A^{b}_{\mu}=0\,,\,\,
\partial_{\mu}A^{i}_{\mu}=0\,,\,\,\mathcal{M}^{ab}(A)>0\}\,.\label{OmegaDef}
\end{equation}
The boundary $\partial\Omega$ of $\Omega$ is the region in field
space where $\mathcal{M}^{ab}$ achieve its first vanishing
eigenvalue. This boundary is often called the {\it first Gribov
horizon}, or simply the {\it Gribov horizon}. Notice also that for
very small values of the coupling constant $g$, corresponding to the perturbative regime, the
Faddeev-Popov operator behaves essentially like the 4-dimensional Laplacian,
$\mathcal{M}^{ab}\approx-\partial^{2}\d^{ab}$, exhibiting in this
case only positive eigenvalues. This means that perturbation
theory is contained within the Gribov region $\Omega$.

\subsection{Some properties of the Gribov region}
\noindent In eq.\eqref{OmegaDef}, we defined the region
$\Omega$ as the set of fields fulfilling the maximal Abelian gauge
conditions and for which the operator ${\cal M}^{ab}$ is positive
definite. Let us now establish some properties of this
region. Let us first take a look at the Faddeev-Popov operator
${\cal M}^{ab}$, which can be written as
\begin{equation}
{\cal M}^{ab}={\cal O}^{ab}_{1}-{\cal
O}^{ab}_{2}-{\cal O}^{ab}_{3}\,,
\end{equation}
where
\begin{equation}
{\cal O}^{ab}_{1}=-D^{ac}_{\mu}D^{cd}_{\mu}\,,\qquad {\cal
O}^{ab}_{2}=gf^{acd}A^{c}_{\mu}D^{db}_{\mu}\,,\qquad {\cal
O}^{ab}_{3}=g^{2}f^{aci}f^{bdi}A^{c}_{\mu}A^{d}_{\mu}\,.  \label{oi}
\end{equation}
As pointed out in \cite{Bruckmann:2000xd,Capri:2008vk}, the operators ${\cal
O}^{ab}_{1}$ and ${\cal O}^{ab}_{3}$ are positive definite. We can
easily show this by the following elementary calculation:
\begin{equation}
\langle\psi|\mathcal{O}_{1}|\psi\rangle =
-\int d^{4}x\,(\psi^{a})^{\dag}D^{ac}_{\mu}D^{cb}_{\mu}\psi^{b}
=\int d^{4}x\,(D^{ac}_{\mu}\psi^{c})^{\dag}D^{ab}_{\mu}\psi^{b}
=\left\|D^{ab}_{\mu}\psi^{b}\right\|^{2}\geq0\,.
\end{equation}
\begin{equation}
\langle\psi|\mathcal{O}_{3}|\psi\rangle =
\int d^{4}x\,(\psi^{a})^{\dag}g^{2}f^{aci}f^{bdi}A^{c}_{\mu}A^{d}_{\mu}\psi^{b}
=\int d^{4}x\,(gf^{aci}A^{c}_{\mu}\psi^{a})^{\dag}
(gf^{bdi}A^{d}_{\mu}\psi^{b})=\left\|gf^{abi}A^{b}_{\mu}\psi^{a}\right\|^{2}
\geq0\,.
\end{equation}
Concerning the operator ${\cal O}^{ab}_{2}$, we note that its trace in color space vanishes, {\it i.e.}
\begin{equation}
\mathrm{tr}\mathcal{O}_{2}\equiv\mathcal{O}^{aa}_{2}=gf^{acd}A^{c}_{\mu}D^{da}_{\mu}
=gf^{acd}A^{c}_{\mu}(\d^{da}\partial_{\mu}-gf^{dai}A^{i}_{\mu})
=-gf^{aab}A^{b}_{\mu}\partial_{\mu}-g^{2}\underbrace{f^{abc}f^{abi}}_{=0}A^{c}_{\mu}A^{i}_{\mu}=0\,.
\end{equation}
As a consequence, ${\cal O}^{ab}_{2}$ has both positive and negative
eigenvalues. As we shall see in next sections, this feature can
be used to establish a few  properties of $\Omega$.

\subsubsection{The region $\Omega$ is unbounded in all diagonal directions}
To prove this statement it is sufficient to observe that the
purely diagonal field configuration $(0,A^{i}_{\mu})$ with
$A^{i}_{\mu}(x)$ transverse, $\partial_{\mu}A^{i}_{\mu}=0$,
fulfills the maximal Abelian gauge condition. Moreover, for this
kind of configuration, the Faddeev-Popov  operator ${\cal M}^{ab}$ reduces to the
covariant Laplacian
\begin{equation}
{\cal M}^{ab} (0,A^{i}) = - D^{ac}_{\mu}(A)D^{cb}_{\mu}(A)\,,
\end{equation}
which is always positive for an arbitrary choice of the transverse
diagonal configuration $A^{i}_{\mu}(x)$. We see thus that one can
freely move along the diagonal directions in field space. The
Faddeev-Popov operator will never become negative, meaning that
$\Omega$ is unbounded in the diagonal directions.

\subsubsection{The region $\Omega$ is bounded in all off-diagonal directions}
To establish this features, we observe that if $(B^{a}_{\mu},B^{i}_{\mu})$ is a field
configuration fulfilling the maximal Abelian conditions, $D^{ab}_{\mu}B^b_{\mu}=0$, $\partial_{\mu}B^i_{\mu}=0$, then the rescaled configuration $(\lambda B^{a}_{\mu},B^{i}_{\mu})$, where $\lambda$ is an
arbitrary positive constant $\lambda$, obeys the same conditions, {\it i.e.}
\begin{equation}
D_{\mu}^{ab}(\lambda B^b_{\mu}) = \lambda D_{\mu}^{ab}B^b_{\mu}= 0\;. \label{same}
\end{equation}
Let now $(A^{a}_{\mu},A^{i}_{\mu})$ be a configuration which
belongs to $\Omega$, namely
\begin{equation}
\langle\psi|{\cal M}(A^{a},A^{i})|\psi\rangle = \langle \psi| {\cal O}_{1} |\psi \rangle -  \langle \psi| {\cal O}_{2}| \psi \rangle - \langle \psi |{\cal O}_{3} |\psi \rangle  > 0 \;, \label{bomega}
\end{equation}
and let us evaluate the quantity
$\langle\psi|{\cal M}|\psi\rangle$ for the rescaled configuration
$(\lambda A^{a}_{\mu}, A^{i}_{\mu})$. From expressions \eqref{oi} one obtains
\begin{equation}
\langle\psi|{\cal M}(\lambda A^{a},A^{i})|\psi\rangle = \langle \psi| {\cal O}_{1} |\psi \rangle -  \lambda \langle \psi| {\cal O}_{2}| \psi \rangle - \lambda^2 \langle \psi |{\cal O}_{3} |\psi \rangle  > 0 \;. \label{om1}
\end{equation}
Two cases have to be considered.
$(i)$ When $0\leq\lambda\leq1$, the quantity above is always
positive since $\langle \psi| {\cal O}_{1} |\psi \rangle>0$  for every configuration and
$(A^{a}_{\mu},A^{i}_{\mu})$ belongs to the Gribov region by
hypothesis, see eq.\eqref{bomega}. $(ii)$ When $\lambda$ is larger than 1, we have to pay attention to the value of
$\langle \psi| {\cal O}_{2} |\psi \rangle$. If this quantity is positive, we just need to take
$\lambda$ as large as we wish and certainly a zero-mode will
be achieved. Notice in fact that the third term in eq.\eqref{om1}  is always positive, $\langle \psi| {\cal O}_{3} |\psi \rangle>0$.  If $\langle \psi| {\cal O}_{2} |\psi \rangle$ turns out to be negative, it would be sufficient to observe that the contribution coming from $\langle \psi| {\cal O}_{2} |\psi \rangle$ is of first-order in
$\lambda$, while the contribution coming from $\langle \psi| {\cal O}_{3} |\psi \rangle$,
which is always positive, is of the order $\lambda^{2}$. Thus, we
conclude that even if $\langle \psi| {\cal O}_{2} |\psi \rangle$ is negative, we shall still
achieve a zero-mode of ${\cal M}^{ab}$ for $\lambda$ sufficiently
large. In other words, moving along the
off-diagonal directions parameterized by the rescaled
configuration $(\lambda A^{a}_{\mu},A^{i}_{\mu})$, with
$(A^{a}_{\mu},A^{i}_{\mu})$ belonging to the Gribov region
$\Omega$, one always encounters a boundary $\partial\Omega$,
$i.\,e.$ the horizon, where the first vanishing eigenvalue of the
Faddeev-Popov operator appears. Beyond $\partial\Omega$, the
operator ${\cal M}^{ab}$ ceases to be positive definite.

\noindent Before ending this section, it is worth to discuss the case of a particular field configuration\footnote{We are grateful to Nicolas Wschebor for having pointed out to us this possibility.} which can be handled by making use of the charge conjugation invariance displayed by the maximal Abelian gauge, see \cite{Fazio:2001rm,Dudal:2004rx,Capri:2005tj,Capri:2008vk,Capri:2006cz,Capri:2008ak} for a detailed account on this symmetry. This field configuration would correspond to the case in which
the eigenvalues of the operators ${\cal O}_{2}$ and ${\cal O}_{3}$ both vanish. If this
particular configuration obeys the maximal Abelian gauge
conditions \eqref{MAG},\eqref{abmag}, and allows for an
arbitrary rescaling of the off-diagonal components of the gauge
field, then it would give rise to off-diagonal unbounded directions in field space.
Luckily, it turns out that  this kind of configuration can be excluded by invoking
the charge conjugation symmetry. In order to illustrate this issue,
let us consider in detail the simpler example of $SU(2)$. In this case, the operator
${\cal O}_{2}$ is absent and the eigenvalues of ${\cal
O}^{ab}_{3}(A)$ are given by the characteristic equation:
\begin{equation}
\displaystyle\left\|\begin{matrix}g^{2}A^{2}_{\mu}A^{2}_{\mu}
-\varepsilon_{3}&-g^{2}A^{1}_{\mu}A^{2}_{\mu}\cr\cr
-g^{2}A^{1}_{\mu}A^{2}_{\mu}&g^{2}A^{1}_{\mu}A^{1}_{\mu}
-\varepsilon_{3}\end{matrix}\right\|=
(g^{2}A^{2}_{\mu}A^{2}_{\mu}
-\varepsilon_{3})(g^{2}A^{1}_{\nu}A^{1}_{\nu}-\varepsilon_{3})
-g^{4}A^{1}_{\mu}A^{2}_{\mu}A^{1}_{\nu}A^{2}_{\nu}=0\,.
\end{equation}
For configurations of the form
\begin{equation}
A^{1}_{\mu}=\lambda\,A^{2}_{\mu}\,,\label{configSU2}
\end{equation}
with arbitrary real $\lambda$, it is easily seen that the
characteristic equation displays in fact a zero eigenvalue. Furthermore,
this configuration can be made to satisfy the maximal Abelian
gauge conditions in a nontrivial way. It would  appear then that by
choosing $\lambda$ as large as we wish we could make  the
Gribov region $\Omega$  unbounded along these particular off-diagonal
directions. However, it turns out that  Yang-Mills theory in the maximal Abelian gauge possesses a charge conjugation invariance, which is also enjoyed by the horizon term implementing the restriction to the region $\Omega$ \cite{Fazio:2001rm,Dudal:2004rx,Capri:2005tj,Capri:2008vk,Capri:2006cz,Capri:2008ak} . This symmetry is defined by
\begin{equation}
A^{1}_{\mu}\to A^{1}_{\mu}\,,\qquad
A^{2}_{\mu}\to-A^{2}_{\mu}\,,\qquad
A^{3}_{\mu}\to-A^{3}_{\mu}\,.\label{chargeSU2}
\end{equation}
and is  clearly not obeyed by the configuration
\eqref{configSU2}. As such, these configurations can be excluded from
the field space relevant for the Gribov region $\Omega$. Although the general case of $SU(N)$ looks more involved, we expect that a similar
reasoning applies as well. Needless to say, the charge conjugation can be also introduced for $SU(N)$.


\subsubsection{The convexity of the region $\Omega$}
Let us now discuss the issue of the convexity of the region
$\Omega$. Due to the nonlinearity of the gauge conditions, this
property will be established for field configurations having the
same diagonal components. Namely, consider two configurations
$(B^{a}_{\mu},A^{i}_{\mu})$ and $(C^{a}_{\mu},A^{i}_{\mu})$ which
obey the gauge conditions
\begin{equation}
D^{ab}_{\mu}(A)B^{b}_{\mu}=0\,,\qquad
D^{ab}_{\mu}(A)C^{b}_{\mu}=0\,,\qquad\partial_{\mu}
A^{i}_{\mu}=0\,,
\end{equation}
and belong to the Gribov region, {\it i.e.}
\begin{equation}
{\cal M}^{ab}(B,A)>0\,,\qquad
{\cal M}^{ab}(C,A)>0\,.
\end{equation}
In order to establish the convexity of the region $\Omega$, we should show that the field configuration $(E^{a}_{\mu},A^{i}_{\mu})$
defined by
\begin{equation}
E^{a}_{\mu}=\alpha\,B^{a}_{\mu}+(1-\alpha)C^{a}_{\mu}\,,\qquad
0\leq\alpha\leq1\,,
\end{equation}
belongs to $\Omega$, $i.\,e.$ ${\cal M}^{ab}(E,A)>0$. In fact, one
can verify straightforwardly that
\begin{equation}
{\cal M}^{ab}(E,A)=\alpha\,{\cal M}^{ab}(B,A) +(1-\alpha)\,{\cal
M}^{ab}(C,A)+\alpha(1-\alpha)\,{\cal O}^{ab}_{3}(B-C)\,.
\end{equation}
Since ${\cal O}^{ab}_{3}$ is positive definite, as already
mentioned, and since both $(B^{a}_{\mu},A^{i}_{\mu})$ and $(C^{a}_{\mu},A^{i}_{\mu})$ belong to $\Omega$, we finally conclude that
\begin{equation}
{\cal M}^{ab}(E,A)>0\,,
\end{equation}
showing that $(E^{a}_{\mu},A^{i}_{\mu})$ belongs to $\Omega$ too. This  proves the convexity of $\Omega$.

\section{Gribov's statement about field configurations close to the horizon}

In \cite{Gribov:1977wm}, Gribov proved, in the case of Landau
gauge, that for any field configuration located within the Gribov
region and close to the Gribov horizon, there exists a nearby copy located on the other
side of the horizon, $i.\,e.$ outside of the Gribov
region. This result was extended in \cite{Capri:2005tj} to the
maximal Abelian gauge in $SU(2)$. Here, we provide the necessary generalization to $SU(N)$. First, let us consider a field configuration $(C^{a}_{\mu},
C^{i}_{\mu})$ lying on the horizon, that is
\begin{equation}
D^{ab}_{\mu}(C)C^{b}_{\mu}=0\,,\qquad\partial_{\mu}C^{i}_{\mu}=0\,,\label{C_gauge}
\end{equation}
and
\begin{equation}
{\cal M}^{ab}(C)\phi^{b}_{0}=0 \;, \label{zm}
\end{equation}
where $\phi^{a}_{0}(x)$ stands for a normalizable zero mode of the Faddeev-Popov operator ${\cal M}^{ab}(C)$. For later purposes, it turns out to be useful to introduce the diagonal components $\phi^{i}_{0}(x)$ defined
according to the second equation of \eqref{copy_existence} as
\begin{equation}
\phi^{i}_{0}=-gf^{abi}\frac{1}{\partial^{2}}\partial_{\mu}(C^{a}_{\mu}\phi^{b}_{0})\,.
\end{equation}
Let $(A^{a}_{\mu},A^{i}_{\mu})$ be a field configuration
belonging to the Gribov region $\Omega$ and located close to the horizon
$\partial\Omega$. For such a configuration, we can write
\begin{equation}
A^{a}_{\mu}=C^{a}_{\mu}+a^{a}_{\mu}\,,\qquad
A^{i}_{\mu}=C^{i}_{\mu}+a^{i}_{\mu}\,,
\end{equation}
where $(a^{a}_{\mu},a^{i}_{\mu})$ are treated as small perturbative components.
As $(A^{a}_{\mu},A^{i}_{\mu})$ obeys the gauge conditions
\eqref{MAG},\eqref{abmag}, from  eq.\eqref{C_gauge} it follows that
\begin{equation}
D^{ab}_{\mu}(C)a^{b}_{\mu}-gf^{abi}a^{i}_{\mu}C^{b}_{\mu}=0\,,\qquad\partial_{\mu}a^{i}_{\mu}=0\,.
\end{equation}
The eigenvalue of the Faddeev-Popov operator corresponding to the field
configuration $(A^{a}_{\mu}, A^{i}_{\mu})$ is easily determined at first
order in the small components $(a^{a}_{\mu},a^{i}_{\mu})$ by using the standard
perturbation theory of quantum mechanics, yielding
\begin{eqnarray}
\varepsilon(A)&\!\!\!=\!\!\!&\varepsilon(C)+\int
d^{4}x\,\phi^{a}_{0}\,\Bigl[\,2gf^{abi}a^{i}_{\mu}D^{bc}_{\mu}(C)\phi^{c}_{0}
-gf^{acd}a^{c}_{\mu}D^{db}_{\mu}(C)\phi^{b}_{0} \nonumber\\
&&+g^{2}f^{acd}f^{dbi}a^{i}_{\mu}C^{c}_{\mu}\phi^{b}_{0}
-g^{2}f^{aci}f^{bdi}(C^{c}_{\mu}a^{d}_{\mu}
+a^{c}_{\mu}C^{d}_{\mu})\phi^{b}_{0}\,\Bigr]\,.\label{eigenvalue}
\end{eqnarray}
The first term in the r.h.s. of \eqref{eigenvalue} vanishes because
$(C^{a}_{\mu}, C^{i}_{\mu})$ lies on the Gribov horizon
$\partial\Omega$.\\\\We can now proceed as in
\cite{Gribov:1977wm,Capri:2005tj}, and introduce a third
configuration $(\widetilde{A}^{a}_{\mu},\widetilde{A}^{i}_{\mu})$, still located close
to the Gribov horizon, defined by
\begin{equation}
\widetilde{A}^{a}_{\mu}=C^{a}_{\mu}+\widetilde{a}^{a}_{\mu}\,,\qquad
\widetilde{A}^{i}_{\mu}=C^{i}_{\mu}+\widetilde{a}^{i}_{\mu}\,,
\end{equation}
where
\begin{eqnarray}
\widetilde{a}^{a}_{\mu}&\!\!\!=\!\!\!&a^{a}_{\mu}
-(D^{ab}_{\mu}(C)\phi^{b}_0+gf^{abc}C^{b}_{\mu}\phi^{c}_0+gf^{abi}C^{b}_{\mu}\phi^{i}_{0})\,,\nonumber\\
\widetilde{a}^{i}_{\mu}&\!\!\!=\!\!\!&a^{i}_{\mu}-(\partial_{\mu}\phi^{i}_0
+gf^{abi}C^{a}_{\mu}\phi^{b}_{0})\,,\label{a_tilde}
\end{eqnarray}
are small as compared to $(C^{a}_{\mu},C^{i}_{\mu})$. It is easy to
verify that this new field configuration was constructed in such a
way that it fulfills the maximal Abelian gauge conditions
\begin{equation}
D^{ab}_{\mu}(\widetilde{A})\widetilde{A}^{b}_{\mu}=0\,,\qquad\partial_{\mu}\widetilde{A}^{i}_{\mu}=0\,.
\end{equation}
Therefore, we can ask ourselves if
$(\widetilde{A}^{a}_{\mu},\widetilde{A}^{i}_{\mu})$ is a Gribov
copy of $({A}^{a}_{\mu},{A}^{i}_{\mu})$, {\it i.e.} if these two field
configurations can be connected by a gauge transformation. In
order to answer this question we shall repeat the argument of
\cite{Gribov:1977wm,Capri:2005tj}, amounting to construct such a
gauge transformation in an iterative way. Let us suppose thus that
it is possible to connect the two configurations by a gauge transformation, so that
\begin{eqnarray}
\widetilde{A}^{a}_{\mu}&\!\!\!=\!\!\!&A^{a}_{\mu}-(D^{ab}_{\mu}\omega^{b} +gf^{abc}A^{b}_{\mu}\omega^{c}
+gf^{abi}A^{b}_{\mu}\omega^{i}) -\frac{g}{2}f^{abc}\omega^{b}(D^{cd}_{\mu}\omega^{d}
+gf^{cde}A^{d}_{\mu}\omega^{e}+gf^{cdi}A^{d}_{\mu}\omega^{i})\nonumber\\
&&-\frac{g}{2}f^{abi}\omega^{b}(\partial_{\mu}\omega^{i} +gf^{cdi}A^{c}_{\mu}\omega^{d})
+\frac{g}{2}f^{abi}\omega^{i}(D^{bc}_{\mu}\omega^{c}
+gf^{bcd}A^{c}_{\mu}\omega^{d}+gf^{bcj}A^{c}_{\mu}\omega^{j})+O(\omega^{3})\,,\nonumber\\
\widetilde{A}^{i}_{\mu}&\!\!\!=\!\!\!&A^{i}_{\mu}-(\partial_{\mu}\omega^{i}+gf^{abi}A^{a}_{\mu}\omega^{b})
-\frac{g}{2}f^{abi}\omega^{a}(D^{bc}_{\mu}\omega^{c}+gf^{bcd}A^{c}_{\mu}\omega^{d}+gf^{bcj}A^{c}_{\mu}\omega^{j})
+O(\omega^{3})\;,\label{gauge_2nd_order}
\end{eqnarray}
where, according to \cite{Gribov:1977wm}, the gauge transformation has been considered till the
second order\footnote{Notice that if we had taken only the first order
gauge transformations, we would get
$$\widetilde{A}^{a}_{\mu}=A^{a}_{\mu}-(D^{ab}_{\mu}\omega^{b} +gf^{abc}A^{b}_{\mu}\omega^{c}
+gf^{abi}A^{b}_{\mu}\omega^{i})\,,\qquad
\widetilde{A}^{i}_{\mu}=A^{i}_{\mu}-(\partial_{\mu}\omega^{i}+gf^{abi}A^{a}_{\mu}\omega^{b})\,.$$
Then, applying the maximal Abelian gauge conditions we would obtain
the following equations:$$\mathcal{M}^{ab}(A)\omega^{b}=0\,,\qquad
\partial_{\mu}(\partial_{\mu}\omega^{i}+gf^{abi}A^{a}_{\mu}\omega^{b})=0\,,$$
which have no solutions since, by hypothesis,
$(A^{a}_{\mu},A^{i}_{\mu})$ is not located on the horizon
$\partial\Omega$.} in the gauge parameters $\omega^i, \omega^a$. From the maximal Abelian conditions it follows that
\begin{eqnarray}
&&\mathcal{M}^{ab}(A)\omega^{b}+D^{ab}_{\mu}(A)\biggl[-\frac{g}{2}f^{bcd}\omega^{c}(D^{de}_{\mu}\omega^{e}
+gf^{def}A^{e}_{\mu}\omega^{f}+gf^{dei}A^{e}_{\mu}\omega^{i})
-\frac{g}{2}f^{bci}\omega^{c}(\partial_{\mu}\omega^{i}+gf^{dei}A^{d}_{\mu}c^{e})\nonumber\\
&&+\frac{g}{2}f^{bci}\omega^{i}(D^{cd}_{\mu}\omega^{d}+gf^{cde}A^{d}_{\mu}\omega^{e}
+gf^{cdj}A^{d}_{\mu}\omega^{j})\biggr]
+\frac{g^{2}}{2}f^{abi}f^{cdi}A^{b}_{\mu}\omega^{c}(D^{de}_{\mu}\omega^{e}
+gf^{def}A^{e}_{\mu}\omega^{f}+gf^{dej}A^{e}_{\mu}\omega^{j})\nonumber\\
&&-gf^{abi}(\partial_{\mu}\omega^{i}+gf^{cdi}A^{c}_{\mu}\omega^{d})
(D^{be}_{\mu}\omega^{e}+gf^{bef}A^{e}_{\mu}\omega^{f}
+gf^{bej}A^{e}_{\mu}\omega^{j})\;=\;0\,,\label{condition_2nd_order}\\\nonumber\\
&&\partial_{\mu}\biggl[\partial_{\mu}\omega^{i}+gf^{abi}A^{a}_{\mu}\omega^{b}
+\frac{g}{2}f^{abi}\omega^{a}(D^{bc}_{\mu}\omega^{c}+gf^{bcd}A^{c}_{\mu}\omega^{d}
+gf^{bcj}A^{c}_{\mu}\omega^{j})\biggr]\;=\;0\,.
\end{eqnarray}
The next step is to express $(\omega^{a},\omega^{i})$ in terms of
$(\phi^{a}_{0},\phi^{i}_{0})$. Let us start by writing
\begin{equation}
\omega^{a}=\phi^{a}_{0}+\eta^{a}_{0}\,,\qquad\omega^{i}=\phi^{i}_{0}+\eta^{i}_{0}\,,
\end{equation}
with $(\eta^{a},\eta^{i})$ small when compared to
$(\phi^{a}_{0},\phi^{i}_{0})$. Equation
\eqref{condition_2nd_order} gives thus the following condition
\begin{equation}
\partial^{2}\eta^{a}=\mathcal{P}^{a}(C,a,\phi_0)+\mathcal{Q}^{ab}(C)\eta^{b}\,,\label{eta_eq}
\end{equation}
with ${\cal{P}}^a$ and ${\cal{Q}}^{ab}$ given by
\begin{eqnarray}
\mathcal{P}^{a}&\!\!\!=\!\!\!&2gf^{abi}a^{i}_{\mu}D^{bc}_{\mu}(C)\phi^{c}_{0}
-gf^{acd}a^{c}_{\mu}D^{ab}_{\mu}(C)\phi^{b}_{0}
+g^{2}f^{acd}f^{dbi}a^{i}_{\mu}C^{c}_{\mu}\phi^{b}_{0}
-g^{2}f^{aci}f^{bdi}(C^{c}_{\mu}a^{d}_{\mu}+a^{c}_{\mu}C^{d}_{\mu})\phi^{b}_{0}\nonumber\\
&&+D^{ab}_{\mu}(C)\biggl[-\frac{g}{2}f^{bcd}\phi^{c}_{0}(D^{de}_{\mu}(C)\phi^{e}_{0}
+gf^{def}C^{e}_{\mu}\phi^{f}_{0}+gf^{dei}C^{e}_{\mu}\phi^{i}_{0})
-\frac{g}{2}f^{bci}\phi^{c}_{0}(\partial_{\mu}\phi^{i}_{0}
+gf^{dei}C^{d}_{\mu}\phi^{e}_{0})\nonumber\\
&&+\frac{g}{2}f^{bci}(D^{cd}_{\mu}(C)\phi^{d}_{0} +gf^{cde}C^{d}_{\mu}\phi^{e}_{0}
+gf^{cdj}C^{d}_{\mu}\phi^{j}_{0})\biggr]
+\frac{g^{2}}{2}f^{abi}f^{cdi}C^{b}_{\mu}\phi^{c}_{0}(D^{de}_{\mu}(C)\phi^{e}_{0}
+gf^{def}C^{e}_{\mu}\phi^{f}_{0}\nonumber\\
&&+gf^{dej}C^{e}_{\mu}\phi^{j}_{0})
-gf^{abi}(\partial_{\mu}\phi_{0}^{i}+gf^{cdi}C^{c}_{\mu}\phi_{0}^{d})
(D^{be}_{\mu}(C)\phi_{0}^{e}+gf^{bef}C^{e}_{\mu}\phi_{0}^{f}
+gf^{bej}C^{e}_{\mu}\phi_{0}^{j})\,,\\\nonumber\\
\mathcal{Q}^{ab}&\!\!\!=\!\!\!&2gf^{abi}C^{i}_{\mu}\partial_{\mu}
-gf^{abc}C^{c}_{\mu}D^{db}_{\mu}-g^{2}f^{aci}f^{cbj}C^{i}_{\mu}C^{j}_{\mu}
-g^{2}f^{aci}f^{bdi}C^{c}_{\mu}C^{d}_{\mu}\,.
\end{eqnarray}
We notice now that equation \eqref{eta_eq} can be solved for $\eta^{a}(x)$
in an iterative way:
\begin{equation}
\eta^{a}=\frac{1}{\partial^{2}}\mathcal{P}^{a}
+\frac{1}{\partial^{2}}\mathcal{Q}^{ab}\frac{1}{\partial^{2}}\mathcal{P}^{b}+\cdots\,,
\end{equation}
allowing us to obtain a recursive expression for the parameters
$(\omega^{a},\omega^{i})$ as well as for the gauge transformations \eqref{gauge_2nd_order} in terms of $(\phi_0^{a},\phi_0^{i})$. We have shown thus  that the configuration
$(\widetilde{A}^{a}_{\mu},\widetilde{A}^{i}_{\mu})$ is indeed a
Gribov copy of $(A^{a}_{\mu},A^{i}_{\mu})$. Furthermore, equation \eqref{eta_eq} can be used to  establish
another relevant property. Recalling
that $\mathcal{M}^{ab}(C)\phi^{b}_{0}=0$ and that
$\mathcal{M}^{ab}$ is Hermitian, we have
\begin{equation}
\int d^{4}x\,\phi^{a}_{0}\mathcal{M}^{ab}(C)\eta^{b}=0\;.
\end{equation}
Therefore, from \eqref{eta_eq}, it follows
\begin{eqnarray}
&&\hspace{-18pt}\int
d^{4}x\,\phi^{a}_{0}\,\biggl\{2gf^{abi}a^{i}_{\mu}D^{bc}_{\mu}(C)\phi^{c}_{0}
-gf^{acd}a^{c}_{\mu}D^{ab}_{\mu}(C)\phi^{b}_{0}
+g^{2}f^{acd}f^{dbi}a^{i}_{\mu}C^{c}_{\mu}\phi^{b}_{0}
-g^{2}f^{aci}f^{bdi}(C^{c}_{\mu}a^{d}_{\mu}+a^{c}_{\mu}C^{d}_{\mu})\phi^{b}_{0}\nonumber\\
&&\hspace{-18pt}+D^{ab}_{\mu}(C)\biggl[-\frac{g}{2}f^{bcd}\phi^{c}_{0}(D^{de}_{\mu}(C)\phi^{e}_{0}
+gf^{def}C^{e}_{\mu}\phi^{f}_{0}+gf^{dei}C^{e}_{\mu}\phi^{i}_{0})
-\frac{g}{2}f^{bci}\phi^{c}_{0}(\partial_{\mu}\phi^{i}_{0}
+gf^{dei}C^{d}_{\mu}\phi^{e}_{0})\nonumber\\
&&\hspace{-18pt}+\frac{g}{2}f^{bci}(D^{cd}_{\mu}(C)\phi^{d}_{0} +gf^{cde}C^{d}_{\mu}\phi^{e}_{0}
+gf^{cdj}C^{d}_{\mu}\phi^{j}_{0})\biggr]
+\frac{g^{2}}{2}f^{abi}f^{cdi}C^{b}_{\mu}\phi^{c}_{0}(D^{de}_{\mu}(C)\phi^{e}_{0}
+gf^{def}C^{e}_{\mu}\phi^{f}_{0}\nonumber\\
&&\hspace{-18pt}+gf^{dej}C^{e}_{\mu}\phi^{j}_{0})
-gf^{abi}(\partial_{\mu}\phi_{0}^{i}+gf^{cdi}C^{c}_{\mu}\phi_{0}^{d})
(D^{be}_{\mu}(C)\phi_{0}^{e}+gf^{bef}C^{e}_{\mu}\phi_{0}^{f}
+gf^{bej}C^{e}_{\mu}\phi_{0}^{j})\biggr\}\;=\;0\,.\label{it_is_a_zero}
\end{eqnarray}
Moreover, the eigenvalue of the Faddeev-Popov operator corresponding to the configuration
$(\widetilde{A}^{a}_{\mu},\widetilde{A}^{i}_{\mu})$ can be
obtained in the same way  as done before for
$({A}^{a}_{\mu},{A}^{i}_{\mu})$, amounting in fact to replace $a_\mu$ by
$\widetilde{a}_{\mu}$. Thus,
\begin{eqnarray}
\varepsilon(\widetilde{A})&\!\!\!=\!\!\!&\int
d^{4}x\,\phi^{a}_{0}\,\Bigl[\,2gf^{abi}\widetilde{a}^{i}_{\mu}D^{bc}_{\mu}(C)\phi^{c}_{0}
-gf^{acd}\widetilde{a}^{c}_{\mu}D^{db}_{\mu}(C)\phi^{b}_{0} \nonumber\\
&&+g^{2}f^{acd}f^{dbi}\widetilde{a}^{i}_{\mu}C^{c}_{\mu}\phi^{b}_{0}
-g^{2}f^{aci}f^{bdi}(C^{c}_{\mu}\widetilde{a}^{d}_{\mu}
+\widetilde{a}^{c}_{\mu}C^{d}_{\mu})\phi^{b}_{0}\,\Bigr]\,.\label{eigenvalue_tilde}
\end{eqnarray}
Finally, from equations \eqref{a_tilde}, \eqref{it_is_a_zero} and
\eqref{eigenvalue_tilde} we obtain that
\begin{equation}
\varepsilon(A)=-\varepsilon(\widetilde{A})\,.
\end{equation}
Since $\varepsilon(A)>0$ by hypothesis, it follows that
$\varepsilon(\widetilde{A})=-\varepsilon(A)<0$, showing that the Gribov copy
$(\widetilde{A}^{a}_{\mu},\widetilde{A}^{i}_{\mu})$ is in fact located outside
of $\Omega$.

\section{Restriction of the domain of integration to the Gribov region $\Omega$}
As we have seen before, field configurations located inside
$\Omega$ and close to the boundary $\partial\Omega$ have copies
outside of $\Omega$. Therefore, restricting the domain of integration in the Feynman integral to the Gribov region
allows us to get rid of a certain amount of copies\footnote{Nowadays, it is known that the restriction to $\Omega$ does
not enable us to get rid of all copies. As already remarked, a further restriction to a smaller region, known as the
fundamental modular region, should be implemented \cite{vanBaal:1991zw}. Though, till now, this seems to be beyond our present
capabilities.}. To implement this restriction we shall follow
\cite{Gribov:1977wm,Capri:2005tj} and introduce in
the partition function of the theory, eq.\eqref{Z}, the factor
$\mathcal{V}(\Omega)$ which formally constrains the domain of integration to the region $\Omega$
\begin{equation}
\mathcal{Z}= \int_{\Omega} d\mu
\,e^{-S_{\mathrm{YM}}}= \int d\mu
\,e^{-S_{\mathrm{YM}}}\,
\mathcal{V}(\Omega)\,,
\end{equation}
where $d\mu$ is the Faddeev-Popov functional measure given in eq.\eqref{Z}
\begin{equation}
d\mu=\mathcal{D}A^{\mbox{\tiny off}}\mathcal{D}A^{\mbox{\tiny diag}}\,
\d(D\cdot A^{\mbox{\tiny off}})\d(\partial\cdot A^{\mbox{\tiny diag}})
(\det\mathcal{M})\,.
\end{equation}
In order to obtain an explicit expression for $\mathcal{V}(\Omega)$, we make use of so called \emph{no-pole condition} \cite{Gribov:1977wm}, which is a condition on the connected two-point function of the off-diagonal Faddeev-Popov ghost fields. As pointed out in \cite{Gribov:1977wm}, see also \cite{Sobreiro:2005ec} for a pedagogical introduction, the no-pole condition stems from the positivity of the Faddeev-Popov operator, ${\cal M}^{ab}>0$, within the Gribov region $\Omega$, according to eq.\eqref{OmegaDef}. As a consequence, within $\Omega$, the operator ${\cal M}^{ab}$ is invertible. Moreover, its inverse, $({\cal M}^{ab})^{-1}$, which is nothing but the connected two-point off-diagonal ghost function evaluated in an external gauge field background:
\begin{equation}
({\mathcal{M}}^{ab}(x,y;A))^{-1}=\frac{\int\mathcal{D}c^{\rm off}\mathcal{D}\bar{c}^{\rm off}\;
\bar{c}^{a}(x)c^{a}(y)\,e^{\int d^{4}z\,\bar{c}^{b}(z)\mathcal{M}^{bc}(z)\,c^{c}(z)}}{\int\mathcal{D}c^{\rm off}\mathcal{D}\bar{c}^{\rm off}\;
\,e^{\int d^{4}z\,\bar{c}^{b}(z)\mathcal{M}^{bc}(z)\,c^{c}(z)}} \;, \label{s1}
\end{equation}
is positive too within $\Omega$, {\it i.e.} $({\cal M}^{ab})^{-1}>0$. Following \cite{Gribov:1977wm}, the next step to achieve the factor $\mathcal{V}(\Omega)$ is that of considering the quantity
\begin{equation}
{\mathcal{G}}(k;A)=\int d^{4}x\;d^{4}y\;\,{\mathcal{G}}(x,y;A)\;e^{ik(x-y)} \;, \label{s2}
\end{equation}
where $\mathcal{G}(x,y;A)$ stands for the trace, in color space, of $({\cal M}^{ab})^{-1}$, namely
\begin{equation}
\mathcal{G}(x,y;A)= \frac{Tr ({\mathcal{M}}^{ab}(x,y;A))^{-1}}{N(N-1)} =\frac{1}{N(N-1)}
\frac{\int\mathcal{D}c^{\rm off}\mathcal{D}\bar{c}^{\rm off}\;
\bar{c}^{a}(x)c^{a}(y)\,e^{\int d^{4}z\,\bar{c}^{b}(z)\mathcal{M}^{bc}(z)\,c^{c}(z)}}{\int\mathcal{D}c^{\rm off}\mathcal{D}\bar{c}^{\rm off}\;
\,e^{\int d^{4}z\,\bar{c}^{b}(z)\mathcal{M}^{bc}(z)\,c^{c}(z)}} \;. \label{ss3}
\end{equation}
Parametrizing thus ${\mathcal{G}}(k;A)$ as in \cite{Gribov:1977wm,Capri:2005tj}
\begin{equation}
\mathcal{G}(k;A)\approx\frac{1}{k^{2}}\frac{1}{1-\sigma(k;A)}+\frac{\mathcal{J}}{k^{4}} \;, \label{s4}
\end{equation}
where $\cal J$ is constant, the no-pole condition can be stated as \cite{Gribov:1977wm}:
\begin{equation}
\sigma(k;A) < 1 \;. \label{npl}
\end{equation}
From this condition, it follows that the off-diagonal ghost propagator has no poles for any finite value of the
momentum $k$, so it cannot change sign by varying $k$. It will always remain positive, meaning that the Gribov
region $\Omega$ will be never crossed. The only allowed pole is at $k=0$, where expression \eqref{s4} becomes
singular, meaning that we are approaching the boundary $\partial \Omega$ where $({\cal M}^{ab})^{-1}$ is in fact singular,
due to the zero modes of ${\cal M}^{ab}$. Moreover, taking into account that $\sigma(k;A)$ is a decreasing function of the
momentum $k$ \cite{Gribov:1977wm}, for the final form of the no-pole condition we might take \cite{Gribov:1977wm}
\begin{equation}
\sigma(0;A) < 1 \;, \label{s5}
\end{equation}
which is very suitable for explicit calculations. Therefore, the factor $\mathcal{V}(\Omega)$ implementing the restriction of the domain of integration in the functional integral to the region $\Omega$ is
\begin{equation}
\mathcal{V}(\Omega) = \theta(1 - \sigma(0;A)) \;, \label{s6}
\end{equation}
where $\theta$ stands for the step function. Consequently, for the partition function we can write
\begin{equation}
{\cal
Z}=\int_{\Omega} d\mu
\,e^{-S_{\mathrm{YM}}}=\int d\mu
\,e^{-S_{\mathrm{YM}}}\,\theta(1-\sigma(0,A))\,.\label{Zrestricted}
\end{equation}

\subsection{Evaluation of $\mathcal{V}(\Omega)$}
\label{sigamasection}

\noindent Let us face now the characterization of the factor
$\mathcal{V}(\Omega)$. To that purpose, we start with expression
\eqref{ss3} and evaluate $\mathcal{G}(x,y;A)$ order-by-order in
perturbation theory. Performing the calculation until the third
order\footnote{In \cite{Gribov:1977wm,Capri:2005tj} the
calculations were performed only till the second order, as this is
sufficient to determine how the restriction to $\Omega$ affects the
form of the tree level propagators. Furthermore, a third order calculation
will be needed in order to characterize the horizon function,
as recently done in \cite{Gomez:2009tj} in the case of the Landau gauge.} we obtain\footnote{More details
of this lenghty calculation can be found in the Appendix \ref{wicks}.}
\begin{eqnarray}
\mathcal{G}(x,y;A)&=&\mathcal{G}_{0}(x-y)-g^{2}\int d^{4}z\,
[N\,A^{i}_{\mu}(z)A^{i}_{\mu}(z)
-A^{a}_{\mu}(z)A^{a}_{\mu}(z)]\,\mathcal{G}_{0}(z-y)\mathcal{G}_{0}(x-z)\nonumber\\
&&-g^{2}\int d^{4}z_1d^{4}z_2\,\left[4N\,A^{i}_{\mu}(z_1)A^{i}_{\nu}(z_2)
+(N-2)A^{a}_{\mu}(z_1)A^{a}_{\nu}(z_2)\right]\times\nonumber\\
&&\times\partial_{\mu}^{z_1}\mathcal{G}_{0}(x-z_1)\,
\mathcal{G}_{0}(z_{2}-y)\,\partial_{\nu}^{z_{2}}\mathcal{G}_{0}(z_1-z_2)\nonumber\\
&&+\frac{g^{3}}{N(N-1)}\int{d^{4}z_{1}d^{4}z_{2}}\,
f^{abc}f^{bde}f^{adi}A^{c}_{\mu}(z_1)A^{e}_{\nu}(z_2)A^{i}_{\nu}(z_2)\nonumber\\
&&\times\Bigl[\Bigl(\partial^{z_1}_{\mu}\mathcal{G}_0(x-z_1)\Bigr)
\mathcal{G}_0(z_2-y)\mathcal{G}_0(z_1-z_2)+\mathcal{G}_0(x-z_2)\,\mathcal{G}_0(z_1-y)
\,\partial_{\mu}^{z_1}\mathcal{G}_0(z_2-z_1)\Bigr]\nonumber\\
&&-\frac{g^{3}}{N(N-1)}\int{d^{4}z_{1}d^{4}z_{2}d^{4}z_{3}}\,
\Bigl(8f^{abi}f^{bcj}f^{cak}A^{i}_{\mu}(z_1)A^{j}_{\nu}(z_2)A^{k}_{\s}(z_3)\nonumber\\
&&+4f^{abi}f^{bdj}f^{dac}A^{i}_{\mu}(z_1)A^{j}_{\nu}(z_2)A^{c}_{\s}(z_3)
+4f^{abi}f^{bdc}f^{daj}A^{i}_{\mu}(z_1)A^{c}_{\nu}(z_2)A^{j}_{\s}(z_3)\nonumber\\
&&+4f^{abc}f^{bdi}f^{daj}A^{c}_{\mu}(z_1)A^{i}_{\nu}(z_2)A^{j}_{\s}(z_3)
+2f^{abi}f^{bdc}f^{dae}A^{i}_{\mu}(z_1)A^{c}_{\nu}(z_2)A^{e}_{\s}(z_3)\nonumber\\
&&+2f^{abc}f^{bdi}f^{dae}A^{c}_{\mu}(z_1)A^{i}_{\nu}(z_2)A^{e}_{\s}(z_3)
+2f^{abc}f^{bde}f^{dai}A^{c}_{\mu}(z_1)A^{e}_{\nu}(z_2)A^{i}_{\s}(z_3)\nonumber\\
&&+f^{abc}f^{bde}f^{daf}A^{c}_{\mu}(z_1)A^{e}_{\nu}(z_2)A^{f}_{\s}(z_3)\Bigr)\nonumber\\
&&\times\Bigl[\Bigl(\partial_{\mu}^{z_1}\mathcal{G}_0(x-z_1)\Bigr)
\mathcal{G}_0(z_2-y) \Bigl(\partial_{\s}^{z_3}\mathcal{G}_0(z_1-z_3)\Bigr)
\partial_{\nu}^{z_2}\mathcal{G}_0(z_3-z_2)\Bigr]\;,
\end{eqnarray}
where
\begin{equation}
\mathcal{G}_{0}(x-y)=\int\frac{d^{4}q}{(2\pi)^{4}}\,\frac{1}{q^{2}}\,e^{-iq(x-y)}\,.\label{G0}
\end{equation}
In momentum space we can write
\begin{eqnarray}
\mathcal{G}(k;A)&=&\int d^{4}xd^{4}y\,\mathcal{G}(x,y;A)e^{ik(x-y)}\nonumber\\
&=&\frac{1}{k^{2}}-\frac{g^{2}}{k^{4}}\int\frac{d^{4}q}{(2\pi)^{4}}\biggl(\frac{A^{i}_{\mu}(q)A^{i}_{\mu}(-q)}{N-1}
-\frac{A^{a}_{\mu}(q)A^{a}_{\mu}(-q)}{N(N-1)}\biggr)\nonumber\\
&&+g^{2}\int\frac{d^{4}q}{(2\pi)^{4}}\frac{k_{\mu}(k-q)_{\nu}}{k^{4}(k-q)^{2}}\biggl(\frac{4}{(N-1)}
A^{i}_{\mu}(q)A^{i}_{\nu}(-q) +\frac{N-2}{N(N-1)}
A^{a}_{\mu}(q)A^{a}_{\nu}(-q)\biggr)\nonumber\\
&&+\frac{g^{3}}{N(N-1)}\int\frac{d^{4}p}{(2\pi)^{4}}\frac{d^{4}q}{(2\pi)^{4}}
\left(\frac{ik_{\mu}}{k^{4}(p+k)^{2}}+\frac{i(k-p)_{\mu}}{k^{4}(k-p)^{2}}\right)
f^{abc}f^{bde}f^{adi}A^{c}_{\mu}(-p)A^{e}_{\nu}(-q)A^{i}_{\nu}(p+q)\nonumber\\
&&+\frac{g^{3}}{N(N-1)}\int\frac{d^{4}p}{(2\pi)^{4}}\frac{d^{4}q}{(2\pi)^{4}}
\frac{ik_{\s}(k-p)_{\mu}(k-p-q)_{\nu}}{k^{4}(k-p)^{2}(k-p-q)^{2}}
\Bigl(8f^{abi}f^{bcj}f^{cak}A^{i}_{\mu}(-p)A^{j}_{\nu}(-q)A^{k}_{\s}(p+q)\nonumber\\
&&+4f^{abi}f^{bdj}f^{dac}A^{i}_{\mu}(-p)A^{j}_{\nu}(-q)A^{c}_{\s}(p+q)
+4f^{abi}f^{bdc}f^{daj}A^{i}_{\mu}(-p)A^{c}_{\nu}(-q)A^{j}_{\s}(p+q)\nonumber\\
&&+4f^{abc}f^{bdi}f^{daj}A^{c}_{\mu}(-p)A^{i}_{\nu}(-q)A^{j}_{\s}(p+q)
+2f^{abi}f^{bdc}f^{dae}A^{i}_{\mu}(-p)A^{c}_{\nu}(-q)A^{e}_{\s}(p+q)\nonumber\\
&&+2f^{abc}f^{bdi}f^{dae}A^{c}_{\mu}(-p)A^{i}_{\nu}(-q)A^{e}_{\s}(p+q)
+2f^{abc}f^{bde}f^{dai}A^{c}_{\mu}(-p)A^{e}_{\nu}(-q)A^{i}_{\s}(p+q)\nonumber\\
&&+f^{abc}f^{bde}f^{daf}A^{c}_{\mu}(-p)A^{e}_{\nu}(-q)A^{f}_{\s}(p+q)\Bigr)\,.
\end{eqnarray}
According to eq.\eqref{s4}, the expression above can be rewritten as
\begin{equation}
\mathcal{G}(k;A)=\frac{1}{k^{2}}(1+\sigma(k;A))+\frac{\mathcal{J}}{k^{4}}\,,
\end{equation}
where the form factor $\s(k;A)$ and $\mathcal{J}$ are given by
\begin{eqnarray}
\sigma(k;A)&=&g^{2}\int\frac{d^{4}q}{(2\pi)^{4}}\frac{k_{\mu}(k-q)_{\nu}}{k^{2}(k-q)^{2}}\biggl(\frac{4}{(N-1)}
A^{i}_{\mu}(q)A^{i}_{\nu}(-q) +\frac{N-2}{N(N-1)}
A^{a}_{\mu}(q)A^{a}_{\nu}(-q)\biggr)\nonumber\\
&&+\frac{g^{3}}{N(N-1)}\int\frac{d^{4}p}{(2\pi)^{4}}\frac{d^{4}q}{(2\pi)^{4}}
\left(\frac{ik_{\mu}}{k^{2}(p+k)^{2}}+\frac{i(k-p)_{\mu}}{k^{2}(k-p)^{2}}\right)
f^{abc}f^{bde}f^{adi}A^{c}_{\mu}(-p)A^{e}_{\nu}(-q)A^{i}_{\nu}(p+q)\nonumber\\
&&+\frac{g^{3}}{N(N-1)}\int\frac{d^{4}p}{(2\pi)^{4}}\frac{d^{4}q}{(2\pi)^{4}}
\frac{ik_{\s}(k-p)_{\mu}(k-p-q)_{\nu}}{k^{2}(k-p)^{2}(k-p-q)^{2}}
\Bigl(8f^{abi}f^{bcj}f^{cak}A^{i}_{\mu}(-p)A^{j}_{\nu}(-q)A^{k}_{\s}(p+q)\nonumber\\
&&+4f^{abi}f^{bdj}f^{dac}A^{i}_{\mu}(-p)A^{j}_{\nu}(-q)A^{c}_{\s}(p+q)
+4f^{abi}f^{bdc}f^{daj}A^{i}_{\mu}(-p)A^{c}_{\nu}(-q)A^{j}_{\s}(p+q)\nonumber\\
&&+4f^{abc}f^{bdi}f^{daj}A^{c}_{\mu}(-p)A^{i}_{\nu}(-q)A^{j}_{\s}(p+q)
+2f^{abi}f^{bdc}f^{dae}A^{i}_{\mu}(-p)A^{c}_{\nu}(-q)A^{e}_{\s}(p+q)\nonumber\\
&&+2f^{abc}f^{bdi}f^{dae}A^{c}_{\mu}(-p)A^{i}_{\nu}(-q)A^{e}_{\s}(p+q)
+2f^{abc}f^{bde}f^{dai}A^{c}_{\mu}(-p)A^{e}_{\nu}(-q)A^{i}_{\s}(p+q)\nonumber\\
&&+f^{abc}f^{bde}f^{daf}A^{c}_{\mu}(-p)A^{e}_{\nu}(-q)A^{f}_{\s}(p+q)\Bigr)\,,\label{first_sigma_expression}
\end{eqnarray}
\begin{equation}
\mathcal{J}=-g^{2}\int\frac{d^{4}q}{(2\pi)^{4}}\biggl(\frac{A^{i}_{\mu}(q)A^{i}_{\mu}(-q)}{N-1}
-\frac{A^{a}_{\mu}(q)A^{a}_{\mu}(-q)}{N(N-1)}\biggr)\,.
\end{equation}
Notice that $\mathcal{J}$ is independent from the external momentum $k$.
Also, expression \eqref{first_sigma_expression} can be simplified
by making use of the maximal Abelian gauge
conditions in momentum space \eqref{MAGmomentum}, namely
\begin{equation}
q_{\mu}A^{i}_{\mu}=0\,,\qquad
q_{\mu}A^{a}_{\mu}(q)=igf^{abi}\int\frac{d^{4}p}{(2\pi)^{4}}\,A^{i}_{\mu}(q-p)A^{b}_{\mu}(p)\,.
\end{equation}
Then, for $\sigma(k;A)$ we obtain
\begin{eqnarray}
\sigma(k;A)&=&g^{2}\frac{k_{\mu}k_{\nu}}{k^{2}}\int\frac{d^{4}q}{(2\pi)^{4}}\frac{1}{(k-q)^{2}}\biggl(\frac{4}{(N-1)}
A^{i}_{\mu}(q)A^{i}_{\nu}(-q) +\frac{N-2}{N(N-1)}A^{a}_{\mu}(q)A^{a}_{\nu}(-q)\biggr)\nonumber\\
&&+\frac{ig^{3}(N-2)}{N(N-1)}\frac{k_{\mu}}{k^{2}}\int\frac{d^{4}p}{(2\pi)^{4}}\frac{d^{4}q}{(2\pi)^{4}}
\frac{1}{(k-q)^{2}}f^{abi}A^{a}_{\mu}(q)A^{i}_{\nu}(p-q)A^{b}_{\nu}(-p)\nonumber\\
&&+\frac{ig^{3}}{N(N-1)}\frac{k_{\mu}}{k^{2}}\int\frac{d^{4}p}{(2\pi)^{4}}\frac{d^{4}q}{(2\pi)^{4}}
\left(\frac{1}{(p+k)^{2}}+\frac{1}{(k-p)^{2}}\right)
f^{abc}f^{bde}f^{adi}A^{c}_{\mu}(-p)A^{e}_{\nu}(-q)A^{i}_{\nu}(p+q)\nonumber\\
&&+\frac{ig^{3}}{N(N-1)}\frac{k_{\s}k_{\mu}}{k^{2}}\int\frac{d^{4}p}{(2\pi)^{4}}\frac{d^{4}q}{(2\pi)^{4}}
\frac{(k-p)_{\nu}}{(k-p)^{2}(k-p-q)^{2}}
\Bigl(8f^{abi}f^{bcj}f^{cak}A^{i}_{\mu}(-p)A^{j}_{\nu}(-q)A^{k}_{\s}(p+q)\nonumber\\
&&+4f^{abi}f^{bdj}f^{dac}A^{i}_{\mu}(-p)A^{j}_{\nu}(-q)A^{c}_{\s}(p+q)
+4f^{abi}f^{bdc}f^{daj}A^{i}_{\mu}(-p)A^{c}_{\nu}(-q)A^{j}_{\s}(p+q)\nonumber\\
&&+4f^{abc}f^{bdi}f^{daj}A^{c}_{\mu}(-p)A^{i}_{\nu}(-q)A^{j}_{\s}(p+q)
+2f^{abi}f^{bdc}f^{dae}A^{i}_{\mu}(-p)A^{c}_{\nu}(-q)A^{e}_{\s}(p+q)\nonumber\\
&&+2f^{abc}f^{bdi}f^{dae}A^{c}_{\mu}(-p)A^{i}_{\nu}(-q)A^{e}_{\s}(p+q)
+2f^{abc}f^{bde}f^{dai}A^{c}_{\mu}(-p)A^{e}_{\nu}(-q)A^{i}_{\s}(p+q)\nonumber\\
&&+f^{abc}f^{bde}f^{daf}A^{c}_{\mu}(-p)A^{e}_{\nu}(-q)A^{f}_{\s}(p+q)\Bigr)
+O(g^{4})\,.\label{sigmaqq}
\end{eqnarray}
Using the relation
\begin{equation}
f^{abc}f^{bde}f^{adi}=-\frac{N-2}{2}f^{cei} \;, \label{fs1}
\end{equation}
and performing the change of variables $p\to-q$ and $q\to p$ in the third line of the r.h.s.
of expression \eqref{sigmaqq}, we get
\begin{equation}
\s(k;A)=g^{2}\frac{k_{\mu}k_{\nu}}{k^{2}}\,I^{(1)}_{\mu\nu}(k)
+\frac{ig^{3}(N-2)}{2N(N-1)}\frac{k_{\mu}}{k^{2}}\,I^{(2)}_{\mu}(k)
+\frac{ig^{3}}{N(N-1)}\biggl(\frac{k_{\mu}k_{\nu}k_{\s}}{k^{2}}\,I^{(3)}_{\mu\nu\s}(k)
-\frac{k_{\mu}k_{\s}}{k^{2}}\,I^{(4)}_{\mu\s}(k)\biggr)+O(g^{4})\,,
\end{equation}
where
\begin{eqnarray}
I^{(1)}_{\mu\nu}(k)&=&\int\frac{d^{4}q}{(2\pi)^{4}}\frac{1}{(k-q)^{2}}\biggl(\frac{4}{(N-1)}
A^{i}_{\mu}(q)A^{i}_{\nu}(-q) +\frac{N-2}{N(N-1)}A^{a}_{\mu}(q)A^{a}_{\nu}(-q)\biggr)\,,\label{I1}\\
I^{(2)}_{\mu}(k)&=&\int\frac{d^{4}p}{(2\pi)^{4}}\frac{d^{4}q}{(2\pi)^{4}}
\left(\frac{1}{(k-q)^{2}}-\frac{1}{(k+q)^{2}}\right)f^{abi}A^{a}_{\mu}(q)A^{i}_{\nu}(p-q)A^{b}_{\nu}(-p)\,,\\
I^{(3)}_{\mu\nu\s}(k)&=&\int\frac{d^{4}p}{(2\pi)^{4}}\frac{d^{4}q}{(2\pi)^{4}}
\frac{1}{(k-p)^{2}(k-p-q)^{2}}
\Bigl(8f^{abi}f^{bcj}f^{cak}A^{i}_{\mu}(-p)A^{j}_{\nu}(-q)A^{k}_{\s}(p+q)\nonumber\\
&&+4f^{abi}f^{bdj}f^{dac}A^{i}_{\mu}(-p)A^{j}_{\nu}(-q)A^{c}_{\s}(p+q)
+4f^{abi}f^{bdc}f^{daj}A^{i}_{\mu}(-p)A^{c}_{\nu}(-q)A^{j}_{\s}(p+q)\nonumber\\
&&+4f^{abc}f^{bdi}f^{daj}A^{c}_{\mu}(-p)A^{i}_{\nu}(-q)A^{j}_{\s}(p+q)
+2f^{abi}f^{bdc}f^{dae}A^{i}_{\mu}(-p)A^{c}_{\nu}(-q)A^{e}_{\s}(p+q)\nonumber\\
&&+2f^{abc}f^{bdi}f^{dae}A^{c}_{\mu}(-p)A^{i}_{\nu}(-q)A^{e}_{\s}(p+q)
+2f^{abc}f^{bde}f^{dai}A^{c}_{\mu}(-p)A^{e}_{\nu}(-q)A^{i}_{\s}(p+q)\nonumber\\
&&+f^{abc}f^{bde}f^{daf}A^{c}_{\mu}(-p)A^{e}_{\nu}(-q)A^{f}_{\s}(p+q)\Bigr)\,,\\\cr
I^{(4)}_{\mu\s}(k)&=&\int\frac{d^{4}p}{(2\pi)^{4}}\frac{d^{4}q}{(2\pi)^{4}}
\frac{p_{\nu}}{(k-p)^{2}(k-p-q)^{2}}
\Bigl(8f^{abi}f^{bcj}f^{cak}A^{i}_{\mu}(-p)A^{j}_{\nu}(-q)A^{k}_{\s}(p+q)\nonumber\\
&&+4f^{abi}f^{bdj}f^{dac}A^{i}_{\mu}(-p)A^{j}_{\nu}(-q)A^{c}_{\s}(p+q)
+4f^{abi}f^{bdc}f^{daj}A^{i}_{\mu}(-p)A^{c}_{\nu}(-q)A^{j}_{\s}(p+q)\nonumber\\
&&+4f^{abc}f^{bdi}f^{daj}A^{c}_{\mu}(-p)A^{i}_{\nu}(-q)A^{j}_{\s}(p+q)
+2f^{abi}f^{bdc}f^{dae}A^{i}_{\mu}(-p)A^{c}_{\nu}(-q)A^{e}_{\s}(p+q)\nonumber\\
&&+2f^{abc}f^{bdi}f^{dae}A^{c}_{\mu}(-p)A^{i}_{\nu}(-q)A^{e}_{\s}(p+q)
+2f^{abc}f^{bde}f^{dai}A^{c}_{\mu}(-p)A^{e}_{\nu}(-q)A^{i}_{\s}(p+q)\nonumber\\
&&+f^{abc}f^{bde}f^{daf}A^{c}_{\mu}(-p)A^{e}_{\nu}(-q)A^{f}_{\s}(p+q)\Bigr)\,.\label{I4}
\end{eqnarray}
In order to be able to evaluate the form factor $\sigma(k;A)$ at zero momentum, eq.\eqref{s5},
\begin{equation}
\s(0;A)\equiv\lim_{k\to0}\s(k;A)\,.
\end{equation}
we need to analyze the limit $k\to 0$ of the integrals
\eqref{I1}\,--\,\eqref{I4}. As these integrals depend explicitly  on the Fourier components of the gauge fields, $A^i_\mu(q), A^a_\mu(q)$, the existence of the zero momentum limit, $k\to 0$, of expressions \eqref{I1}\,--\,\eqref{I4} relies on the infrared behavior of the fields $A^i_\mu(q)$ and  $A^a_\mu(q)$. For example, considering \eqref{I1}, it is apparent that the zero momentum integral
\begin{equation}
I^{(1)}_{\mu\nu}(0)= \int\frac{d^{4}q}{(2\pi)^{4}}\frac{1}{q^{2}}\biggl(\frac{4}{(N-1)}
A^{i}_{\mu}(q)A^{i}_{\nu}(-q) +\frac{N-2}{N(N-1)}A^{a}_{\mu}(q)A^{a}_{\nu}(-q)\biggr)\,,\label{I1f}
\end{equation}
does exist provided the quantities $A^{i}_{\mu}(q)A^{i}_{\mu}(-q)$, $A^{a}_{\mu}(q)A^{a}_{\mu}(-q)$, are not singular at $q \approx 0$, {\it i.e.}
\begin{eqnarray}
A^{i}_{\mu}(q)A^{i}_{\mu}(-q)\Big|_{q^2 \approx 0} &{ \approx}& (q^2)^\alpha \;, \qquad \alpha > -2 \nonumber \;, \\
A^{a}_{\mu}(q)A^{a}_{\mu}(-q)\Big|_{q^2 \approx 0} & \approx & (q^2)^\beta \;, \qquad \beta > -2 \;, \label{ab}
\end{eqnarray}
so that expression \eqref{I1f} is integrable at $q \approx 0$. At the present stage, a formal proof of the infrared behavior expressed by \eqref{ab} cannot be given. An explicit check of the validity of \eqref{ab} can be done only after the evaluation of the gluon propagators. In other words, as done in the original paper by Gribov \cite{Gribov:1977wm},  \eqref{ab} is assumed as working hypothesis in order to ensure the existence of the zero momentum limit of expressions \eqref{I1}\,--\,\eqref{I4}. Afterwards, this assumption has to be checked out by looking at the behavior of the diagonal and off-diagonal gluon propagators at the origin. Nevertheless, we can provide a physical justification for assuming \eqref{ab}, which follows from the implementation of the restriction to the Gribov region $\Omega$. As also observed in \cite{Gribov:1977wm} in the case of the Landau gauge, the restriction to the region  $\Omega$ amounts to put a boundary in field space, namely the Gribov horizon, which manifests itself through the appearance of a dynamical mass parameter $\gamma$, which is known as the Gribov mass \cite{Gribov:1977wm}. As we shall see in the next sections, such a parameter will provide a natural infrared scale for the field configurations, resulting in a gluon propagator which is not singular at the origin, thus justifying the starting working hypothesis \eqref{ab}. In fact, looking at the expressions for the diagonal and off-diagonal propagators given in eqs.  \eqref{offdgln}, \eqref{dgln}, one sees that they are not singular at the origin. \\\\The integral $I^{(2)}_{\mu}(k)$ is
automatically vanishing when $k=0$; the integrals $I^{(1)}_{\mu\nu}(0)$
and $I^{(4)}_{\mu\nu}(0)$ must be proportional to $\d_{\mu\nu}$ in the limit
$k\to 0$, since $\d_{\mu\nu}$  is the only Lorentz invariant second rank
tensor available; and finally $I^{(3)}_{\mu\nu\s}(k)$ must vanish when
$k\to0$, as in four dimensions there is no Lorentz invariant third
order rank tensor. Summarizing, we have
\begin{equation}
\begin{tabular}{lcll}
$I^{(1)}_{\mu\nu}(0)$&=&$\displaystyle\frac{I^{(1)}_{\lambda\lambda}(0)}{4}\,\d_{\mu\nu}\,,$&(by Lorentz covariance)\,,\vspace{3pt}\cr
$I^{(2)}_{\mu}(0)$&=&$0\,,$&(automatically)\,,\vspace{3pt}\cr
$I^{(3)}_{\mu\nu\s}(0)$&=&$0\,,$&(by Lorentz covariance)\,,\vspace{3pt}\cr
$I^{(4)}_{\mu\nu}(0)$&=&$\displaystyle\frac{I^{(4)}_{\lambda\lambda}(0)}{4}\,\d_{\mu\nu}\,,$&(by Lorentz covariance)\,.
\end{tabular}
\end{equation}
Thus, taking the limit $k\to0$ of  $\s(k;A)$, for the form factor $\s(0;A)$ we get
\begin{eqnarray}
\s(0;A)&\equiv&\lim_{k\to0}\s(k;A)\nonumber\\
&=&g^{2}\,\frac{I^{(1)}_{\lambda\lambda}(0)}{4}
-\frac{ig^{3}}{N(N-1)}
\,\frac{I^{(4)}_{\lambda\lambda}(0)}{4}+O(g^{4})\nonumber\\
&=&g^{2}\int\frac{d^{4}q}{(2\pi)^{4}}\biggl(\frac{1}{(N-1)}
\frac{A^{i}_{\lambda}(q)A^{i}_{\lambda}(-q)}{q^{2}} +\frac{N-2}{4N(N-1)}
\frac{A^{a}_{\lambda}(q)A^{a}_{\lambda}(-q)}{q^{2}}\biggr)\nonumber\\
&&-\frac{ig^{3}}{N(N-1)}\int\frac{d^{4}p}{(2\pi)^{4}}\frac{d^{4}q}{(2\pi)^{4}}
\frac{p_{\mu}}{p^{2}(p+q)^{2}}
\biggl(2f^{abi}f^{bcj}f^{cak}A^{i}_{\lambda}(-p)A^{j}_{\mu}(-q)A^{k}_{\lambda}(p+q)\nonumber\\
&&+f^{abi}f^{bdj}f^{dac}A^{i}_{\lambda}(-p)A^{j}_{\mu}(-q)A^{c}_{\lambda}(p+q)
+f^{abi}f^{bdc}f^{daj}A^{i}_{\lambda}(-p)A^{c}_{\mu}(-q)A^{j}_{\lambda}(p+q)\nonumber\\
&&+f^{abc}f^{bdi}f^{daj}A^{c}_{\lambda}(-p)A^{i}_{\mu}(-q)A^{j}_{\lambda}(p+q)
+\frac{1}{2}f^{abi}f^{bdc}f^{dae}A^{i}_{\lambda}(-p)A^{c}_{\mu}(-q)A^{e}_{\lambda}(p+q)\nonumber\\
&&+\frac{1}{2}f^{abc}f^{bdi}f^{dae}A^{c}_{\lambda}(-p)A^{i}_{\mu}(-q)A^{e}_{\lambda}(p+q)
+\frac{1}{2}f^{abc}f^{bde}f^{dai}A^{c}_{\lambda}(-p)A^{e}_{\mu}(-q)A^{i}_{\lambda}(p+q)\nonumber\\
&&+\frac{1}{4}f^{abc}f^{bde}f^{daf}A^{c}_{\lambda}(-p)A^{e}_{\mu}(-q)A^{f}_{\lambda}(p+q)\biggr)\,.
\label{sigmafinal}
\end{eqnarray}
As a useful check, we notice here that, for $N=2$,  the previous results
obtained in \cite{Capri:2005tj} are recovered.
\subsection{Gribov's quadratic approximation and the gap equation}
Having characterized the factor $\sigma(0;A)$, eq.\eqref{sigmafinal}, we start to analyze the consequences on the theory of the restriction of the domain of integration to the Gribov region, as stated by eq.\eqref{Zrestricted}. As we shall see, this will amount to modify the starting gauge field theory by the addition of a nonperturbative term, usually referred as to the horizon function  \cite{Zwanziger:1989mf,Zwanziger:1992qr,Capri:2006cz}. To illustrate this point, we discuss first the form factor $\sigma(0;A)$ in the so called quadratic approximantion \cite{Gribov:1977wm}, {\it i.e.} we shall first consider $\sigma(0;A)$ till the second order in the gauge fields. Thus
\begin{eqnarray}
\s(0;A)
&=&g^{2}\frac{1}{V}\sum_{q}\biggl(\frac{1}{(N-1)}
\frac{A^{i}_{\lambda}(q)A^{i}_{\lambda}(-q)}{q^{2}} +\frac{N-2}{4N(N-1)}
\frac{A^{a}_{\lambda}(q)A^{a}_{\lambda}(-q)}{q^{2}}\biggr)\, ,
\end{eqnarray}
where, according to \cite{Gribov:1977wm}, a finite volume $V$ has been considered. As in the case of $SU(2)$
\cite{Capri:2005tj}, the quadratic approximation will enable us to obtain the first term of the horizon function, which will play an important role for its all order extension. \\\\Using the integral representation of the step function,
\begin{equation}
\theta(x)=\int^{\epsilon+i\infty}_{\epsilon-i\infty}\frac{d\zeta}{2\pi i\beta}\,e^{\zeta x}\,,
\end{equation}
for the  partition function \eqref{Zrestricted} we get the following expression
\begin{equation}
\mathcal{Z}=\int\frac{d\zeta}{2\pi i\zeta}d\mu
\,e^{-S_{\mathrm{YM}}+\zeta(1-\s(0;A))}\,.
\end{equation}
Moreover, in the quadratic approximation, $\mathcal{Z} \rightarrow \mathcal{Z}_{\mbox{\tiny quad}}$:
\begin{eqnarray}
\mathcal{Z}_{\mbox{\tiny quad}}&=&\lim_{{\alpha\to0}\atop{\beta\to0}}
\int\frac{d\zeta}{2\pi i\zeta} e^{\zeta}
\int\mathcal{D}A^{\mbox{\tiny off}}\mathcal{D}A^{\mbox{\tiny diag}}\,e^{-\frac{1}{2}\sum_{k}
\left(A^{a}_{\mu}(k)\mathcal{P}^{ab}_{\mu\nu}(k;\zeta,\alpha)A^{b}_{\nu}(-k)
+A^{i}_{\mu}(k)\mathcal{Q}^{ij}_{\mu\nu}(k;\zeta,\beta)A^{j}_{\nu}(-k)\right)}\,,\nonumber\\
\mathcal{P}^{ab}_{\mu\nu}(k;\zeta,\alpha)&=&\biggl[\biggl(k^{2}+\frac{g^{2}(N-2)\zeta}{2N(N-1)Vk^{2}}\biggr)\d_{\mu\nu}
-\biggl(1-\frac{1}{\alpha}\biggr)k_{\mu}k_{\nu}\biggr]\d^{ab}\,,\nonumber\\
\mathcal{Q}^{ij}_{\mu\nu}(k;\zeta,\beta)&=&\biggl[\biggl(k^{2}+\frac{2g^{2}\zeta}{(N-1)Vk^{2}}\biggr)\d_{\mu\nu}
-\biggl(1-\frac{1}{\beta}\biggr)k_{\mu}k_{\nu}\biggr]\d^{ij}\,,
\label{Zquad}
\end{eqnarray}
where use has been made of the relations:
\begin{eqnarray}
\d(D\cdot A^{\mbox{\tiny off}})&\varpropto&\lim_{\alpha\to0}e^{-\frac{1}{2\alpha}\int d^{4}x\,(D^{ab}_{\mu}A^{b}_{\mu})^{2}}\,,\nonumber\\
\d(\partial\cdot A^{\mbox{\tiny diag}})&\varpropto&\lim_{\beta\to0}e^{-\frac{1}{2\beta}\int d^{4}x\,(\partial_{\mu}A^{i}_{\mu})^{2}}\,,\nonumber\\
A^{a,i}_{\mu}(x)&=&\frac{1}{V^{1/2}}\sum_{k}A^{a,i}_{\mu}(k)e^{-ikx}\,,\nonumber\\
\d_{k,k'}&=&\frac{1}{V}\int_V d^{4}x\,e^{ix(k-k')}\,.  \label{gps}
\end{eqnarray}
Integrating over the gauge fields gives
\begin{eqnarray}
\mathcal{Z}_{\mbox{\tiny quad}}&=&\lim_{{\alpha\to0}\atop{\beta\to0}}\int\frac{d \zeta}{2\pi i}\,e^{f(\zeta;\alpha,\beta)}\,,\nonumber\\
f(\zeta;\alpha,\beta)&=&\zeta-\ln\zeta-\frac{1}{2}\ln\det\mathcal{P}^{ab}_{\mu\nu}(k;\zeta,\alpha)
-\frac{1}{2}\ln\det\mathcal{Q}^{ij}_{\mu\nu}(k;\zeta,\beta)\nonumber\\
&=&\zeta-\ln\zeta-\frac{3N(N-1)}{2}\sum_{k}\left[\ln\biggl(k^{2}+\frac{g^{2}(N-2)\zeta}{2N(N-1)Vk^{2}}\biggr)
+\frac{1}{3}\ln\biggl(\frac{k^{2}}{\alpha}+\frac{g^{2}(N-2)\zeta}{2N(N-1)Vk^{2}}\biggr)\right]\nonumber\\
&&-\frac{3(N-1)}{2}\sum_{k}\left[\ln\biggl(k^{2}+\frac{2g^{2}\zeta}{(N-1)Vk^{2}}\biggr)
+\frac{1}{3}\ln\biggl(\frac{k^{2}}{\beta}+\frac{2g^{2}\zeta}{(N-1)Vk^{2}}\biggr)\right]\,.
\end{eqnarray}
The integral over $\zeta$ can be evaluated in a saddle point
approximation \cite{Gribov:1977wm}
\begin{equation}
\mathcal{Z}_{\mbox{\tiny quad}}\approx\lim_{{\alpha\to0}\atop{\beta\to0}}\,e^{f(\zeta^{\star};\alpha,\beta)}\,,
\qquad\frac{df(\zeta)}{d\zeta}\biggl|_{\zeta=\zeta^{\star}}=0\,. \label{sp}
\end{equation}
Condition \eqref{sp} gives us the following relation\footnote{Notice that the gauge parameters $\alpha, \beta$ have been set to zero, see eq.\eqref{gps}.}
\begin{equation}
1-\frac{1}{\zeta^{\star}}-\frac{1}{V}\sum_{k}\frac{3g^{2}}{k^{4}+\frac{2g^{2}\zeta^{\star}}{(N-1)V}}
-\frac{1}{V}\sum_{k}\frac{3(N-2)g^{2}/4}{k^{4}+\frac{(N-2)g^{2}\zeta^{\star}}{2N(N-1)V}}=0
\end{equation}
As in  \cite{Gribov:1977wm}, in order to take the thermodynamic limit, $V\rightarrow\infty$, we introduce the so called Gribov parameter $\gamma$
\begin{equation}
\gamma^{4}= \lim_{V\to\infty} \frac{\zeta^{\star}}{2V}\,. \label{gbpm}
\end{equation}
Therefore, in the infinite volume limit, $V\to\infty$, we get
\begin{equation}
\int\frac{d^{4}k}{(2\pi)^{4}}\frac{3g^{2}}{k^{4}+\frac{4g^{2}\gamma^{4}}{N-1}}
+\int\frac{d^{4}k}{(2\pi)^{4}}\frac{3(N-2)g^{2}/4}{k^{4}+\frac{(N-2)g^{2}\gamma^{4}}{N(N-1)}}=1\,,
\label{gapeq}
\end{equation}
where we have neglected the term $1/\zeta^{\star}$. As observed in  \cite{Gribov:1977wm}, the Gribov parameter is not an independent parameter of the theory. It turns out to be determined at the quantum level in a self-consistent way as a function of the coupling constant. In fact, eq.\eqref{gapeq} is nothing but the gap equation enabling us to express $\gamma$ as a function of the coupling constant $g$ and of the invariant scale $\Lambda_{QCD}$, see ref.\cite{Dudal:2008sp}. \\\\Coming back to the partition function in the saddle
point approximation, taking the form factor $\s(0;k)$ in the quadratic  approximation,  and performing  the
thermodynamic limit, we can write
\begin{equation}
\mathcal{Z}=\mathcal{N}'\int d{\mu}\,e^{-(S_{\mathrm{YM}}+S_{\mathrm{G}})}\,. \label{add}
\end{equation}
Here, $\mathcal{N}'$ is an irrelevant normalization factor and
$S_{\mathrm{G}}$ is a nonlocal term, given by
\begin{eqnarray}
S_{\mathrm{G}}&=&\lim_{V\to\infty}\zeta^{\star}\s(0;A)\nonumber\\
&=&g^{2}\gamma^{4}\int{d^{4}xd^{4}y}\,\left(\frac{2}{N-1}\,
A^{i}_{\mu}(x)\mathcal{G}_{0}(x-y)A^{i}_{\mu}(y)
+\frac{N-2}{2N(N-1)}\,
A^{a}_{\mu}(x)\mathcal{G}_{0}(x-y)A^{a}_{\mu}(y)\right) \nonumber \\
&=&g^{2}\gamma^{4}\int{d^{4}x}\,\left(\frac{2}{N-1}\,
A^{i}_{\mu}\left( \frac{1}{-\partial^2} \right)A^{i}_{\mu}
+\frac{N-2}{2N(N-1)}\,
A^{a}_{\mu}\left( \frac{1}{-\partial^2} \right) A^{a}_{\mu} \right) \;. \label{GribovTerm}
\end{eqnarray}
Equation \eqref{add} shows that the implementation of the restriction of the domain of integration to the Gribov region $\Omega$ can be achieved by adding to the Yang-Mills action a nonlocal term, given in eq.\eqref{GribovTerm}. Expression \eqref{GribovTerm} represents only the first term of what is usually called the horizon function, whose final expression will be discussed in details in the
next section.
\section{Characterization of  the horizon function}

\subsection{A few words on the horizon function in the landau gauge and in $SU(2)$ maximal Abelian  gauge}
Before entering into the technical details of the determination of the horizon function for $SU(N)$, it is useful to give a short survey of what has been already achieved in other cases. In the Landau gauge, it has been shown by  Zwanziger  \cite{Zwanziger:1989mf,Zwanziger:1992qr}  that the horizon function
implementing the restriction to the Gribov region $\Omega$\footnote{We remind here that, analogously to the case of the maximal Abelian gauge, the Gribov region in the Landau gauge is defined through the positivity of the Faddeev-Popov operator, {\it i.e.}
\begin{equation}
\Omega = \{ \; A^A_\mu;\; \partial_\mu A^A_{\mu}=0;\; -\partial_{\mu}(\partial_{\mu}\d^{AB}-gf^{ABC}A^{C}_{\mu})>0 \;\} \ . \label{landaug}
\end{equation}
} is
\begin{equation}
H_{\mathrm{Landau}}=g^{2}\int{d^{4}x\;d^{4}y} \;D_{\mu}^{AE}(x)
\left[(-\partial\cdot D)^{-1}\right]^{AB}(x,y)D_{\mu}^{BE}(y)\,,\label{Hlandau}
\end{equation}
where $[(-\partial\cdot D)^{-1}]^{AB}(x,y)$ is the inverse of the
Faddeev-Popov operator
\begin{equation}
-(\partial\cdot D)^{AB}=-\partial_{\mu}D^{AB}_{\mu}=-\partial_{\mu}(\partial_{\mu}\d^{AB}-gf^{ABC}A^{C}_{\mu})\,.
\end{equation}
Accordingly, for the partition function implementing the restriction of the domain of integration to the Gribov region, we write
\begin{equation}
\mathcal{Z}_{\mathrm{Landau}}=\int_{\Omega}d\mu\,e^{-S_{\mathrm{YM}}}
=\int d\mu\,e^{-(S_{\mathrm{YM}}+\gamma^{4}H_{\mathrm{Landau}})}\,,
\end{equation}
where, to the first order, the Gribov parameter  $\gamma$ is defined by the gap equation  \cite{Zwanziger:1989mf,Zwanziger:1992qr}
\begin{equation}
1=\frac{3Ng^{2}}{4}\int\frac{d^{4}k}{(2\pi)^{4}}\,\frac{1}{k^{4}+2Ng^{2}\gamma^{4}}\,.
\end{equation}
One should observe that expression \eqref{Hlandau} is nonlocal. Willing to give a quantum field theory meaning to such an expression, one should be able to cast it into a local and renormalizable form. We mention here that a consistent, local and renormalizable framework for expression \eqref{Hlandau} is in fact available, see \cite{Dudal:2010fq} for an updated discussion of this nontrivial issue.  Moreover, it is worth to point out  that Zwanziger horizon function  \eqref{Hlandau} turns out to be equivalent to Gribov's ghost form factor, as shown in \cite{Gomez:2009tj} by comparing the expansion of both expressions till the third in the gauge field.\\\\In the case of the $SU(2)$ Yang-Mills theory in the maximal
Abelian gauge, the results found in \cite{Capri:2006cz} are
very similar to the ones obtained by Zwanziger in the Landau gauge. Summarizing, the form of the horizon function obtained by requiring that it can be cast in local form, that it can be renormalizable and that it reduces to Gribov's ghost form factor in the quadratic approximation, is given by
\begin{equation}
H_{\mathrm{MAG}}^{SU(2)}=g^{2}\int d^{4}xd^{4}x\, \varepsilon^{ac}A_{\mu}(x)
(\mathcal{M}^{-1})^{ab}(x,y)\varepsilon^{bc}A_{\mu}(y)\,,\label{Hsu2}
\end{equation}
where $(\mathcal{M}^{-1})^{ab}(x,y)$ is the inverse of the Faddeev-Popov operator
\begin{equation}
\mathcal{M}^{ab}=-D^{ac}_{\mu}D^{cb}_{\mu}
-g^{2}\varepsilon^{ac}\varepsilon^{bd}A^{c}_{\mu}A^{d}_{\mu}\,.
\end{equation}

\subsection{Candidates for the horizon function in $SU(N)$}
Let us face now the problem of finding the horizon term for the general case of $SU(N)$. The first step for accomplishing this task is that of selecting the possible candidates which are compatible with the following requirements:
\begin{enumerate}
\item{it should give back to the Gribov term \eqref{GribovTerm} in
    the quadratic approximation;}
\item{it should  reproduce the particular case \eqref{Hsu2} when
    $N=2$;}
\item{it should be localizable and compatible with the requirement of power counting renormalizability.}
\end{enumerate}
Five possible candidates which are compatible with the aforementioned requirements have been found, namely
\begin{eqnarray}
H^{1\mathrm{st}}&\!\!\!=\!\!\!&\frac{2g^{2}}{N(N-1)}\int_{x,y}\hspace{-5pt}f^{aci}A^{i}_{\mu,x}({\cal
M}^{-1})^{ab}_{x,y}f^{bcj}A^{j}_{\mu,y}
+\frac{g^{2}}{2N(N-1)}\int_{x,y}\hspace{-5pt}f^{acd}A^{d}_{\mu,x}({\cal
M}^{-1})^{ab}_{x,y}f^{bce}A^{e}_{\mu,y} \label{1st}\\\cr\cr
H^{2\mathrm{nd}}&\!\!\!=\!\!\!&\frac{2g^{2}}{N(N-1)}\int_{x,y}\Bigl(f^{aci}A^{i}_{\mu}+\frac{1}{2}f^{acd}A^{d}_{\mu}\Bigr)_{x}({\cal
M}^{-1})^{ab}_{x,y}\Bigl(f^{bcj}A^{j}_{\mu}+\frac{1}{2}f^{bce}A^{e}_{\mu}\Bigr)_{y}\,,\label{2nd}\\\cr\cr
H^{3\mathrm{rd}}&\!\!\!=\!\!&\frac{2g^{2}}{N(N-1)}\int_{x,y}\Bigl(f^{aci}A^{i}_{\mu}-\frac{1}{2}f^{acd}A^{d}_{\mu}\Bigr)_{x}({\cal
M}^{-1})^{ab}_{x,y}\Bigl(f^{bcj}A^{j}_{\mu}-\frac{1}{2}f^{bce}A^{e}_{\mu}\Bigr)_{y}\,,\label{3rd}\\\cr\cr
H^{4\mathrm{th}}&\!\!\!=\!\!\!&\frac{2g^{2}}{N(N-1)}\int_{x,y} f^{aci}A^{i}_{\mu,x}({\cal
M}^{-1})^{ab}_{x,y}f^{bcj}A^{j}_{\mu,y}
+\xi\,g^{2}\int_{x,y}f^{acd}A^{d}_{\mu,x}({\cal
M}^{-1})^{ab}_{x,y}f^{bce}A^{e}_{\mu,y}\nonumber\\
&&+(N-2)\zeta\,g^{2}
\int_{x,y} f^{aci}A^{c}_{\mu,x}(\mathcal{M}^{-1})^{ab}_{x,y}f^{bdi}A^{d}_{\mu,y}\,,\label{4th}\\\cr\cr
H^{5\mathrm{th}}&\!\!\!=\!\!\!&\frac{2g^{2}}{N(N-1)}\int_{x,y}\Bigl(f^{aci}A^{i}_{\mu}+\frac{\alpha}{2}f^{acd}A^{d}_{\mu}\Bigr)_{x}({\cal
M}^{-1})^{ab}_{x,y}\Bigl(f^{bcj}A^{j}_{\mu}+\frac{\alpha}{2}f^{bce}A^{e}_{\mu}\Bigr)_{y}\nonumber\\
&&+\frac{(N-2)g^{2}}{2N(N-1)}\beta
\int_{x,y} f^{aci}A^{c}_{\mu,x}(\mathcal{M}^{-1})^{ab}_{x,y}f^{bdi}A^{d}_{\mu,y}\,.\label{5th}
\end{eqnarray}
In all  expressions \eqref{1st} -- \eqref{5th}, the nonlocal
operator $(\mathcal{M}^{-1})^{ab}_{x,y}$ is the inverse of the
Faddeev-Popov operator \eqref{FPop} and the symbol $\int_{x,y}$ stands for $\int d^4x d^4y$.  In eq.\eqref{4th}, the parameters $\xi$ and $\zeta$ are  positive and  obey the following constraint:
\begin{equation}
\xi+\zeta=\frac{1}{2N(N-1)}\,.
\end{equation}
In expression \eqref{5th}, the parameter $\beta$ is positive and is related to $\alpha$ according to
\begin{equation}
\alpha^{2}+(N-2)\beta=N-2\,.
\end{equation}
Let us now see how these five candidates fulfill the  three
requirements. The first criterion claims that, in the quadratic approximation, the horizon function
has to reduce to the Gribov term defined in \eqref{GribovTerm}.
In the quadratic approximation, the inverse of the Faddeev-Popov
can be taken as
\begin{equation}
({\cal M}^{-1})^{ab}\approx\frac{\d^{ab}}{-\partial^{2}}\,.
\end{equation}
Moreover, using the following relations between the structure constants
\begin{equation}
f^{abi}f^{abc}=0\,,\qquad f^{abi}f^{abj}=N\d^{ij}\,,\qquad
f^{abc}f^{abd}=(N-2)\d^{cd}\,,\qquad f^{aic}f^{aid}=\d^{cd}\,,
\end{equation}
one can easily check that all five candidates \eqref{1st} --
\eqref{5th} obey the first requirement.\\\\ The second requirement
states that, when $N=2$,  the horizon function must equal expression
\eqref{Hsu2}. It is easy to check that this requirement is also fulfilled by observing
that in the case of $SU(2)$ we have:
\begin{eqnarray}
f^{abi}\to f^{ab3}=\varepsilon^{ab}\,,\qquad f^{abc}\to 0\,,\qquad A^{i}_{\mu}\to A^{3}_{\mu}=A_{\mu}\,.
\end{eqnarray}
Besides, there are terms proportional to $(N-2)$ which
automatically vanish for $N=2$.\\\\The third requirement is about the possibility of the nonlocal
horizon terms be cast in a local form by introducing a suitable set of auxiliary
fields, in a way compatible with power counting renormalizability. Let us analyze each case separately. The first candidate
\eqref{1st} consists of the sum of two different terms which can be localized
independently as follows:
\begin{eqnarray}
e^{-\gamma^{4}H^{1\mathrm{st}}}&\!\!\!=\!\!\!&\int\mathcal{D}\varphi\mathcal{D}\bar{\varphi}
\,(\det\mathcal{M})^{4N(N-1)}\,\exp\left[-\int d^{4}x\left(\bar\varphi^{ac}_{\mu}\mathcal{M}^{ab}\varphi^{bc}_{\mu}
+\theta^{2}\,f^{abi}A^{i}_{\mu}(\varphi-\bar\varphi)^{ab}_{\mu}\right)\right]\nonumber\\
&&\times\int\mathcal{D}\lambda\mathcal{D}\bar{\lambda}\,(\det\mathcal{M})^{4N(N-1)}\,
\exp\left[-\int d^{4}x\left(\bar\lambda^{ac}_{\mu}\mathcal{M}^{ab}\lambda^{bc}_{\mu}
+\frac{\theta^{2}}{2}\,f^{abc}A^{c}_{\mu}(\lambda-\bar\lambda)^{ab}_{\mu}\right)\right]\,,
\end{eqnarray}
where $\theta^{2}$ stands for
\begin{equation}
\theta^{2}=\sqrt{\frac{2g^{2}\gamma^{4}}{N(N-1)}}\,.\label{theta}
\end{equation}
In this expression $(\varphi^{ab}_{\mu},\bar\varphi^{ab}_{\mu})$
and $(\lambda^{ab}_{\mu},\bar\lambda^{ab}_{\mu})$ are two pairs of
complex commuting auxiliary fields. The determinant,  $(\det\mathcal{M})^{4N(N-1)}$, can be
exponentiated by employing a pair of complex anticommuting fields $(\bar\omega^{ac}_{\mu}, \omega^{bc}_{\mu})$:
\begin{equation}
(\det\mathcal{M})^{4N(N-1)}=\int\mathcal{D}\omega\mathcal{D}\bar{\omega}\,
e^{\,\int d^{4}x\,\bar\omega^{ac}_{\mu}\mathcal{M}^{ab}\omega^{bc}_{\mu}}\,.
\end{equation}
Thus, the local version of $H^{1\mathrm{st}}$ is
\begin{eqnarray}
H^{1\mathrm{st}}_{\mathrm{Local}}&=&\int d^{4}x\,\Bigl[\bar\varphi^{ac}_{\mu}\mathcal{M}^{ab}\varphi^{bc}_{\mu}
-\bar\omega^{ac}_{\mu}\mathcal{M}^{ab}\omega^{bc}_{\mu}
+\theta^{2}\,f^{abi}A^{i}_{\mu}(\varphi-\bar\varphi)^{ab}_{\mu}\nonumber\\
&&+\bar\lambda^{ac}_{\mu}\mathcal{M}^{ab}\lambda^{bc}_{\mu}
-\bar\eta^{ac}_{\mu}\mathcal{M}^{ab}\eta^{bc}_{\mu}
+\frac{\theta^{2}}{2}\,f^{abc}A^{c}_{\mu}(\lambda-\bar\lambda)^{ab}_{\mu}\Bigl]\,, \label{h1l}
\end{eqnarray}
where $(\eta^{ab}_{\mu},\bar\eta^{ab}_{\mu})$ is another pair of
complex anticommuting fields similar to
$(\omega^{ab}_{\mu},\bar\omega^{ab}_{\mu})$. As one can easily see, the set of fields
$(\varphi,\bar\varphi,\omega,\bar\omega)$ is responsible for the
localization of  the diagonal sector, while the set
$(\lambda,\bar\lambda,\eta,\bar\eta)$ allows us to localize
the off-diagonal sector.\\\\In a similar way, we can
localize the remaining candidates. To localize expressions \eqref{2nd} and
\eqref{3rd} it is necessary to employ only one set of localizing fields, $(\varphi,\bar\varphi,\omega,\bar\omega)$:
\begin{eqnarray}
H^{2\mathrm{nd}}_{\mathrm{Local}}&=&\int d^{4}x\,\Bigl[\bar\varphi^{ac}_{\mu}\mathcal{M}^{ab}\varphi^{bc}_{\mu}
-\bar\omega^{ac}_{\mu}\mathcal{M}^{ab}\omega^{bc}_{\mu}
+\theta^{2}\,\Bigl(f^{abi}A^{i}_{\mu}+\frac{1}{2}f^{abc}A^{c}_{\mu}\Bigr)
(\varphi-\bar\varphi)^{ab}_{\mu}\Bigr]\,,\label{2ndLocal}\\
H^{3\mathrm{rd}}_{\mathrm{Local}}&=&\int d^{4}x\,\Bigl[\bar\varphi^{ac}_{\mu}\mathcal{M}^{ab}\varphi^{bc}_{\mu}
-\bar\omega^{ac}_{\mu}\mathcal{M}^{ab}\omega^{bc}_{\mu}
+\theta^{2}\,\Bigl(f^{abi}A^{i}_{\mu}-\frac{1}{2}f^{abc}A^{c}_{\mu}\Bigr)
(\varphi-\bar\varphi)^{ab}_{\mu}\Bigr]\,. \label{h3l}
\end{eqnarray}
In the case of expression  \eqref{4th} we find
\begin{eqnarray}
H^{4\mathrm{th}}_{\mathrm{Local}}&=&\int d^{4}x\,\Bigl[\bar\varphi^{ac}_{\mu}\mathcal{M}^{ab}\varphi^{bc}_{\mu}
-\bar\omega^{ac}_{\mu}\mathcal{M}^{ab}\omega^{bc}_{\mu}
+\theta^{2}\,f^{abi}A^{i}_{\mu}
(\varphi-\bar\varphi)^{ab}_{\mu}\nonumber\\
&&+\bar\lambda^{ac}_{\mu}\mathcal{M}^{ab}\lambda^{bc}_{\mu}
-\bar\eta^{ac}_{\mu}\mathcal{M}^{ab}\eta^{bc}_{\mu}
+\xi^{1/2}\gamma^{2}\,gf^{abc}A^{c}_{\mu}(\lambda-\bar\lambda)^{ab}_{\mu}\nonumber\\
&&+\bar\psi^{ai}_{\mu}\mathcal{M}^{ab}\psi^{bi}_{\mu}
-\bar\chi^{ai}_{\mu}\mathcal{M}^{ab}\chi^{bi}_{\mu}
+\sqrt{(N-2)\zeta}\,\gamma^{2}\,gf^{abi}A^{a}_{\mu}(\psi-\bar\psi)^{bi}_{\mu}\Bigr]\,. \label{h4l}
\end{eqnarray}
Notice that, here,  a new set of fields has been employed:
$(\psi^{ai}_{\mu},\bar\psi^{ai}_{\mu})$ stands for a pair of
complex commuting fields and
$(\chi^{ai}_{\mu},\bar\chi^{ai}_{\mu})$ for a pair of complex
anticommuting ones. We also observe that these fields carry both off-diagonal and diagonal indices. \\\\ Finally, the local
version of \eqref{5th} is
\begin{eqnarray}
H^{5\mathrm{th}}_{\mathrm{Local}}&=&\int d^{4}x\,\Bigl[\bar\varphi^{ac}_{\mu}\mathcal{M}^{ab}\varphi^{bc}_{\mu}
-\bar\omega^{ac}_{\mu}\mathcal{M}^{ab}\omega^{bc}_{\mu}
+\theta^{2}\,\Bigl(f^{abi}A^{i}_{\mu}+\frac{\alpha}{2}f^{abc}A^{c}_{\mu}\Bigr)
(\varphi-\bar\varphi)^{ab}_{\mu}\nonumber\\
&&+\bar\psi^{ai}_{\mu}\mathcal{M}^{ab}\psi^{bi}_{\mu}
-\bar\chi^{ai}_{\mu}\mathcal{M}^{ab}\chi^{bi}_{\mu}
+\sqrt{(N-2)\beta}\,\frac{\theta^{2}}{2}\,f^{abi}A^{a}_{\mu}(\psi-\bar\psi)^{bi}_{\mu}\Bigr]\,. \label{h5l}
\end{eqnarray}
It is worth to point out that all local expressions, eqs.\eqref{h1l}, \eqref{2ndLocal}, \eqref{h3l},  \eqref{h4l}, \eqref{h5l}, are power counting renormalizable. All five candidates, \eqref{1st} --
\eqref{5th}, fulfill the three requirements given at the beginning of this section. \\\\A further requirement is thus needed
in order to select only one candidate. The natural set up is that of strengthening the requirement of the equivalence between Gribov's form factor $\sigma(0;A)$ and the possible horizon terms, as done in  \cite{Gomez:2009tj} in the case of the Landau gauge. This amounts to expand the horizon terms, eqs.\eqref{1st} --
\eqref{5th}, till the third order in the gauge field and compare the resulting expressions with that which we have already obtained for $\sigma(0;A)$, see equation \eqref{sigmafinal}.

\subsection{Selecting only one candidate for the horizon function}
In this section we show that the only candidate which is compatible with  the expression of  $\sigma(0;A)$ is $H^{2\mathrm{nd}}$, eq.~\eqref{2nd}. To that purpose,  the expansion till the first order in the gauge field of the inverse, $(\mathcal{M}^{-1})^{ab}$, of the Faddeev-Popov operator
\begin{equation}
\mathcal{M}^{ab}(x)(\mathcal{M}^{-1})^{bc}(x,y)=\d^{ac}\d(x-y)\,,
\label{uno}
\end{equation}
is needed. Recalling that $\mathcal{M}^{ab}(x)$ is defined by expression \eqref{FPop},  we can write
\begin{equation}
\mathcal{M}^{ab}=-\d^{ab}\partial^{2}+2gf^{abi}A^{i}_{\mu}\partial_{\mu}
+gf^{abc}A^{c}_{\mu}\partial_{\mu} +O(g^{2})\,.\label{inverse}
\end{equation}
To evaluate the inverse of $\mathcal{M}^{ab}$ order-by-order, we set
\begin{equation}
(\mathcal{M}^{-1})^{ab}(x,y)=\mathcal{G}^{ab}_{0}(x-y)+g\,\mathcal{G}^{ab}_{1}(x,y)+O(g^{2})\,,
\end{equation}
so that, to the first order, we have to solve the equation
\begin{equation}
\mathcal{M}^{ac}(x)\Bigr(\mathcal{G}^{ab}_{0}(x-y)+g\,\mathcal{G}^{ab}_{1}(x,y)+O(g^{2})\Bigl)=\d^{ab}\d(x-y)\,,
\end{equation}
where $\mathcal{G}_{0}^{ab}(x-y)$ is the free ghost propagator:
\begin{equation}
\mathcal{G}_{0}^{ab}(x-y)=\frac{\d^{ab}}{|x-y|^{2}}=\d^{ab}\int\frac{d^{4}q}{(2\pi)^{4}}\,\frac{1}{q^{2}}\,
e^{-iq(x-y)}\,.
\end{equation}
To the first order, we have
\begin{equation}
-\partial^{2}_{x}\mathcal{G}_{1}^{ab}(x,y)=-2f^{aci}A^{i}_{\mu}(x)\,\partial^{x}_{\mu}\mathcal{G}_{0}^{cb}(x-y)
-f^{acd}A^{d}_{\mu}(x)\,\partial_{\mu}^{x}\mathcal{G}_{0}^{cb}(x-y)\,,
\end{equation}
which gives
\begin{equation}
\mathcal{G}_{1}^{ab}(x,y)=-2\int{d^{4}z}\,\frac{1}{|x-z|^{2}}\,\biggl(f^{abi}A^{i}_{\mu}(z)
+\frac{1}{2}f^{abc}A^{c}_{\mu}(z)\biggr)\partial_{\mu}^{z}\frac{1}{|z-y|^{2}}\,.
\end{equation}
Therefore,
\begin{equation}
(\mathcal{M}^{-1})^{ab}(x,y)=\frac{\d^{ab}}{|x-y|^{2}}
-2\int{d^{4}z}\,\frac{1}{|x-z|^{2}}\,\biggl(f^{abi}A^{i}_{\mu}(z)
+\frac{1}{2}f^{abc}A^{c}_{\mu}(z)\biggr)\partial_{\mu}^{z}\frac{1}{|z-y|^{2}}
+O(g^{2})\,.\label{Minverse}
\end{equation}
We are now ready to expand the horizon function \eqref{2nd} in powers of the gauge field. Introducing the combination
\begin{equation}
B^{ab}_{\mu}=f^{abi}A^{i}_{\mu}+\frac{1}{2}f^{abc}A^{c}_{\mu}\,,\label{Bfield}
\end{equation}
one can write
\begin{equation}
H^{2\mathrm{nd}} =\frac{2g^{2}}{N(N-1)}\int{d^{4}xd^{4}y}\,B^{ac}_{\mu}(x)
(\mathcal{M}^{-1})^{ab}(x,y)B^{bc}_{\mu}(y)\,,\label{HorFunc}
\end{equation}
which, by means of eqs. \eqref{Minverse} and \eqref{Bfield}, becomes
\begin{eqnarray}
H^{2\mathrm{nd}} &=&\frac{2g^{2}}{N(N-1)}\int{d^{4}xd^{4}y}\,B^{ab}_{\mu}(x)\frac{1}{|x-y|^{2}}B^{ab}_{\mu}(y)
\nonumber\\
&&-\frac{4g^{3}}{N(N-1)}\int{d^{4}xd^{4}y}{d^{4}z}\,
B^{ac}_{\mu}(x)\,\frac{1}{|x-z|^{2}}\,B^{ab}_{\nu}(z)\,\partial_{\nu}^{z}\frac{1}{|z-y|^{2}}\,B^{bc}_{\mu}(y)
+O(g^{4})\,.
\end{eqnarray}
Explicitly,  we have:
\begin{eqnarray}
H^{2\mathrm{nd}} &=&\frac{2g^{2}}{N-1}\int{d^{4}xd^{4}y}\,\frac{A^{i}_{\mu}(x)A^{i}_{\mu}(y)}{|x-y|^{2}}
+\frac{(N-2)g^{2}}{2N(N-1)}\int{d^{4}xd^{4}y}\,\frac{A^{a}_{\mu}(x)A^{a}_{\mu}(y)}{|x-y|^{2}}\nonumber\\
&&-\frac{4g^{3}}{N(N-1)}\int{d^{4}xd^{4}y}{d^{4}z}\,
\frac{1}{|x-z|^{2}}\partial^{z}_{\nu}\frac{1}{|z-y|^{2}}\,\biggl(f^{aci}f^{abj}f^{bck}A^{i}_{\mu}(x)A^{j}_{\nu}(z)A^{k}_{\mu}(y)
\nonumber\\
&&+\frac{1}{2}f^{aci}f^{abj}f^{bcf}A^{i}_{\mu}(x)A^{j}_{\nu}(z)A^{f}_{\mu}(y)
+\frac{1}{2}f^{aci}f^{abe}f^{bck}A^{i}_{\mu}(x)A^{e}_{\nu}(z)A^{k}_{\mu}(y)\nonumber\\
&&+\frac{1}{4}f^{aci}f^{abe}f^{bcf}A^{i}_{\mu}(x)A^{e}_{\nu}(z)A^{f}_{\mu}(y)
+\frac{1}{2}f^{acd}f^{abj}f^{bck}A^{d}_{\mu}(x)A^{j}_{\nu}(z)A^{k}_{\mu}(y)\nonumber\\
&&+\frac{1}{4}f^{acd}f^{abj}f^{bcf}A^{d}_{\mu}(x)A^{j}_{\nu}(z)A^{f}_{\mu}(y)
+\frac{1}{4}f^{acd}f^{abe}f^{bck}A^{d}_{\mu}(x)A^{e}_{\nu}(z)A^{k}_{\mu}(y)\nonumber\\
&&+\frac{1}{8}f^{acd}f^{abe}f^{bcf}A^{d}_{\mu}(x)A^{e}_{\nu}(z)A^{f}_{\mu}(y)\biggr)+O(g^{4})\,.
\end{eqnarray}
Taking the Fourier transform
\begin{eqnarray}
H^{2\mathrm{nd}} &=&\frac{2g^{2}}{(N-1)}\int\frac{d^{4}q}{(2\pi)^{4}}
\frac{A^{i}_{\lambda}(q)A^{i}_{\lambda}(-q)}{q^{2}}
+\frac{(N-2)g^{2}}{2N(N-1)}\int\frac{d^{4}q}{(2\pi)^{4}}
\frac{A^{a}_{\lambda}(q)A^{a}_{\lambda}(-q)}{q^{2}}\nonumber\\
&&-\frac{2ig^{3}}{N(N-1)}\int\frac{d^{4}p}{(2\pi)^{4}}\frac{d^{4}q}{(2\pi)^{4}}\,
\frac{p_\mu}{p^{2}(p+q)^{2}}\,\biggl(2f^{abi}f^{bcj}f^{cak}A^{i}_{\mu}(-p)A^{j}_{\nu}(-q)A^{k}_{\mu}(p+q)
\nonumber\\
&&+f^{abi}f^{bdj}f^{dac}A^{i}_{\mu}(-p)A^{j}_{\nu}(-q)A^{c}_{\mu}(p+q)
+f^{abi}f^{bdc}f^{daj}A^{i}_{\mu}(-p)A^{c}_{\nu}(-q)A^{j}_{\mu}(p+q)\nonumber\\
&&+f^{abc}f^{bdi}f^{daj}A^{c}_{\mu}(-p)A^{i}_{\nu}(-q)A^{j}_{\mu}(p+q)
+\frac{1}{2}f^{bcd}f^{bde}f^{dai}A^{c}_{\mu}(-p)A^{e}_{\nu}(-q)A^{i}_{\mu}(p+q)\nonumber\\
&&+\frac{1}{2}f^{abc}f^{bdi}f^{dae}A^{c}_{\mu}(-p)A^{i}_{\nu}(-q)A^{e}_{\mu}(p+q)
+\frac{1}{2}f^{abi}f^{bdc}f^{dae}A^{i}_{\mu}(-p)A^{c}_{\nu}(-q)A^{e}_{\mu}(p+q)\nonumber\\
&&+\frac{1}{4}f^{abc}f^{bde}f^{daf}A^{d}_{\mu}(-p)A^{e}_{\nu}(-q)A^{f}_{\mu}(p+q)\biggr)+O(g^{4})\,,
\end{eqnarray}
and comparing with Gribov's form factor $\sigma(0;A)$, eq.\eqref{sigmafinal}, it is apparent that
\begin{eqnarray}
\fbox{$\displaystyle \sigma(0,A)=\frac{1}{2}  H^{2\mathrm{nd}} +O(A^{4})\,.$}
\end{eqnarray}
The same procedure can be easily repeated for the other expressions, $H^{1\mathrm{st}}, H^{3\mathrm{st}}, H^{4\mathrm{st}}, H^{5\mathrm{st}}$. The corresponding expansions do not coincide with equation \eqref{sigmafinal}. Therefore, from now on,  we shall  consider expression  \eqref{HorFunc} as the correct
horizon function we were looking for. Consequently, for the partition function defining the theory we shall take
\begin{equation}
\mathcal{Z}=\int d\mu\,e^{-(S_{\mathrm{YM}}+\gamma^{4}\, H^{2\mathrm{nd}} )}\,. \label{zz}
\end{equation}
In the next section, we  shall construct a complete local action out of $H^{2\mathrm{nd}} $ and
we shall study the possible dimension two operators which can be
introduced. The effect of the condensation of these operators on the  tree level gluon and ghost propagators will be also worked out.
\section{Local dimension two operators and the tree level gluon and ghost propagators}
\subsection{A local and invariant action}
According to expression \eqref{zz}, the general local action we can construct for the $SU(N)$
Euclidean Yang-Mills theory in the maximal Abelian gauge reads
\begin{eqnarray}
\mathcal{Z}&=&\int \mathcal{D}\Phi\,e^{-S_{0}}\,,\nonumber\\
S_{0}&=&S_{\mathrm{YM}}+S_{\mathrm{MAG}}+S_{\mathrm{Local}}\,,
\end{eqnarray}
where
\begin{equation}
\mathcal{D}\Phi\equiv\mathcal{D}A^{a}\mathcal{D}A^{i}\mathcal{D}b^{a}\mathcal{D}b^{i}
\mathcal{D}c^{a}\mathcal{D}\bar{c}^{a}\mathcal{D}c^{i}\mathcal{D}\bar{c}^{i}
\mathcal{D}\varphi\mathcal{D}\bar{\varphi}
\mathcal{D}\omega\mathcal{D}\bar{\omega}\,,
\end{equation}
and $S_{\mathrm{Local}}\equiv H^{2\mathrm{nd}}_{\mathrm{Local}}$
given by eq.\eqref{2ndLocal}. Explicitly,  we have:
\begin{eqnarray}
S_{0}&=&\int d^{4}x\,\biggl(\frac{1}{4}F^{a}_{\mu\nu}F^{a}_{\mu\nu}
+\frac{1}{4}F^{i}_{\mu\nu}F^{i}_{\mu\nu}
+ib^{a}\,D^{ab}_{\mu}A^{b}_{\mu}
-\bar{c}^{a}\mathcal{M}^{ab}c^{b}
-(D^{ad}_{\mu}A^{d}_{\mu})\Bigl(gf^{abc}\bar{c}^{b}c^{c}
+gf^{abi}\bar{c}^{b}c^{i}\Bigr)\nonumber\\
&&\phantom{\int d^{4}x\,\biggl(}+ib^{i}\,\partial_{\mu}A^{i}_{\mu}
+\bar{c}^{i}\,\partial_{\mu}(\partial_{\mu}c^{i}+gf^{abi}A^{a}_{\mu}c^{b})
+\bar\varphi^{ac}_{\mu}\mathcal{M}^{ab}\varphi^{bc}_{\mu}
-\bar\omega^{ac}_{\mu}\mathcal{M}^{ab}\omega^{bc}_{\mu}\nonumber\\
&&\phantom{\int d^{4}x\,\biggl(}+\sqrt{\frac{2\gamma^{4}}{N(N-1)}}\,
\Bigl(gf^{abi}A^{i}_{\mu}+\frac{g}{2}f^{abc}A^{c}_{\mu}\Bigr)
(\varphi^{ab}_{\mu}-\bar\varphi^{ab}_{\mu})\biggr)\,.
\end{eqnarray}
Having at our disposal a local action, we may investigate its symmetry content. Let us start with the BRST symmetry, already obtained for the gauge fixed theory, see eqs.\eqref{brst_fields}. As established in the case of the Landau gauge   \cite{Zwanziger:1989mf,Zwanziger:1992qr,Dudal:2008sp} and $SU(2)$ maximal Abelian gauge \cite{Capri:2008ak}, the auxiliary fields $(\varphi,\bar\varphi,\omega,\bar\omega)$ transform as a BRST quartet, {\it i.e.}
\begin{eqnarray}
s\varphi^{ab}_{\mu}&\!\!\!=\!\!\!&\omega^{ab}_{\mu}\,,\qquad s\omega^{ab}_{\mu}=0\,,\nonumber\\
s\bar\omega^{ab}_{\mu}&\!\!\!=\!\!\!&\bar\varphi^{ab}_{\mu}\,,\qquad s\bar\varphi^{ab}_{\mu}=0\;, \nonumber \\
s^2&=&0 \;. \label{qt}
\end{eqnarray}
As a consequence, we get
\begin{eqnarray}
sS_{\mathrm{Local}}&=&sS^{(1)}+sS^{(2)}\,,\nonumber\\
sS^{(1)}&=&s\int d^{4}x\,(\bar\varphi^{ac}_{\mu}\mathcal{M}^{ab}\varphi^{bc}_{\mu}
-\bar\omega^{ac}_{\mu}\mathcal{M}^{ab}\omega^{bc}_{\mu})
=\int d^{4}x\,(\bar\varphi^{ac}_{\mu}\mathcal{N}^{ab}\varphi^{bc}_{\mu}
+\bar\omega^{ac}_{\mu}\mathcal{N}^{ab}\omega^{bc}_{\mu})\,,\nonumber\\
sS^{(2)}&=&\theta^{2}\,s\int
d^{4}x\,B^{ab}_{\mu}
(\varphi^{ab}_{\mu}-\bar\varphi^{ab}_{\mu})
=\theta^{2}\int
d^{4}x\,[(sB^{ab}_{\mu})
(\varphi^{ab}_{\mu}-\bar\varphi^{ab}_{\mu})+
B^{ab}_{\mu}\omega^{ab}_{\mu}]\,.
\end{eqnarray}
where $\theta^{2}$ is given in eq.\eqref{theta}, $B^{ab}_{\mu}(x)$ in
eq.\eqref{Bfield}, and the operator $\mathcal{N}^{ab}$ is defined as
\begin{eqnarray}
\mathcal{N}^{ab}\Phi^{b}&=&s(\mathcal{M}^{ab}\Phi^{b})-\mathcal{M}^{ab}\,s\Phi^{b}\nonumber\\
&=&\Bigl[\,2gf^{abi}(sA^{i}_{\mu})\partial_{\mu} +gf^{abc}(sA^{c}_{\mu})\partial_{\mu}
+gf^{abi}\partial_{\mu}(sA^{i}_{\mu})
+2g^{2}f^{aci}f^{bcj}(sA^{i}_{\mu})A^{j}_{\mu}\nonumber\\
&&g^{2}f^{adc}f^{bdi}s(A^{c}_{\mu}A^{i}_{\mu})
-g^{2}(f^{aci}f^{bdi}+f^{adi}f^{bci})(sA^{c}_{\mu})A^{d}_{\mu}\,\Bigr]\,\Phi^{b}\nonumber\\
&=&-\Bigr[\,2gf^{aci}(\partial_{\mu}c^{i}+gf^{cdi}A^{c}_{\mu}c^{d})D^{cb}_{\mu}
-gf^{acd}(D^{ce}_{\mu}c^{e}+gf^{cef}A^{e}_{\mu}c^{f}+gf^{cei}A^{e}_{\mu}c^{i})D^{db}_{\mu}\nonumber\\
&&+gf^{abi}\partial_{\mu}(\partial_{\mu}c^{i}+gf^{cdi}A^{c}_{\mu}c^{d})
+g^{2}f^{adc}f^{bdi}A^{c}_{\mu}(\partial_{\mu}c^{i}+gf^{efi}A^{e}_{\mu}c^{f})\nonumber\\
&&-g^{2}(f^{aci}f^{bdi}+f^{adi}f^{bci})(D^{ce}_{\mu}c^{e}+gf^{cef}A^{e}_{\mu}c^{f}+gf^{cej}A^{e}_{\mu}c^{j})
A^{d}_{\mu}\,\Bigr]\,\Phi^{b}\,,
\end{eqnarray}
with $\Phi^{a}$ representing all the off-diagonal fields of the theory
\begin{equation}
\Phi^{a}\equiv\{ A^{a}_{\mu},b^{a},c^{a},\bar{c}^{a},\varphi^{ab}_{\mu},\bar\varphi^{ab}_{\mu},
\omega^{ab}_{\mu},\bar\omega^{ab}_{\mu}\}\,.
\end{equation}
The first term $S^{(1)}$ can be rewritten in a BRST
invariant way by making a linear shift in the variable
$\omega^{ab}_{\mu}(x)$  \cite{Zwanziger:1989mf,Zwanziger:1992qr,Dudal:2008sp,Capri:2008ak},
\begin{equation}
\omega^{ab}_{\mu}(x)\to\omega^{ab}_{\mu}(x)+\int d^{4}y\,(\mathcal{M}^{-1})^{ac}(x,y)\mathcal{N}^{cd}(y)\varphi^{db}_{\mu}(y)\,.
\end{equation}
Thus,
\begin{eqnarray}
S^{(1)}\to S^{(1)}_{\mathrm{inv}}&=&\int d^{4}x\,\,(\bar\varphi^{ac}_{\mu}\mathcal{M}^{ab}\varphi^{bc}_{\mu}
-\bar\omega^{ac}_{\mu}\mathcal{M}^{ab}\omega^{bc}_{\mu}
-\bar\omega^{ac}_{\mu}\mathcal{N}^{ab}\varphi^{bc}_{\mu})\nonumber\\
&=&\int d^{4}x\,[\bar\varphi^{ac}_{\mu}\mathcal{M}^{ab}\varphi^{bc}_{\mu}
-\bar\omega^{ac}_{\mu}\mathcal{M}^{ab}\omega^{bc}_{\mu}
-\bar\omega^{ac}_{\mu}(s(\mathcal{M}^{ab}\varphi^{bc}_{\mu})-\mathcal{M}^{ab}\omega^{bc}_{\mu})]\nonumber\\
&=&\int d^{4}x\,[\bar\varphi^{ac}_{\mu}\mathcal{M}^{ab}\varphi^{bc}_{\mu}
-\bar\omega^{ac}_{\mu}s(\mathcal{M}^{ab}\varphi^{bc}_{\mu})]\nonumber\\
&=&s\int d^{4}x\,\bar\omega^{ac}_{\mu}\mathcal{M}^{ab}\varphi^{bc}_{\mu}\,.
\end{eqnarray}
In order to deal with the second  term $S^{(2)}$, let us
notice that
\begin{equation}
B^{ab}_{\mu}\bar\varphi^{ab}_{\mu}=B^{ab}_{\mu}s\bar\omega^{ab}_{\mu}=s(B^{ab}_{\mu}\bar\omega^{ab}_{\mu})
-(sB^{ab}_{\mu})\,\bar\omega^{ab}_{\mu}\,,
\end{equation}
then,
\begin{equation}
S^{(2)}=\theta^{2}\int d^{4}x\,[B^{ab}_{\mu}\varphi^{ab}_{\mu}-s(B^{ab}_{\mu}\bar\omega^{ab}_{\mu})
+(sB^{ab}_{\mu})\,\bar\omega^{ab}_{\mu}]\,.\label{S(2)}
\end{equation}
The last term of the r.h.s. of \eqref{S(2)} can be eliminated by
performing a second linear shift in $\omega^{ab}_{\mu}(x)$ \cite{Zwanziger:1989mf,Zwanziger:1992qr,Dudal:2008sp,Capri:2008ak}
\begin{equation}
\omega^{ab}_{\mu}(x)\to\omega^{ab}_{\mu}(x)+\theta^{2}\int d^{4}y\,(\mathcal{M}^{-1})^{ac}(x,y)\,sB^{cb}_{\mu}(y)\,.
\end{equation}
Therefore, the local version of the horizon function might be taken as
\begin{equation}
S_{\mathrm{Local}}=s\int d^{4}x\,\bar\omega^{ac}_{\mu}\mathcal{M}^{ab}\varphi^{bc}_{\mu}
+\theta^{2}\int d^{4}x\,[B^{ab}_{\mu}\varphi^{ab}_{\mu}-s(B^{ab}_{\mu}\bar\omega^{ab}_{\mu})]\,.
\end{equation}
Consequently, the action $S_{0}$ becomes
\begin{eqnarray}
S_{0}&=&\int d^{4}x\,\biggl(\frac{1}{4}F^{a}_{\mu\nu}F^{a}_{\mu\nu}
+\frac{1}{4}F^{i}_{\mu\nu}F^{i}_{\mu\nu}
+ib^{a}\,D^{ab}_{\mu}A^{b}_{\mu}
-\bar{c}^{a}\mathcal{M}^{ab}c^{b}
-(D^{ad}_{\mu}A^{d}_{\mu})\Bigl(gf^{abc}\bar{c}^{b}c^{c}
+gf^{abi}\bar{c}^{b}c^{i}\Bigr)\nonumber\\
&&\phantom{\int d^{4}x\,\biggl(}+ib^{i}\,\partial_{\mu}A^{i}_{\mu}
+\bar{c}^{i}\,\partial_{\mu}(\partial_{\mu}c^{i}+gf^{abi}A^{a}_{\mu}c^{b}) \biggr)  \nonumber \\
&&\qquad + \; s\int d^{4}x\,\bar\omega^{ac}_{\mu}\mathcal{M}^{ab}\varphi^{bc}_{\mu}
+\theta^{2}\int d^{4}x\,[B^{ab}_{\mu}\varphi^{ab}_{\mu}-s(B^{ab}_{\mu}\bar\omega^{ab}_{\mu})]\,. \label{sof}
\end{eqnarray}
As it happens in the case of the Landau gauge \cite{Zwanziger:1989mf,Zwanziger:1992qr,Dudal:2008sp}  and $SU(2)$ maximal Abelian gauge \cite{Capri:2008ak}, the action $S_0$ does not possess exact BRST invariance, which turns out to be broken by soft terms proportional to the Gribov parameter, {\it i.e.} to $\theta^2$.  In fact
\begin{equation}
s S_0 = \theta^2 \Delta_{break} \;, \label{brk}
\end{equation}
where
\begin{equation}
\Delta_{break}= \int d^4x\; \left[ B^{ab}_\mu \omega^{ab}_\mu - \left( f^{abi}(\partial_\mu c^i + g f^{cdi}A^c_\mu c^d) +
\frac{1}{2} f^{abc} (D^{cm}_\mu c^m + g f^{cmn} A^m_\mu c^n + g f^{cmi} A^m_\mu c^i) \right) \varphi^{ab}_\mu \right]  \\ \label{brk1}
\end{equation}
Being of dimension two, the breaking term $\Delta_{break}$ is soft. As extensively analyzed in  \cite{Dudal:2008sp,Baulieu:2009ha}, the existence of such a breaking does not jeopardize the renormalizability of the theory as well as the introduction of meaningful operators which display good analyticity  properties. Of course, the fact that the breaking is soft is of pivotal importance here. As underlined in  \cite{Dudal:2008sp,Sorella:2009vt}, the existence of such a breaking is linked to the presence of a boundary in field space, namely the Gribov horizon $\partial \Omega$.
\subsection{Embedding the theory into a more general one}
The standard way of dealing with soft breaking terms is that of introducing them into the starting action as composite operators coupled to a suitable set of external sources. In doing so, a more general action displaying exact BRST invariance is obtained. Furthermore, the starting action $S_0$ and its breaking term $\Delta_{break}$ are recovered by demanding that the external sources attain a particular value, usually referred as to the physical value \cite{Zwanziger:1989mf,Zwanziger:1992qr,Dudal:2008sp}.  In order to construct such a generalized action, we introduce external sources $(V^{ab}_{\mu\nu}, U^{ab}_{\mu\nu}, \bar{V}^{ab}_{\mu\nu}, \bar{U}^{ab}_{\mu\nu})$ transforming as
\begin{eqnarray}
sV^{ab}_{\mu\nu}&\!\!\!=\!\!\!&U^{ab}_{\mu\nu}\,,\qquad sU^{ab}_{\mu\nu}=0\,,\nonumber\\
s\bar{U}^{ab}_{\mu\nu}&\!\!\!=\!\!\!&\bar{V}^{ab}_{\mu\nu}\,,\qquad s\bar{V}^{ab}_{\mu\nu}=0\,,
\end{eqnarray}
and the invariant action $S^{\mathrm{inv}}_{\mathrm{Local}}$, given by
\begin{eqnarray}
S^{\mathrm{inv}}_{\mathrm{Local}}&\!\!\!=\!\!\!&s\int d^{4}x\,\bar\omega^{ac}_{\mu}\mathcal{M}^{ab}\varphi^{bc}_{\mu}
+s\int d^{4}x\,\biggl[\bar{U}^{ac}_{\mu\nu}\Bigl(D^{ab}_{\mu}-\frac{g}{2}f^{abd}A^{d}_{\mu}\Bigr)\varphi^{bc}_{\nu}
+V^{ac}_{\mu\nu}\Bigl(D^{ab}_{\mu}-\frac{g}{2}f^{abd}A^{d}_{\mu}\Bigr)\bar\omega^{bc}_{\nu}\biggr]\nonumber\\
&\!\!\!=\!\!\!&\int
d^{4}x\,\biggl\{\bar\varphi^{ac}_{\mu}{\cal M}^{ab}\varphi^{bc}_{\mu}
-\bar\omega^{ac}_{\mu}{\cal M}^{ab}\omega^{bc}_{\mu}
-\bar\omega^{ac}_{\mu}{\cal N}^{ab}\varphi^{bc}_{\mu}
+V^{ac}_{\mu\nu}\biggl[\Bigl(D^{ab}_{\mu}-\frac{g}{2}f^{abd}A^{d}_{\mu}\Bigr)\bar\varphi^{bc}_{\nu}
\nonumber\\
&&+gf^{abi}(\partial_{\mu}c^{i}+gf^{dei}A^{d}_{\mu}c^{e})\bar\omega^{bc}_{\nu}
+\frac{g}{2}f^{abd}(D^{de}_{\mu}c^{e} +gf^{def}A^{e}_{\mu}c^{f}
+gf^{dei}A^{e}_{\mu}c^{i})\bar\omega^{bc}_{\nu}\biggr]\nonumber\\
&&+\bar{V}^{ac}_{\mu\nu}\Bigl(D^{ab}_{\mu}
-\frac{g}{2}f^{abd}A^{d}_{\mu}\Bigr)\varphi^{bc}_{\nu}
-\bar{U}^{ac}_{\mu\nu}\biggl[\Bigr(D^{ab}_{\mu}-\frac{g}{2}f^{abd}A^{d}_{\mu}\Bigr)\omega^{bc}_{\nu}
+gf^{abi}(\partial_{\mu}c^{i}+gf^{dei}A^{d}_{\mu}c^{e})\varphi^{bc}_{\nu}
\nonumber\\
&&+\frac{g}{2}f^{abd}(D^{de}_{\mu}c^{e} +gf^{def}A^{e}_{\mu}c^{f}
+gf^{dei}A^{e}_{\mu}c^{i})\varphi^{bc}_{\nu}\biggr]
+{U}^{ac}_{\mu\nu}\Bigl(D^{ab}_{\mu}
-\frac{g}{2}f^{abd}A^{d}_{\mu}\Bigr)\bar\omega^{bc}_{\nu}
\biggr\}\,.
\end{eqnarray}
The expression $S_{\mathrm{Local}}$ is thus recovered from the generalized one  $S^{\mathrm{inv}}_{\mathrm{Local}}$
when the sources attain their physical values:
\begin{equation}
V^{ab}_{\mu\nu}\Bigl|_{\mathrm{phys}}=-\bar{V}^{ab}_{\mu\nu}\Bigl|_{\mathrm{phys}}=-\sqrt{\frac{2\gamma^{4}}{N(N-1)}}\,\,\d^{ab}\d_{\mu\nu}\,,\qquad
U^{ab}_{\mu\nu}\Bigl|_{\mathrm{phys}}=\bar{U}^{ab}_{\mu\nu}\Bigl|_{\mathrm{phys}}=0\,,
\end{equation}
so that
\begin{equation}
S^{\mathrm{inv}}_{\mathrm{Local}}\Bigl|_{\rm phys} = S_{\mathrm{Local} }\;. \label{plim}
\end{equation}
We end up thus with a generalized invariant action \footnote{The complete invariant action must contain a generalized version of the term $S_\alpha$ in \eqref{alpha} in order to establish its renormalizability. We refer the reader to \cite{Capri:2006cz}, where this analysis is done in detail for the $SU(2)$ case. However, in the present section, we are ultimately interested only in the computation of the tree level propagators of the maximal Abelian gauge; therefore, we can set $\alpha =0$ at this level.}
\begin{equation}
S_{\mathrm{inv}}=S_{\mathrm{YM}}+S_{\mathrm{MAG}}+S_{\mathrm{Local}}^{\mathrm{inv}}\,,
\end{equation}
\begin{equation}
 sS_{\mathrm{inv}}=0\,, \label{invact}
\end{equation}
which reduces to $S_0$ in the physical limit
\begin{equation}
S_{\mathrm{inv}} \Bigl|_{\rm phys} = S_0 \;. \label{is1}
\end{equation}
The BRST invariance enjoyed by $S_{\mathrm{inv}}$ can be translated in a powerful functional identity, the Slavnov-Taylor identity, which is the starting point for the analysis of the all order renormalizability, which we shall present separately in a forthcoming work.  To derive such identity,
we need to introduce more external sources coupled to
the nonlinear BRST transformations of the fields \cite{Piguet:1995er}. To that purpose, we notice that the BRST transformation of the off-diagonal
component of the gauge field $A^{a}_{\mu}(x)$ can be split in
three nonlinear parts:
\begin{equation}
sA^{a}_{\mu}=\mathcal{P}^{a}_{\mu}+\mathcal{Q}^{a}_{\mu}+\mathcal{R}^{a}_{\mu}\,,
\end{equation}
\begin{equation}
\mathcal{P}^{a}_{\mu}=-D^{ab}_{\mu}c^{b}\,,\qquad
\mathcal{Q}^{a}_{\mu}=-gf^{abc}A^{b}_{\mu}c^{c}\,,\qquad
\mathcal{R}^{a}_{\mu}=-gf^{abi}A^{b}_{\mu}c^{i}\,.
\end{equation}
These nonlinear terms can be defined separately by introducing  the
following set of external sources
\begin{eqnarray}
&s\xi^{a}_{\mu}=K^{a}_{\mu}-\Omega^{a}_{\mu}\,,\qquad
sK^{a}_{\mu}=s\Omega^{a}_{\mu}=0\,,&\nonumber\\
&s\vartheta^{a}_{\mu}=\Upsilon^{a}_{\mu}-\Omega^{a}_{\mu}\,,\qquad
s\Upsilon^{a}_{\mu}=s\Omega^{a}_{\mu}=0\,,&
\end{eqnarray}
and writing
\begin{equation}
S_{\mathrm{ext}}^{(1)}=\int d^{4}x\,[\Omega^{a}_{\mu}\,\mathcal{P}^{a}_{\mu}
+K^{a}_{\mu}\,\mathcal{Q}^{a}_{\mu}
+\Upsilon^{a}_{\mu}\,\mathcal{R}^{a}_{\mu}
+\xi^{a}_{\mu}\,(s\mathcal{Q}^{a}_{\mu})
+\vartheta^{a}_{\mu}\,(s\mathcal{R}^{a}_{\mu})]\,.
\end{equation}
Since
\begin{equation}
s^{2}A^{a}_{\mu}=s\mathcal{P}^{a}+s\mathcal{Q}^{a}+s\mathcal{R}^{a}=0\,,
\end{equation}
one immediately verifies that $sS^{(1)}_{\mathrm{ext}}=0$. The
remaining nonlinear transformations can be defined by
\begin{equation}
S_{\mathrm{ext}}^{(2)}=\int d^{4}x\,\left[\Omega^{i}_{\mu}\,(sA^{i}_{\mu})
+L^{a}\,(sc^{a})+L^{i}\,(sc^{i})\right]\,,
\end{equation}
with
\begin{equation}
s\Omega^{i}_{\mu}=sL^{i}=sL^{a}=0\,.
\end{equation}
Therefore, the complete action $\S_0$ defined by
\begin{equation}
\S_0=S_{\mathrm{inv}}+S^{(1)}_{\mathrm{ext}}+S^{(2)}_{\mathrm{ext}}\,,\label{SigmaZero}
\end{equation}
obeys the following Slavnov-Taylor identity:
\begin{eqnarray}
\mathcal{S}(\S_{0})&=&\int d^{4}x\,\biggl[\biggl(\frac{\d\S_{0}}{\d\Omega^{a}_{\mu}}+\frac{\d\S_{0}}{\d{K}^{a}_{\mu}}
+\frac{\d\S_{0}}{\d\Upsilon^{a}_{\mu}}\biggr)\frac{\d\S_{0}}{\d{A}^{a}_{\mu}}
+\frac{\d\S_{0}}{\d\Omega^{i}_{\mu}}\frac{\d\S_{0}}{\d{A}^{i}_{\mu}}
+\frac{\d\S_{0}}{\d{L}^{a}}\frac{\d\S_{0}}{\d{c}^{a}}
+\frac{\d\S_{0}}{\d{L}^{i}}\frac{\d\S_{0}}{\d{c}^{i}}
\nonumber\\
&&+ib^{a}\frac{\d\S_{0}}{\d\bar{c}^{a}}
+ib^{i}\frac{\d\S_{0}}{\d\bar{c}^{i}}
+\omega^{ab}_{\mu}\frac{\d\S_{0}}{\d\varphi^{ab}_{\mu}}
+\bar\varphi^{ab}_{\mu}\frac{\d\S_{0}}{\d\bar\omega^{ab}_{\mu}}
+U^{ab}_{\mu\nu}\frac{\d\S_{0}}{\d{V}^{ab}_{\mu\nu}}
+\bar{V}^{ab}_{\mu\nu}\frac{\d\S_{0}}{\d\bar{U}^{ab}_{\mu\nu}}\nonumber\\
&&+(K^{a}_{\mu}-\Omega^{a}_{\mu})\frac{\d\S_{0}}{\d\xi^{a}_{\mu}}
+(\Upsilon^{a}_{\mu}-\Omega^{a}_{\mu})\frac{\d\S_{0}}{\d\vartheta^{a}_{\mu}}
\,\biggr]=0\,. \label{sti}
\end{eqnarray}
Besides the Slavnov-Taylor identity, eq.\eqref{sti}, the  action $\S_{0}$ displays a rather rich
symmetry content. There exist in fact  several symmetries involving
the exchange between the Faddeev-Popov  ghosts $(c^{a},\bar{c}^{a})$ and the localizing auxiliary fields
$(\varphi,\bar\varphi,\omega,\bar\omega)$, namely
\begin{equation}
\d^{a}_{\mu}\S_0=0\,,\qquad\bar\d^{a}_{\mu}\S_0=0\,,\qquad d^{a}_{\mu}\S_0=0\,,\qquad
\bar{d}^{a}_{\mu}\S_0=0\,,
\end{equation}
where the operators
$(\d^{a}_{\mu},\bar\d^{a}_{\mu},d^{a}_{\mu},\bar{d}^{a}_{\mu})$
act on the fields and sources as:
\begin{itemize}
\item{The $\d^{a}_{\mu}$-transformation:
\begin{eqnarray}
\d^{c}_{\nu}\bar{c}^{a}&=&\varphi^{ac}_{\nu}\,,\nonumber\\
\d^{c}_{\nu}\bar\varphi^{ab}_{\mu}&=&\d_{\mu\nu}\d^{bc}\,c^{a}\,,\nonumber\\
\d^{c}_{\nu}b^{a}&=&-igf^{abi}\varphi^{bc}_{\nu}c^{i}\,,\nonumber\\
\d^{c}_{\nu}\Omega^{a}_{\mu}&=&V^{ac}_{\mu\nu}\,,\nonumber\\
\d^{c}_{\nu}K^{a}_{\mu}&=&\frac{1}{2}\,V^{ac}_{\mu\nu}\,,\label{delta}
\end{eqnarray}}
\item{The $\bar\d^{a}_{\mu}$-transformation:
\begin{eqnarray}
\bar\d^{c}_{\nu}\bar{c}^{a}&=&\bar\omega^{ac}_{\nu}\,,\nonumber\\
\bar\d^{c}_{\nu}\omega^{ab}_{\mu}&=&-\d_{\mu\nu}\d^{bc}\,c^{a}\,,\nonumber\\
\bar\d^{c}_{\nu}b^{a}&=&-igf^{abd}\bar\omega^{bc}_{\nu}c^{d}
-igf^{abi}\bar\omega^{bc}_{\nu}c^{i}\,,\nonumber\\
\bar\d^{c}_{\nu}\Omega^{a}_{\mu}&=&\bar{U}^{ac}_{\mu\nu}\,,\nonumber\\
\bar\d^{c}_{\nu}K^{a}_{\mu}&=&\frac{1}{2}\,\bar{U}^{ac}_{\mu\nu}\,,
\end{eqnarray}}
\item{The $d^{a}_{\mu}$-transformation:
\begin{eqnarray}
d^{c}_{\nu}\bar{c}^{a}&=&\omega^{ac}_{\nu}+gf^{abi}\varphi^{bc}_{\nu}c^{i}\,,\nonumber\\
d^{c}_{\nu}\bar\omega^{ab}_{\mu}&=&\d_{\mu\nu}\d^{bc}\,c^{a}\,,\nonumber\\
d^{c}_{\nu}\bar\varphi^{ab}_{\mu}&=&-\d_{\mu\nu}\d^{bc}\Bigl(gf^{adi}c^{d}c^{i}
+\frac{g}{2}f^{ade}c^{d}c^{e}\Bigr)\,,\nonumber\\
d^{c}_{\nu}b^{a}&=&-igf^{abi}\omega^{bc}_{\nu}c^{i}
-i\frac{g^{2}}{2}f^{abi}f^{dei}\omega^{bc}_{\nu}c^{d}c^{e}\,,\nonumber\\
d^{c}_{\nu}\Omega^{a}_{\mu}&=&{U}^{ac}_{\mu\nu}\,,\nonumber\\
d^{c}_{\nu}K^{a}_{\mu}&=&\frac{1}{2}\,{U}^{ac}_{\mu\nu}\,,\nonumber\\
d^{c}_{\nu}\xi^{a}_{\mu}&=&-\frac{1}{2}\,{V}^{ac}_{\mu\nu}\,,\nonumber\\
d^{c}_{\nu}\vartheta^{a}_{\mu}&=&-{V}^{ac}_{\mu\nu}\,,
\end{eqnarray}}
\item{The $\bar{d}^{a}_{\mu}$-transformation:
\begin{eqnarray}
\bar{d}^{c}_{\nu}\bar{c}^{a}&=&-\bar\varphi^{ac}_{\nu}+gf^{abd}\bar\omega^{bc}_{\nu}c^{d}
+gf^{abi}\bar\omega^{bc}_{\nu}c^{i}\,,\nonumber\\
\bar{d}^{c}_{\nu}\varphi^{ab}_{\mu}&=&-\d_{\mu\nu}\d^{bc}\,c^{a}\,,\nonumber\\
\bar{d}^{c}_{\nu}\omega^{ab}_{\mu}&=&\d_{\mu\nu}\d^{bc}\Bigl(gf^{adi}c^{d}c^{i}
+\frac{g}{2}f^{ade}c^{d}c^{e}\Bigr)\,,\nonumber\\
\bar{d}^{c}_{\nu}b^{a}&=&igf^{abd}\bar\varphi^{bc}_{\nu}c^{c}
+igf^{abi}\bar\varphi^{bc}_{\nu}c^{i}
-ig^{2}f^{abd}f^{dei}\bar\omega^{bc}_{\nu}c^{e}c^{i}\nonumber\\
&&-i\frac{g^{2}}{2}f^{abi}f^{dei}\bar\omega^{bc}_{\nu}c^{d}c^{e}
-i\frac{g^{2}}{2}f^{abd}f^{def}\bar\omega^{bc}_{\nu}c^{e}c^{f}\,,\nonumber\\
\bar{d}^{c}_{\nu}\Omega^{a}_{\mu}&=&-\bar{V}^{ac}_{\mu\nu}\,,\nonumber\\
\bar{d}^{c}_{\nu}K^{a}_{\mu}&=&-\frac{1}{2}\,\bar{V}^{ac}_{\mu\nu}\,,\nonumber\\
\bar{d}^{c}_{\nu}\xi^{a}_{\mu}&=&-\frac{1}{2}\,\bar{U}^{ac}_{\mu\nu}\,,\nonumber\\
\bar{d}^{c}_{\nu}\vartheta^{a}_{\mu}&=&-\bar{U}^{ac}_{\mu\nu}\,.\label{dbar}
\end{eqnarray}}
\end{itemize}
We can also show that:
\begin{equation}
\{\d^{a}_{\mu},s\}=d^{a}_{\mu}\,,\qquad [\bar\d^{a}_{\mu},s]=\bar{d}^{a}_{\mu}\,,\qquad
[d^{a}_{\mu},s]=0\,,\qquad\{\bar{d}^{a}_{\mu},s\}=0\,.
\end{equation}
The symmetries of $\S_0$ will help us to determine a suitable
set of dimension two operators which can be consistently introduced in the theory.
This will be the subject of the following section.

\subsection{Dimension two condensates}
In this section we shall spend  a few words about the subject of the
dimension two condensates, which are the result of the condensation
of local dimension two operators. One should notice that the
introduction of the horizon function in its localized form,
expression \eqref{2ndLocal}, entails the introduction of a
dimension two condensate. In fact, the gap equation \eqref{gapeq},
implies that the dimension two operator
$(B^{ab}_{\mu}(\varphi^{ab}_{\mu}-\bar\varphi^{ab}_{\mu}))$
acquires a nonvanishing expectation value, {\it i.e.} $\langle
B^{ab}_{\mu}(\varphi^{ab}_{\mu}-\bar\varphi^{ab}_{\mu})\rangle\neq0$.
An analogous condensate is found in the Landau gauge
\cite{Zwanziger:1989mf,Zwanziger:1992qr,Dudal:2005na,Dudal:2007cw,Dudal:2008sp},
where the gap equation for the Gribov parameter $\gamma$ implies
that $\langle
f^{ABC}(\phi^{AB}_{\mu}-\bar\phi^{AB}_{\mu})A^{C}_{\mu}\rangle\neq0$,
where $(\phi^{AB}_{\mu},\bar\phi^{AB}_{\mu})$ are the auxiliary
fields needed for the localization of the horizon function in the
Landau gauge.\\\\Furthermore, in complete analogy with the case of the
Landau gauge \cite{Dudal:2005na,Dudal:2007cw,Dudal:2008sp}, other
dimension two condensates have to be taken into account in the
maximal Abelian gauge, see also \cite{Capri:2008ak} for the
particular case of $SU(2)$. More precisely, the following dimension
two operators can be introduced in a way which preserves
renormalizability of the theory as well as its symmetry content:
\begin{eqnarray}
&\mathcal{O}_{A^{2}}=A^{a}_{\mu}A^{a}_{\mu}\,,&\label{A2}\\\cr
&\mathcal{O}^{i}_{\bar{c}\times{c}}=gf^{abi}c^{a}c^{b}\,,&\label{ghostop}\\\cr
&\mathcal{O}_{\bar{f}f}=\bar\varphi^{ab}_{\mu}\varphi^{ab}_{\mu}-\bar\omega^{ab}_{\mu}\omega^{ab}_{\mu}
-\bar{c}^{a}c^{a}\,.&\label{ff}
\end{eqnarray}
The operator \eqref{A2} is related to the dynamical mass
generation for the off-diagonal gluons, a feature which supports the
Abelian dominance hypothesis. Its condensation has been
established in \cite{Dudal:2004rx}, where a dynamical off-diagonal
gluon mass has been reported. The ghost operator \eqref{ghostop} is needed in
order to account for the dynamical breaking of the
$SL(2,\mathbb{R})$ symmetry present in the ghost sector of the
maximal Abelian gauge:
\begin{equation}
\d\S_{0}=0\,,\qquad\d\bar{c}^{a}=c^{a}\,,\qquad\d b^{a}=gf^{abi}c^{b}c^{i}+\frac{g}{2}f^{abc}c^{b}c^{c}\,.
\end{equation}
Its condensation has been analyzed recently in
\cite{Capri:2007hw}. Concerning now the third operator,
eq.\eqref{ff}, we notice that it depends on the auxiliary fields
$(\varphi,\bar\varphi,\omega,\bar\omega)$. It is in fact needed to
account for the nontrivial dynamics developed by those fields.
Besides, one can see that this operator is invariant under the
transformations \eqref{delta}--\eqref{dbar}. An analogous operator
has been found in the Landau gauge
\cite{Dudal:2005na,Dudal:2007cw,Dudal:2008sp}, where it has allowed to
reconcile the Gribov-Zwanziger framework with the most recent
lattice data on the gluon and ghost propagators
\cite{Cucchieri:2007rg,Cucchieri:2008fc}. \\\\Let us briefly show how these operators can be introduced
in the theory, by taking the example of the operator $\mathcal{O}_{A^{2}}$. The other operators can be handled by following an analogous path. Let us introduce a BRST doublet of external sources:
\begin{equation}
s\lambda=J\,,\qquad sJ=0\,,
\end{equation}
and let us define the following BRST invariant term
\begin{eqnarray}
S_{A^{2}}&=&s\int{d^{4}x}\,\left(\frac{1}{2}\,\lambda\,\mathcal{O}_{A^{2}}
-\frac{\zeta}{2}\,\lambda J\right)\nonumber\\
&=&\int{d^{4}x}\,\left(\frac{1}{2}\,J\,\mathcal{O}_{A^{2}}
+\frac{1}{2}\,\lambda\,s\mathcal{O}_{A^{2}}-\frac{\zeta}{2}\,J^{2}\right)\,, \label{aa2}
\end{eqnarray}
where $\zeta$ is a dimensionless constant parameter necessary to
account for the ultraviolet divergences affecting the correlation function
\begin{equation}
\langle\mathcal{O}_{A^{2}}(x)\mathcal{O}_{A^{2}}(y)\rangle\,.
\end{equation}
As discussed in \cite{Dudal:2004rx}, the parameter $\zeta$ is uniquely determined by the renormalization group equations. Expression \eqref{aa2} is thus added to the action $\S_{0}$, giving
\begin{equation}
\S_1=\S_0 +S_{A^{2}}\,.
\end{equation}
Keeping the source $J(x)$ and setting all other external sources to their respective physical
values\footnote{We notice that the external sources
$(\Omega^{a,i}_{\mu}, K^{a}_{\mu}, \Upsilon^{a}_{\mu},
\xi^{a}_{\mu}, \vartheta^{a}_{\mu}, L^{a,i})$ and $\lambda(x)$
carry nonvanishing ghost number, so that their physical values vanish.}, we can introduce  the generating functional  $\mathcal{W}[J]$ according to
\begin{equation}
e^{-\mathcal{W}[J]}=\int \mathcal{D}\Phi\,e^{-\S_{0}-\int{d^{4}x}\,\left(\frac{1}{2}\,J\mathcal{O}_{A^{2}}
-\frac{\zeta}{2}\,J^{2}\right)}\,.\label{W[J]}
\end{equation}
The vacuum expectation value of  the operator $\mathcal{O}_{A^{2}}$ is
then obtained by differentiating with respect to $J$:
\begin{equation}
\frac{\d\mathcal{W}[J]}{\d{J}}\biggl|_{J=0}=-\frac{1}{2}\,\langle\mathcal{O}_{A^{2}}\rangle\,. \label{defj}
\end{equation}
In practice, to solve \eqref{defj} results in a difficult task. A shortcut is usually employed, amounting to make use of the Hubbard-Stratonovich field $\eta(x)$. To introduce this field in the theory, one inserts
the unity
\begin{equation}
1=\mathcal{N}\int\mathcal{D}\eta\,e^{-\frac{1}{2\zeta}\int{d^{4}x}\,\left(\frac{\eta}{g}
+\frac{1}{2}\,\mathcal{O}_{A^{2}}-\zeta J\right)^{2}}\,,
\end{equation}
where $\mathcal{N}$ is a normalization factor, so that expression \eqref{W[J]} becomes
\begin{eqnarray}
e^{-\mathcal{W}[J]}&=&\int \mathcal{D}\Phi\mathcal{D}\eta\,e^{-\S_{\eta}
+\int{d^{4}x\,\frac{\eta}{g}J}}\,,\nonumber\\
\S_{\eta}&=&\S_{0}+\int{d^{4}x}\,\biggl(\frac{\eta^{2}}{2g^{2}\zeta}
+\frac{\eta}{2g\zeta}\,\mathcal{O}_{A^{2}}
+\frac{1}{8\zeta}\,(\mathcal{O}_{A^{2}})^{2}\biggr)\,. \label{seta}
\end{eqnarray}
With $\mathcal{W}[J]$ written in this way, we can achieve the
following relation
\begin{equation}
\langle\mathcal{O}_{A^{2}}\rangle=-\frac{2}{g}\,\langle\,\eta\,\rangle\,, \label{vaa2}
\end{equation}
which easily follows from expression \eqref{seta} upon differentiation with respect to the source $J$. The advantage of having introduced the Hubbard-Stratonovich field $\eta$ relies on the fact that the quadratic term $J^2$  in eq.\eqref{W[J]} has been replaced by the term  $\S_{\eta}$,  eq.\eqref{seta}. Also, from eq.\eqref{vaa2} one observes that a nonvanishing vacuum expectation value of the Hubbard-Stratonovich field $\eta$ gives a nonvanishing condensate $\langle\mathcal{O}_{A^{2}}\rangle$. It remains thus to find out whether the field $\eta$ acquires a nonvanishing vacuum expectation value, a task which can be accomplished by evaluating the effective potential corresponding to the action  $\S_{\eta}$. A detailed account of the analysis of the effective potential can be found in \cite{Dudal:2004rx}, where a nonvanishing vacuum expectation value for $\eta$ has emerged. Setting
\begin{eqnarray}
\eta(x)&=&\langle\,\eta\,\rangle+\tilde{\eta}(x)\,,\nonumber\\
\langle\,\tilde{\eta}\,\rangle&=&0\,,
\end{eqnarray}
we get
\begin{equation}
\S_{\eta}=\S_{0}+\int{d^{4}x}\,\biggl(\frac{\langle\,\eta\,\rangle^{2}}{2g^{2}\zeta}
+\frac{{\tilde{\eta}}^{2}}{2g^{2}\zeta}
+\frac{\langle\,\eta\,\rangle}{2g\zeta}\,\mathcal{O}_{A^{2}}
+\frac{\tilde\eta}{2g\zeta}\,\mathcal{O}_{A^{2}}
+\frac{1}{8\zeta}\,(\mathcal{O}_{A^{2}})^{2}\biggr)\,.
\end{equation}
Introducing thus the gluon mass
\begin{equation}
m^{2}=\frac{\langle\,\eta\,\rangle}{g\zeta}\,,
\end{equation}
we can also write
\begin{equation}
\S_{\eta}=\S_{0}+\int{d^{4}x}\,\biggl(
\frac{\zeta m^{4}}{2}
+\frac{{\tilde{\eta}}^{2}}{2g^{2}\zeta}
+\frac{m^{2}}{2}\,\mathcal{O}_{A^{2}}
+\frac{\tilde\eta}{2g\zeta}\,\mathcal{O}_{A^{2}}
+\frac{1}{8\zeta}\,(\mathcal{O}_{A^{2}})^{2}\biggr)\,,
\end{equation}
from which one sees that the condensation of the operator
$\mathcal{O}_{A^{2}}$ results in the dynamical generation of a gluon mass, {\it i.e.}
$\frac{m^{2}}{2}\,A^{a}_{\mu}A^{a}_{\mu}$.  This term
will affect the tree level off-diagonal gluon propagator.
In much the same way, the other operators
$\mathcal{O}_{\bar{c}\times{c}}^{i}$ and $\mathcal{O}_{\bar{f}f}$
will affect the propagators of theory even at the tree level,
as we shall see in the next section.
\subsection{Tree level gluon and ghost propagators}
In this section we will establish the qualitative behavior of the
gluon and ghost propagators by taking into account the effects of the
restriction to the Gribov region and of the condensation of the
dimension two operators \eqref{A2}--\eqref{ff}, encoded in the following dynamical parameters
\begin{equation}
\langle\mathcal{O}_{A^{2}}\rangle\sim m^{2}\,,\qquad
\langle\mathcal{O}_{\bar{c}\times{c}}^{i}\rangle\sim v^{i}\,,\qquad
\langle\mathcal{O}_{\bar{f}f}\rangle\sim \mu^{2}\,.
\end{equation}
Such parameters will appear in the resulting action as
\begin{equation}
\S=\S_{0}+\int{d^{4}x}\,\Bigl(\,\frac{m^{2}}{2}\,\mathcal{O}_{A^{2}}
+v^{i}\,\mathcal{O}_{\bar{c}\times{c}}^{i}
+\mu^{2}\,\mathcal{O}_{\bar{f}f}+\mathrm{``interaction\,terms"}\Bigr)\,,
\end{equation}
where $\S$ is the complete action containing all  condensates. In order to evaluate the propagators,
it is sufficient to consider only the quadratic terms of $\S$:
\begin{eqnarray}
\S_{\mathrm{quad}}&\!\!\!=\!\!\!&\lim_{{\alpha\to0}\atop{\beta\to0}}\int{d^{4}x}\biggl[
\frac{1}{2}A^{a}_{\mu}\Bigl((-\partial^{2}+m^{2})\d_{\mu\nu}
-\frac{1-\alpha}{\alpha}\,\partial_{\mu}\partial_{\nu}\Bigr)A^{a}_{\nu}
+\frac{1}{2}A^{i}_{\mu}\Bigl(-\partial^{2}\d_{\mu\nu}
-\frac{1-\beta}{\beta}\,\partial_{\mu}\partial_{\nu}\Bigr)A^{i}_{\nu}\nonumber\\
&&-\bar{c}^{a}((-\partial^{2}+\mu^{2})\d^{ab}-v^{i}gf^{abi})c^{b}-\bar{c}^{i}(-\partial^{2})c^{i}
+\bar{\varphi}^{ab}_{\mu}(-\partial^{2}+\mu^{2})\varphi^{ab}_{\mu}
-\bar{\omega}^{ab}_{\mu}(-\partial^{2}+\mu^{2})\omega^{ab}_{\mu}\nonumber\\
&&+\sqrt{\frac{2g^{2}\gamma^{4}}{N(N-1)}}\,\Bigl(f^{abi}A^{i}_{\mu}
+\frac{1}{2}f^{abc}A^{c}_{\mu}\Bigr)(\varphi^{ab}_{\mu}-\bar\varphi^{ab}_{\mu})\biggr]\,,
\end{eqnarray}
where we have already integrated out the Lagrange multipliers $(b^{a},b^{i})$ and where we have taken
the physical values of the sources $(V,\bar{V},U,\bar{U})$.
A further integration over the auxiliary localizing fields
$(\varphi,\bar\varphi,\omega,\bar\omega)$ gives the following expression  in momentum
space
\begin{eqnarray}
\S_{\mathrm{quad}}&=&\lim_{{\alpha\to0}\atop{\beta\to0}}
\int\frac{d^{4}k}{(2\pi)^{4}}\,\biggl(
\frac{1}{2}\,A^{a}_{\mu}(k)\,\mathcal{P}^{ab}_{\mu\nu}(k;\alpha)\,A^{b}_{\nu}(-k)
+\frac{1}{2}\,A^{i}_{\mu}(k)\,\mathcal{Q}^{ij}_{\mu\nu}(k;\beta)\,A^{j}_{\nu}(-k)\nonumber\\
&&-c^{a}(k)\,\mathcal{R}^{ab}(k)\,c^{b}(-k)
-c^{i}(k)\,\d^{ij}k^{2}\,c^{j}(-k)\biggr)\,,
\end{eqnarray}
where
\begin{eqnarray}
\mathcal{P}^{ab}_{\mu\nu}(k;\alpha)&=&\d^{ab}\biggl(\d_{\mu\nu}\,\frac{(k^{2}+m^{2})(k^{2}+\mu^{2})
+(N-2)g^{2}\gamma^{4}/N(N-1)}{k^{2}+\mu^{2}}
-\frac{1-\alpha}{\alpha}\,k_{\mu}k_{\nu}\biggr)\,,\nonumber\\
\mathcal{Q}^{ij}_{\mu\nu}(k;\beta)&=&\d^{ij}\biggl(\d_{\mu\nu}\,\frac{k^{2}(k^{2}+\mu^{2})
+4g^{2}\gamma^{4}/(N-1)}{k^{2}+\mu^{2}}
-\frac{1-\beta}{\beta}\,k_{\mu}k_{\nu}\biggr)\,,\nonumber\\
\mathcal{R}^{ab}_{\mu\nu}(k)&=&\d^{ab}(k^{2}+\mu^{2}) -gf^{abi}v^{i}\,.
\end{eqnarray}
The tree level propagators of the theory are thus given by:
\begin{itemize}
\item{{\bf The off-diagonal gluon propagator:}
\begin{equation}
\langle A^{a}_{\mu}(k)A^{b}_{\nu}(-k)\rangle
=\frac{k^{2}+\mu^{2}}{(k^{2}+m^{2})(k^{2}+\mu^{2})+\frac{(N-2)g^{2}\gamma^{4}}{N(N-1)}}\,
\biggl(\d_{\mu\nu}-\frac{k_{\mu}k_{\nu}}{k^{2}}\biggr)\d^{ab}\,; \label{offdgln}
\end{equation} }
\item{{\bf The diagonal gluon propagator:}
\begin{equation}
\langle A^{i}_{\mu}(k)A^{j}_{\nu}(-k)\rangle
=\frac{k^{2}+\mu^{2}}{k^{2}(k^{2}+\mu^{2})+\frac{4g^{2}\gamma^{4}}{(N-1)}}\,
\biggl(\d_{\mu\nu}-\frac{k_{\mu}k_{\nu}}{k^{2}}\biggr)\d^{ij}\,; \label{dgln}
\end{equation} }
\item{{\bf The symmetric off-diagonal ghost propagator:}
\begin{equation}
\langle \bar{c}^{a}(k)c^{b}(-k)\rangle_{\mathrm{symm}}
=\frac{k^{2}+\mu^{2}}{(k^{2}+\mu^{2})^{2}+\frac{g^{2}v^{2}}{(N-1)}}\,\d^{ab}\,;\label{symmghostprop}
\end{equation} }
\item{{\bf The antisymmetric off-diagonal ghost propagator:}
\begin{equation}
\langle \bar{c}^{a}(k)c^{b}(-k)\rangle_{\mathrm{antisymm}}
=\frac{gf^{abi}v^{i}}{(k^{2}+\mu^{2})^{2}+\frac{g^{2}v^{2}}{(N-1)}}\,;\label{antisymmghostprop}
\end{equation} }
\item{{\bf The diagonal ghost propagator:}
\begin{equation}
\langle \bar{c}^{i}(k)c^{j}(-k)\rangle
=\frac{1}{k^{2}}\,\d^{ij}\,. \label{dgh}
\end{equation} }
\end{itemize}
With the exception of the diagonal ghost propagator, eq.\eqref{dgh}, which exhibits a free behavior, we  observe that all remaining propagators turn out to be suppressed in the infrared, a result which can be seen as a direct generalization of what has been found in the case of $SU(2)$ \cite{Capri:2008ak}.\\\\The off-diagonal gluon propagator deserves a little comment. Notice that it can be written as
\begin{equation}
\langle A^{a}_{\mu}(k)A^{b}_{\nu}(-k)\rangle=\biggl(\frac{1}{k^{2}+m^{2}}-\rho_{N}(k)\biggr)
\biggl(\d_{\mu\nu}-\frac{k_{\mu}k_{\nu}}{k^2}\biggr)\d^{ab}\,,
\end{equation}
where $\rho_{N}(k)$ represents  the deviation from the Yukawa type behavior, and is
given by
\begin{equation}
\rho_{N}(k)=\frac{\frac{(N-2)g^{2}\gamma^{4}}{N(N-1)}}{(k^{2}+m^{2})
\Bigl[(k^{2}+m^{2})(k^{2}+\mu^{2})+\frac{(N-2)g^{2}\gamma^{4}}{N(N-1)}\Bigr]}\,.
\end{equation}
The factor $\rho_{N}(k)$ vanishes for $N=2$, so that the  off-diagonal
gluon propagator behaves exactly like the Yukawa propagator, as
obtained in \cite{Capri:2005tj,Capri:2008ak} for the particular $SU(2)$ case. However, for $N>2$, it seems to deviate from a pure Yukawa behavior, a feature which would be worth to investigate by lattice simulations in
the relevant case of  $SU(3)$.
\section{Conclusions}
In this work we have studied a few aspects of the issue of the Gribov copies in $SU(N)$ Euclidean Yang-Mills
theories in the maximal Abelian gauge. The so-called Gribov region $\Omega$ has been introduced and
some of its properties have been established.  Summarizing, the region
$\Omega$ is convex, bounded in all off-diagonal directions, and
unbounded in all diagonals ones. \\\\The implementation of the restriction of the domain of integration in the functional integral to the region $\Omega$ has been considered.  A careful study of the horizon function has been provided. In particular, we have shown that use of  Gribov's no pole condition allows us to select only one candidate, given by expression \eqref{2nd}. A  local
action from the restriction to the region $\Omega$, eq.\eqref{2ndLocal}, has been constructed and its symmetry content established.
\\\\It is worth mentioning  that the results obtained in
\cite{Capri:2005tj,Capri:2008vk,Capri:2008ak} for the particular
case of $SU(2)$ can be completely recovered by setting $N=2$, which provides a very good check. \\\\A  detailed analysis of the propagators
of the theory has been performed. The general case of $SU(N)$, $N>2$, displays
differences with respect to the case of $SU(2)$.
In particular, as can be observed from expression \eqref{offdgln},
the off-diagonal component of the gluon propagator turns out to be affected
by the restriction to $\Omega$ as well as by the condensation
of the operators \eqref{A2}--\eqref{ff}. The diagonal gluon propagator, eq.\eqref{dgln},
exhibits a Gribov-Stingl type behavior depending on the Gribov
parameter $\gamma$ and on the parameter $\mu$ corresponding to the
condensation of the operator \eqref{ff}. The symmetric
off-diagonal ghost propagator, eq.\eqref{symmghostprop},  turns out to be dependent
from the parameters $\mu$ and $v^{2}=v^{i}v^{i}$ while, as expected,  the antisymmetric off-diagonal ghost
propagator, eq.\eqref{antisymmghostprop}, turns out to be directly proportional to the parameter $v^i$ stemming from the ghost condensate $\langle\mathcal{O}^{i}_{\mathrm{ghost}}\rangle\sim v^{i}$. \\\\All propagators  are seen to be suppressed in the infrared.  Moreover, they are
nonvanishing at $k=0$. Although  numerical studies of the gluon and ghost propagators in the maximal Abelian gauge have been performed only in the case of $SU(2)$, these features seem to be in very good agreement with the most recent numerical data  \cite{Mendes:2006kc}. From this point of view, it would be rather interesting to perform a numerical study of $SU(3)$ in order to check our prevision.

\section*{Acknowledgments}

The Conselho Nacional de Desenvolvimento Cient\'{\i}fico e
Tecnol\'{o}gico (CNPq-Brazil), the Faperj, Funda{\c{c}}{\~{a}}o de
Amparo {\`{a}} Pesquisa do Estado do Rio de Janeiro, the Latin
American Center for Physics (CLAF), the SR2-UERJ and the
Coordena{\c{c}}{\~{a}}o de Aperfei{\c{c}}oamento de Pessoal de
N{\'{\i}}vel Superior (CAPES) are gratefully acknowledged for
financial support.

\begin{appendix}
\section{Details on the calculation of the third order off-diagonal ghost correlation function}
\label{wicks}
In order to evaluate the off-diagonal diagonal ghost correlation function $\langle\bar{c}^{a}(x)c^{b}(y)\rangle$,
let us start with the usual Gell-Mann \& Low formula
theory:
\begin{equation}
\langle\bar{c}^{a}(x)c^{b}(y)\rangle=\langle0|\bar{c}^{a}_{0}(x)c^{b}_{0}(y)
\,e^{-S_{\mathrm{int}}[\bar{c}_{0},c_{0}]}|0\rangle\;, \label{gm}
\end{equation}
where the fields appearing in the right hand side of eq.\eqref{gm} are free fields and $S_{\mathrm{int}}[\bar{c},c]$ stands for  the interaction term given by:
\begin{eqnarray}
S_{\mathrm{int}}[\bar{c},c]&=&-\int{d^{4}x}\,\Bigl(2gf^{abi}A^{i}_{\mu}\bar{c}^{a}\partial_{\mu}c^{b}
+gf^{abc}A^{c}_{\mu}\bar{c}^{a}\partial_{\mu}c^{b}
+g^{2}f^{aci}f^{bcj}A^{i}_{\mu}A^{j}_{\mu}\bar{c}^{a}c^{b}\nonumber\\
&&+g^{2}f^{adc}f^{bdi}A^{c}_{\mu}A^{i}_{\mu}\bar{c}^{a}c^{b}
-g^{2}f^{cai}f^{dbi}A^{c}_{\mu}A^{d}_{\mu}\bar{c}^{a}c^{b}\Bigr)\,.
\end{eqnarray}
To evaluate the aforementioned  two-point correlation function we expand the
term $e^{-S_{\mathrm{int}}}$ till the third order, so that
\begin{eqnarray}
\langle\bar{c}^{a}(x)c^{b}(y)\rangle&=&\langle0|\bar{c}^{a}(x)c^{b}(y)\,\biggl(1-S_{\mathrm{int}}
+\frac{1}{2}\,S_{\mathrm{int}}^{2}-\frac{1}{6}\,S_{\mathrm{int}}^{3}+\cdots\biggr)|0\rangle\nonumber\\
&=&\langle0|\bar{c}^{a}(x)c^{b}(y)|0\rangle
-\langle0|\bar{c}^{a}(x)c^{b}(y)\,S_{\mathrm{int}}|0\rangle
+\frac{1}{2}\,\langle0|\bar{c}^{a}(x)c^{b}(y)\,S_{\mathrm{int}}^{2}|0\rangle\nonumber\\
&&-\frac{1}{6}\,\langle0|\bar{c}^{a}(x)c^{b}(y)\,S_{\mathrm{int}}^{3}|0\rangle+\cdots\,,
\end{eqnarray}
where
\begin{equation}
\langle0|\bar{c}^{a}(x)c^{b}(y)|0\rangle=\d^{ab}G_{0}(x-y)\,,\label{G_0}
\end{equation}
with
\begin{equation}
G_{0}(x-y)=\int\frac{d^{4}q}{(2\pi)^{4}}\frac{1}{q^{2}}\,e^{iq(x-y)}\,.
\end{equation}
We are interested in terms of order three in the gauge field $A$ or,
equivalently,  of order $g^{3}$. These terms appear in  the expression for
$S_{\mathrm{int}}^{2}$ and $S_{\mathrm{int}}^{3}$. Therefore, for the third order correlation function, $G^{ab}_{3}(x,y;A)$, one writes
{\footnotesize
\begin{eqnarray}
G^{ab}_{3}(x,y;A)&\!\!\!=\!\!\!&\frac{1}{2}\biggl\langle\bar{c}^{a}(x)c^{b}(y)\int{d^{4}x_{1}d^{4}x_{2}}\,
\Bigl(4g^{3}f^{a_1b_1i_1}f^{a_2c_2i_2}f^{b_2c_2j_2}A^{i_1}_{\mu}(x_1)A^{i_2}_{\nu}(x_2)A^{j_2}_{\nu}(x_2)\nonumber\\
&&+4g^{3}f^{a_1b_1i_1}f^{a_2d_2c_2}f^{b_2d_2i_2}A^{i_1}_{\mu}(x_1)A^{c_2}_{\nu}(x_2)A^{i_2}_{\nu}(x_2)
-4g^{3}f^{a_1b_1i_1}f^{c_2a_2i_2}f^{d_2b_2i_2}A^{i_1}_{\mu}(x_1)A^{c_2}_{\nu}(x_2)A^{d_2}_{\nu}(x_2)\nonumber\\
&&+2g^{3}f^{a_1b_1c_1}f^{a_2c_2i_2}f^{b_2c_2j_2}A^{c_1}_{\mu}(x_1)A^{i_2}_{\nu}(x_2)A^{j_2}_{\nu}(x_2)
+2g^{3}f^{a_1b_1c_1}f^{a_2d_2c_2}f^{b_2d_2i_2}A^{c_1}_{\mu}(x_1)A^{c_2}_{\nu}(x_2)A^{i_2}_{\nu}(x_2)\nonumber\\
&&-2g^{3}f^{a_1b_1c_1}f^{c_2a_2i_2}f^{d_2b_2i_2}A^{c_1}_{\mu}(x_1)A^{c_2}_{\nu}(x_2)A^{d_2}_{\nu}(x_2)\Bigr)
\bar{c}^{a_1}(x_1)\,\partial^{x_1}_{\mu}c^{b_1}(x_1)\,\bar{c}^{a_2}(x_2)c^{b_2}(x_2)\biggr\rangle\nonumber\\
&&+\frac{1}{6}\biggl\langle\bar{c}^{a}(x)c^{b}(y)\int{d^{4}x_{1}d^{4}x_{2}d^{4}x_{3}}\,
\Bigl(8g^{3}f^{a_1b_1i_1}f^{a_2b_2i_2}f^{a_3b_3i_3}A^{i_1}_{\mu}(x_1)A^{i_2}_{\nu}(x_2)A^{i_3}_{\s}(x_3)\nonumber\\
&&+12g^{3}f^{a_1b_1i_1}f^{a_2b_2i_2}f^{a_3b_3c_3}A^{i_1}_{\mu}(x_1)A^{i_2}_{\nu}(x_2)A^{c_3}_{\s}(x_3)
+6g^{3}f^{a_1b_1i_1}f^{a_2b_2c_2}f^{a_3b_3c_3}A^{i_1}_{\mu}(x_1)A^{c_2}_{\nu}(x_2)A^{c_3}_{\s}(x_3)\nonumber\\
&&+g^{3}f^{a_1b_1c_1}f^{a_2b_2c_2}f^{a_3b_3c_3}A^{c_1}_{\mu}(x_1)A^{c_2}_{\nu}(x_2)A^{c_3}_{\s}(x_3)\Bigr)\nonumber\\
&&\times\bar{c}^{a_1}(x_1)\,\partial^{x_1}_{\mu}c^{b_1}(x_1)\,
\bar{c}^{a_2}(x_2)\,\partial^{x_2}_{\mu}c^{b_2}(x_2)\,
\bar{c}^{a_3}(x_3)\,\partial^{x_3}_{\mu}c^{b_3}(x_3)\biggr\rangle\,.
\end{eqnarray}}
Performing the Wick contractions we get
{\footnotesize
\begin{eqnarray}
G^{ab}_{3}(x,y;A)&\!\!\!=\!\!\!&\frac{1}{2}\int{d^{4}x_{1}d^{4}x_{2}}\,
\Bigl(4g^{3}f^{a_1b_1i_1}f^{a_2c_2i_2}f^{b_2c_2j_2}A^{i_1}_{\mu}(x_1)A^{i_2}_{\nu}(x_2)A^{j_2}_{\nu}(x_2)\nonumber\\
&&+4g^{3}f^{a_1b_1i_1}f^{a_2d_2c_2}f^{b_2d_2i_2}A^{i_1}_{\mu}(x_1)A^{c_2}_{\nu}(x_2)A^{i_2}_{\nu}(x_2)
-4g^{3}f^{a_1b_1i_1}f^{c_2a_2i_2}f^{d_2b_2i_2}A^{i_1}_{\mu}(x_1)A^{c_2}_{\nu}(x_2)A^{d_2}_{\nu}(x_2)\nonumber\\
&&+2g^{3}f^{a_1b_1c_1}f^{a_2c_2i_2}f^{b_2c_2j_2}A^{c_1}_{\mu}(x_1)A^{i_2}_{\nu}(x_2)A^{j_2}_{\nu}(x_2)
+2g^{3}f^{a_1b_1c_1}f^{a_2d_2c_2}f^{b_2d_2i_2}A^{c_1}_{\mu}(x_1)A^{c_2}_{\nu}(x_2)A^{i_2}_{\nu}(x_2)\nonumber\\
&&-2g^{3}f^{a_1b_1c_1}f^{c_2a_2i_2}f^{d_2b_2i_2}A^{c_1}_{\mu}(x_1)A^{c_2}_{\nu}(x_2)A^{d_2}_{\nu}(x_2)\Bigr)\nonumber\\
&&\times\Bigl[\Bigl(\partial^{x_1}_{\mu}\langle\bar{c}^{a}(x)c^{b_1}(x_1)\rangle\Bigr)
\langle\bar{c}^{a_2}(x_2)c^{b}(y)\rangle \langle\bar{c}^{a_1}(x_1)c^{b_2}(x_2)\rangle
\nonumber\\
&&+\langle\bar{c}^{a}(x)c^{b_2}(x_2)\rangle
\langle\bar{c}^{a_1}(x_1)c^{b}(y)\rangle\,\partial_{\mu}^{x_1}\langle\bar{c}^{a_2}(x_2)c^{b_1}(x_1)\rangle\Bigr]\nonumber\\
&&-\frac{1}{6}\int{d^{4}x_{1}d^{4}x_{2}d^{4}x_{3}}\,
\Bigl(8g^{3}f^{a_1b_1i_1}f^{a_2b_2i_2}f^{a_3b_3i_3}A^{i_1}_{\mu}(x_1)A^{i_2}_{\nu}(x_2)A^{i_3}_{\s}(x_3)\nonumber\\
&&+12g^{3}f^{a_1b_1i_1}f^{a_2b_2i_2}f^{a_3b_3c_3}A^{i_1}_{\mu}(x_1)A^{i_2}_{\nu}(x_2)A^{c_3}_{\s}(x_3)
+6g^{3}f^{a_1b_1i_1}f^{a_2b_2c_2}f^{a_3b_3c_3}A^{i_1}_{\mu}(x_1)A^{c_2}_{\nu}(x_2)A^{c_3}_{\s}(x_3)\nonumber\\
&&+g^{3}f^{a_1b_1c_1}f^{a_2b_2c_2}f^{a_3b_3c_3}A^{c_1}_{\mu}(x_1)A^{c_2}_{\nu}(x_2)A^{c_3}_{\s}(x_3)\Bigr)\nonumber\\
&&\times\Bigl[\Bigl(\partial_{\mu}^{x_1}\langle\bar{c}^{a}(x)c^{b_1}(x_1)\rangle\Bigr)
\langle\bar{c}^{a_2}(x_2)c^{b}(y)\rangle \Bigl(\partial_{\s}^{x_3}\langle\bar{c}^{a_1}(x_1)c^{b_3}(x_3)\rangle\Bigr)
\partial_{\nu}^{x_2}\langle\bar{c}^{a_3}(x_3)c^{b_2}(x_2)\rangle\nonumber\\
&&+\Bigl(\partial_{\mu}^{x_1}\langle\bar{c}^{a}(x)c^{b_1}(x_1)\rangle\Bigr)
\langle\bar{c}^{a_3}(x_3)c^{b}(y)\rangle \Bigl(\partial_{\nu}^{x_2}\langle\bar{c}^{a_1}(x_1)c^{b_2}(x_2)\rangle\Bigr)
\partial_{\s}^{x_3}\langle\bar{c}^{a_2}(x_2)c^{b_3}(x_3)\rangle\nonumber\\
&&+\Bigl(\partial_{\nu}^{x_2}\langle\bar{c}^{a}(x)c^{b_2}(x_2)\rangle\Bigr)
\langle\bar{c}^{a_1}(x_1)c^{b}(y)\rangle \Bigl(\partial_{\mu}^{x_1}\langle\bar{c}^{a_3}(x_3)c^{b_1}(x_1)\rangle\Bigr)
\partial_{\s}^{x_3}\langle\bar{c}^{a_2}(x_2)c^{b_3}(x_3)\rangle\nonumber\\
&&+\Bigl(\partial_{\nu}^{x_2}\langle\bar{c}^{a}(x)c^{b_2}(x_2)\rangle\Bigr)
\langle\bar{c}^{a_3}(x_3)c^{b}(y)\rangle \Bigl(\partial_{\s}^{x_3}\langle\bar{c}^{a_1}(x_1)c^{b_3}(x_3)\rangle\Bigr)
\partial_{\mu}^{x_1}\langle\bar{c}^{a_2}(x_2)c^{b_1}(x_1)\rangle\nonumber\\
&&+\Bigl(\partial_{\s}^{x_3}\langle\bar{c}^{a}(x)c^{b_3}(x_3)\rangle\Bigr)
\langle\bar{c}^{a_1}(x_1)c^{b}(y)\rangle \Bigl(\partial_{\mu}^{x_1}\langle\bar{c}^{a_2}(x_2)c^{b_1}(x_1)\rangle\Bigr)
\partial_{\nu}^{x_2}\langle\bar{c}^{a_3}(x_3)c^{b_2}(x_2)\rangle\nonumber\\
&&+\Bigl(\partial_{\s}^{x_3}\langle\bar{c}^{a}(x)c^{b_3}(x_3)\rangle\Bigr)
\langle\bar{c}^{a_2}(x_2)c^{b}(y)\rangle \Bigl(\partial_{\nu}^{x_2}\langle\bar{c}^{a_1}(x_1)c^{b_2}(x_2)\rangle\Bigr)
\partial_{\mu}^{x_1}\langle\bar{c}^{a_3}(x_3)c^{b_1}(x_1)\rangle\Bigr]\,.
\end{eqnarray}}
Renaming now some dummy indices, it follows that
{\footnotesize
\begin{eqnarray}
G^{ab}_{3}(x,y;A)&\!\!\!=\!\!\!&\frac{1}{2}\int{d^{4}x_{1}d^{4}x_{2}}\,
\Bigl(4g^{3}f^{a_1b_1i_1}f^{a_2c_2i_2}f^{b_2c_2j_2}A^{i_1}_{\mu}(x_1)A^{i_2}_{\nu}(x_2)A^{j_2}_{\nu}(x_2)\nonumber\\
&&+4g^{3}f^{a_1b_1i_1}f^{a_2d_2c_2}f^{b_2d_2i_2}A^{i_1}_{\mu}(x_1)A^{c_2}_{\nu}(x_2)A^{i_2}_{\nu}(x_2)
-4g^{3}f^{a_1b_1i_1}f^{c_2a_2i_2}f^{d_2b_2i_2}A^{i_1}_{\mu}(x_1)A^{c_2}_{\nu}(x_2)A^{d_2}_{\nu}(x_2)\nonumber\\
&&+2g^{3}f^{a_1b_1c_1}f^{a_2c_2i_2}f^{b_2c_2j_2}A^{c_1}_{\mu}(x_1)A^{i_2}_{\nu}(x_2)A^{j_2}_{\nu}(x_2)
+2g^{3}f^{a_1b_1c_1}f^{a_2d_2c_2}f^{b_2d_2i_2}A^{c_1}_{\mu}(x_1)A^{c_2}_{\nu}(x_2)A^{i_2}_{\nu}(x_2)\nonumber\\
&&-2g^{3}f^{a_1b_1c_1}f^{c_2a_2i_2}f^{d_2b_2i_2}A^{c_1}_{\mu}(x_1)A^{c_2}_{\nu}(x_2)A^{d_2}_{\nu}(x_2)\Bigr)\nonumber\\
&&\times\Bigl[\Bigl(\partial^{x_1}_{\mu}\langle\bar{c}^{a}(x)c^{b_1}(x_1)\rangle\Bigr)
\langle\bar{c}^{a_2}(x_2)c^{b}(y)\rangle \langle\bar{c}^{a_1}(x_1)c^{b_2}(x_2)\rangle
\nonumber\\
&&+\langle\bar{c}^{a}(x)c^{b_2}(x_2)\rangle
\langle\bar{c}^{a_1}(x_1)c^{b}(y)\rangle\,\partial_{\mu}^{x_1}\langle\bar{c}^{a_2}(x_2)c^{b_1}(x_1)\rangle\Bigr]\nonumber\\
&&-\frac{1}{6}\int{d^{4}x_{1}d^{4}x_{2}d^{4}x_{3}}\,
\Bigl(48g^{3}f^{a_1b_1i_1}f^{a_2b_2i_2}f^{a_3b_3i_3}A^{i_1}_{\mu}(x_1)A^{i_2}_{\nu}(x_2)A^{i_3}_{\s}(x_3)\nonumber\\
&&+24g^{3}f^{a_1b_1i_1}f^{a_2b_2i_2}f^{a_3b_3c_3}A^{i_1}_{\mu}(x_1)A^{i_2}_{\nu}(x_2)A^{c_3}_{\s}(x_3)
+24g^{3}f^{a_1b_1i_1}f^{a_2b_2c_2}f^{a_3b_3i_3}A^{i_1}_{\mu}(x_1)A^{c_2}_{\nu}(x_2)A^{i_3}_{\s}(x_3)\nonumber\\
&&+24g^{3}f^{a_1b_1c_1}f^{a_2b_2i_2}f^{a_3b_3i_3}A^{c_1}_{\mu}(x_1)A^{i_2}_{\nu}(x_2)A^{i_3}_{\s}(x_3)
+12g^{3}f^{a_1b_1i_1}f^{a_2b_2c_2}f^{a_3b_3c_3}A^{i_1}_{\mu}(x_1)A^{c_2}_{\nu}(x_2)A^{c_3}_{\s}(x_3)\nonumber\\
&&+12g^{3}f^{a_1b_1c_1}f^{a_2b_2i_2}f^{a_3b_3c_3}A^{c_1}_{\mu}(x_1)A^{i_2}_{\nu}(x_2)A^{c_3}_{\s}(x_3)
+12g^{3}f^{a_1b_1c_1}f^{a_2b_2c_2}f^{a_3b_3i_3}A^{c_1}_{\mu}(x_1)A^{c_2}_{\nu}(x_2)A^{i_3}_{\s}(x_3)\nonumber\\
&&+6g^{3}f^{a_1b_1c_1}f^{a_2b_2c_2}f^{a_3b_3c_3}A^{c_1}_{\mu}(x_1)A^{c_2}_{\nu}(x_2)A^{c_3}_{\s}(x_3)\Bigr)\nonumber\\
&&\times\Bigl[\Bigl(\partial_{\mu}^{x_1}\langle\bar{c}^{a}(x)c^{b_1}(x_1)\rangle\Bigr)
\langle\bar{c}^{a_2}(x_2)c^{b}(y)\rangle \Bigl(\partial_{\s}^{x_3}\langle\bar{c}^{a_1}(x_1)c^{b_3}(x_3)\rangle\Bigr)
\partial_{\nu}^{x_2}\langle\bar{c}^{a_3}(x_3)c^{b_2}(x_2)\rangle\Bigr]\,.
\end{eqnarray}}
It turns out to be convenient to introduce the following quantity
\begin{equation}
\mathcal{G}^{(3)}(x,y;A):=\frac{1}{N(N-1)}\,G_{3}^{aa}(x,y;A)\,.
\end{equation}
Using expression \eqref{G_0},  we get {\footnotesize
\begin{eqnarray}
\mathcal{G}^{(3)}(x,y;A)&=&-\frac{1}{N(N-1)}\int{d^{4}x_{1}d^{4}x_{2}d^{4}x_{3}}\,
\Bigl(8g^{3}f^{abi}f^{bcj}f^{cak}A^{i}_{\mu}(x_1)A^{j}_{\nu}(x_2)A^{k}_{\s}(x_3)\nonumber\\
&&+4g^{3}f^{abi}f^{bdj}f^{dac}A^{i}_{\mu}(x_1)A^{j}_{\nu}(x_2)A^{c}_{\s}(x_3)
+4g^{3}f^{abi}f^{bdc}f^{daj}A^{i}_{\mu}(x_1)A^{c}_{\nu}(x_2)A^{j}_{\s}(x_3)\nonumber\\
&&+4g^{3}f^{abc}f^{bdi}f^{daj}A^{c}_{\mu}(x_1)A^{i}_{\nu}(x_2)A^{j}_{\s}(x_3)
+2g^{3}f^{abi}f^{bdc}f^{dae}A^{i}_{\mu}(x_1)A^{c}_{\nu}(x_2)A^{e}_{\s}(x_3)\nonumber\\
&&+2g^{3}f^{abc}f^{bdi}f^{dae}A^{c}_{\mu}(x_1)A^{i}_{\nu}(x_2)A^{e}_{\s}(x_3)
+2g^{3}f^{abc}f^{bde}f^{dai}A^{c}_{\mu}(x_1)A^{e}_{\nu}(x_2)A^{i}_{\s}(x_3)\nonumber\\
&&+g^{3}f^{abc}f^{bde}f^{daf}A^{c}_{\mu}(x_1)A^{e}_{\nu}(x_2)A^{f}_{\s}(x_3)\Bigr)\nonumber\\
&&\times\Bigl[\Bigl(\partial_{\mu}^{x_1}G_0(x-x_1)\Bigr)
G_0(x_2-y) \Bigl(\partial_{\s}^{x_3}G_0(x_1-x_3)\Bigr)
\partial_{\nu}^{x_2}G_0(x_3-x_2)\Bigr]\nonumber\\
&&+\frac{1}{N(N-1)}\int{d^{4}x_{1}d^{4}x_{2}}\,
\Bigl(2g^{3}f^{abi}f^{bcj}f^{ack}A^{i}_{\mu}(x_1)A^{j}_{\nu}(x_2)A^{k}_{\nu}(x_2)\nonumber\\
&&+2g^{3}f^{abi}f^{bdc}f^{adj}A^{i}_{\mu}(x_1)A^{c}_{\nu}(x_2)A^{j}_{\nu}(x_2)
-2g^{3}f^{abi}f^{cbj}f^{daj}A^{i}_{\mu}(x_1)A^{c}_{\nu}(x_2)A^{d}_{\nu}(x_2)\nonumber\\
&&+g^{3}f^{abc}f^{bdi}f^{adj}A^{c}_{\mu}(x_1)A^{i}_{\nu}(x_2)A^{j}_{\nu}(x_2)
+g^{3}f^{abc}f^{bde}f^{adi}A^{c}_{\mu}(x_1)A^{e}_{\nu}(x_2)A^{i}_{\nu}(x_2)\nonumber\\
&&-g^{3}f^{abc}f^{dbi}f^{eai}A^{c}_{\mu}(x_1)A^{d}_{\nu}(x_2)A^{e}_{\nu}(x_2)\Bigr)\nonumber\\
&&\times\Bigl[\Bigl(\partial^{x_1}_{\mu}G_0(x-x_1)\Bigr)
G_0(x_2-y)G_0(x_1-x_2)
+G_0(x-x_2)G_0(x_1-y)\,\partial_{\mu}^{x_1}G_0(x_2-x_1)\Bigr]
\end{eqnarray}}
Let us now show that almost all the terms in the double integrals
vanish, {\it i.e}
\begin{eqnarray}
f^{abi}f^{bcj}f^{ack}A^{i}_{\mu}(x_1)A^{j}_{\nu}(x_2)A^{k}_{\nu}(x_2)&=&
-f^{abi}f^{acj}f^{bck}A^{i}_{\mu}(x_1)A^{j}_{\nu}(x_2)A^{k}_{\nu}(x_2)\nonumber\\
&=&-f^{abi}f^{ack}f^{bcj}A^{i}_{\mu}(x_1)A^{k}_{\nu}(x_2)A^{j}_{\nu}(x_2)\nonumber\\
&=&-f^{abi}f^{ack}f^{bcj}A^{i}_{\mu}(x_1)A^{j}_{\nu}(x_2)A^{k}_{\nu}(x_2)\nonumber\\
&=&0\,.
\end{eqnarray}
\begin{eqnarray}
f^{abi}f^{bdc}f^{adj}A^{i}_{\mu}(x_1)A^{c}_{\nu}(x_2)A^{j}_{\nu}(x_2)&=&
-f^{abi}f^{adc}f^{bdj}A^{i}_{\mu}(x_1)A^{c}_{\nu}(x_2)A^{j}_{\nu}(x_2)\nonumber\\
&=&-f^{abi}(-f^{adb}f^{jdc}-f^{adj}f^{cdb})A^{i}_{\mu}(x_1)A^{c}_{\nu}(x_2)A^{j}_{\nu}(x_2)\nonumber\\
&=&(\underbrace{f^{abi}f^{adb}}_{-f^{abi}f^{abd}=0}f^{jdc}+f^{abi}f^{adj}f^{cdb})
A^{i}_{\mu}(x_1)A^{c}_{\nu}(x_2)A^{j}_{\nu}(x_2)\nonumber\\
&=&f^{abi}f^{adj}f^{cdb}A^{i}_{\mu}(x_1)A^{c}_{\nu}(x_2)A^{j}_{\nu}(x_2)\nonumber\\
&=&-f^{abi}f^{adj}f^{bdc}A^{i}_{\mu}(x_1)A^{c}_{\nu}(x_2)A^{j}_{\nu}(x_2)\nonumber\\
&=&0\,,
\end{eqnarray}
\begin{eqnarray}
f^{abi}f^{cbj}f^{daj}A^{i}_{\mu}(x_1)A^{c}_{\nu}(x_2)A^{d}_{\nu}(x_2)&=&
-f^{abi}f^{caj}f^{dbj}A^{i}_{\mu}(x_1)A^{c}_{\nu}(x_2)A^{d}_{\nu}(x_2)\nonumber\\
&=&-f^{abi}f^{daj}f^{cbj}A^{i}_{\mu}(x_1)A^{d}_{\nu}(x_2)A^{c}_{\nu}(x_2)\nonumber\\
&=&-f^{abi}f^{daj}f^{cbj}A^{i}_{\mu}(x_1)A^{c}_{\nu}(x_2)A^{d}_{\nu}(x_2)\nonumber\\
&=&0\,,
\end{eqnarray}
\begin{eqnarray}
f^{abc}f^{bdi}f^{adj}A^{c}_{\mu}(x_1)A^{i}_{\nu}(x_2)A^{j}_{\nu}(x_2)&=&
-f^{abc}f^{adi}f^{bdj}A^{c}_{\mu}(x_1)A^{i}_{\nu}(x_2)A^{j}_{\nu}(x_2)\nonumber\\
&=&-f^{abc}f^{adj}f^{bdi}A^{c}_{\mu}(x_1)A^{j}_{\nu}(x_2)A^{i}_{\nu}(x_2)\nonumber\\
&=&-f^{abc}f^{adj}f^{bdi}A^{i}_{\mu}(x_1)A^{i}_{\nu}(x_2)A^{j}_{\nu}(x_2)\nonumber\\
&=&0\,,
\end{eqnarray}
\begin{eqnarray}
f^{abc}f^{dbi}f^{eai}A^{c}_{\mu}(x_1)A^{d}_{\nu}(x_2)A^{e}_{\nu}(x_2)&=&
-f^{abc}f^{dai}f^{ebi}A^{c}_{\mu}(x_1)A^{d}_{\nu}(x_2)A^{e}_{\nu}(x_2)\nonumber\\
&=&-f^{abc}f^{eai}f^{dbi}A^{c}_{\mu}(x_1)A^{e}_{\nu}(x_2)A^{d}_{\nu}(x_2)\nonumber\\
&=&-f^{abc}f^{eai}f^{dbi}A^{c}_{\mu}(x_1)A^{d}_{\nu}(x_2)A^{e}_{\nu}(x_2)\nonumber\\
&=&0\,.
\end{eqnarray}
The only term that does not vanish is $f^{abc}f^{bde}f^{adi}
A^{c}_{\mu}(x_1)A^{e}_{\nu}(x_2)A^{i}_{\nu}(x_2)$. Thus, we can finally
write $\mathcal{G}^{(3)}$ as {\footnotesize
\begin{eqnarray}
\mathcal{G}^{(3)}(x,y;A)&=&-\frac{1}{N(N-1)}\int{d^{4}x_{1}d^{4}x_{2}d^{4}x_{3}}\,
\Bigl(8g^{3}f^{abi}f^{bcj}f^{cak}A^{i}_{\mu}(x_1)A^{j}_{\nu}(x_2)A^{k}_{\s}(x_3)\nonumber\\
&&+4g^{3}f^{abi}f^{bdj}f^{dac}A^{i}_{\mu}(x_1)A^{j}_{\nu}(x_2)A^{c}_{\s}(x_3)
+4g^{3}f^{abi}f^{bdc}f^{daj}A^{i}_{\mu}(x_1)A^{c}_{\nu}(x_2)A^{j}_{\s}(x_3)\nonumber\\
&&+4g^{3}f^{abc}f^{bdi}f^{daj}A^{c}_{\mu}(x_1)A^{i}_{\nu}(x_2)A^{j}_{\s}(x_3)
+2g^{3}f^{abi}f^{bdc}f^{dae}A^{i}_{\mu}(x_1)A^{c}_{\nu}(x_2)A^{e}_{\s}(x_3)\nonumber\\
&&+2g^{3}f^{abc}f^{bdi}f^{dae}A^{c}_{\mu}(x_1)A^{i}_{\nu}(x_2)A^{e}_{\s}(x_3)
+2g^{3}f^{abc}f^{bde}f^{dai}A^{c}_{\mu}(x_1)A^{e}_{\nu}(x_2)A^{i}_{\s}(x_3)\nonumber\\
&&+g^{3}f^{abc}f^{bde}f^{daf}A^{c}_{\mu}(x_1)A^{e}_{\nu}(x_2)A^{f}_{\s}(x_3)\Bigr)\nonumber\\
&&\times\Bigl[\Bigl(\partial_{\mu}^{x_1}G_0(x-x_1)\Bigr)
G_0(x_2-y) \Bigl(\partial_{\s}^{x_3}G_0(x_1-x_3)\Bigr)
\partial_{\nu}^{x_2}G_0(x_3-x_2)\Bigr]\nonumber\\
&&+\frac{1}{N(N-1)}\int{d^{4}x_{1}d^{4}x_{2}}\,
g^{3}f^{abc}f^{bde}f^{adi}A^{c}_{\mu}(x_1)A^{e}_{\nu}(x_2)A^{i}_{\nu}(x_2)\nonumber\\
&&\times\Bigl[\Bigl(\partial^{x_1}_{\mu}G_0(x-x_1)\Bigr)
G_0(x_2-y)G_0(x_1-x_2)+G_0(x-x_2)G_0(x_1-y)\,\partial_{\mu}^{x_1}G_0(x_2-x_1)\Bigr]\,.
\end{eqnarray}}

\end{appendix}


\begin{thebibliography}{99}

\bibitem{Gribov:1977wm}
  V.~N.~Gribov,
  Nucl.\ Phys.\  B {\bf 139}, 1 (1978).


\bibitem{vanBaal:1991zw}
  P.~van Baal,
  Nucl.\ Phys.\  B {\bf 369}, 259 (1992).

\bibitem{Dell'Antonio:1991xt}
  G.~Dell'Antonio and D.~Zwanziger,
  Commun.\ Math.\ Phys.\  {\bf 138} (1991) 291.


\bibitem{Zwanziger:1982na}
  D.~Zwanziger,
  Nucl.\ Phys.\  B {\bf 209}, 336 (1982).

\bibitem{Dell'Antonio:1989jn}
  G.~Dell'Antonio and D.~Zwanziger,
  Nucl.\ Phys.\  B {\bf 326} (1989) 333.

\bibitem{Zwanziger:1988jt}
  D.~Zwanziger,
  Nucl.\ Phys.\  B {\bf 321}, 591 (1989).


\bibitem{Zwanziger:1989mf}
  D.~Zwanziger,
  Nucl.\ Phys.\  B {\bf 323}, 513 (1989).

\bibitem{Zwanziger:1992qr}
  D.~Zwanziger,
  Nucl.\ Phys.\  B {\bf 399}, 477 (1993).



\bibitem{Maggiore:1993wq}
  N.~Maggiore and M.~Schaden,
  Phys.\ Rev.\  D {\bf 50}, 6616 (1994)
  [arXiv:hep-th/9310111].

\bibitem{Dudal:2005na}
  D.~Dudal, R.~F.~Sobreiro, S.~P.~Sorella and H.~Verschelde,
  Phys.\ Rev.\  D {\bf 72}, 014016 (2005)
  [arXiv:hep-th/0502183].


\bibitem{Dudal:2010fq}
  D.~Dudal, S.~P.~Sorella and N.~Vandersickel,
  arXiv:1001.3103 [hep-th].




\bibitem{Dudal:2007cw}
  D.~Dudal, S.~P.~Sorella, N.~Vandersickel and H.~Verschelde,
  Phys.\ Rev.\  D {\bf 77}, 071501 (2008)
  [arXiv:0711.4496 [hep-th]].



\bibitem{Dudal:2008sp}
  D.~Dudal, J.~A.~Gracey, S.~P.~Sorella, N.~Vandersickel and H.~Verschelde,
  Phys.\ Rev.\  D {\bf 78}, 065047 (2008)
  [arXiv:0806.4348 [hep-th]].


\bibitem{Dudal:2008xd}
  D.~Dudal, S.~P.~Sorella, N.~Vandersickel and H.~Verschelde,
  Phys.\ Lett.\  B {\bf 680}, 377 (2009)
  [arXiv:0808.3379 [hep-th]].




\bibitem{Dudal:2008rm}
  D.~Dudal, J.~A.~Gracey, S.~P.~Sorella, N.~Vandersickel and H.~Verschelde,
  Phys.\ Rev.\  D {\bf 78}, 125012 (2008)
  [arXiv:0808.0893 [hep-th]].




\bibitem{Aguilar:2004sw}
  A.~C.~Aguilar and A.~A.~Natale,
  JHEP {\bf 0408}, 057 (2004)
  [arXiv:hep-ph/0408254].


\bibitem{Aguilar:2008xm}
  A.~C.~Aguilar, D.~Binosi and J.~Papavassiliou,
  Phys.\ Rev.\  D {\bf 78}, 025010 (2008)
  [arXiv:0802.1870 [hep-ph]].

\bibitem{Boucaud:2008ky}
  Ph.~Boucaud, J.~P.~Leroy, A.~Le Yaouanc, J.~Micheli, O.~Pene and J.~Rodriguez-Quintero,
  JHEP {\bf 0806}, 099 (2008)
  [arXiv:0803.2161 [hep-ph]].


\bibitem{Fischer:2008uz}
  C.~S.~Fischer, A.~Maas and J.~M.~Pawlowski,
  Annals Phys.\  {\bf 324}, 2408 (2009)
  [arXiv:0810.1987 [hep-ph]].



\bibitem{Cucchieri:2007rg}
  A.~Cucchieri and T.~Mendes,
  Phys.\ Rev.\ Lett.\  {\bf 100} (2008) 241601
  [arXiv:0712.3517 [hep-lat]].


\bibitem{Cucchieri:2008fc}
  A.~Cucchieri and T.~Mendes,
  Phys.\ Rev.\  D {\bf 78}, 094503 (2008)
  [arXiv:0804.2371 [hep-lat]].



\bibitem{Cucchieri:2008mv}
  A.~Cucchieri and T.~Mendes,
  PoS {\bf CONFINEMENT8}, 040 (2008)
  [arXiv:0812.3261 [hep-lat]].


\bibitem{Cucchieri:2009zt}
  A.~Cucchieri and T.~Mendes,
  arXiv:0904.4033 [hep-lat].

\bibitem{Bogolubsky:2009dc}
  I.~L.~Bogolubsky, E.~M.~Ilgenfritz, M.~Muller-Preussker and A.~Sternbeck,
  Phys.\ Lett.\  B {\bf 676}, 69 (2009)
  [arXiv:0901.0736 [hep-lat]].


\bibitem{Bogolubsky:2009qb}
  I.~L.~Bogolubsky, E.~M.~Ilgenfritz, M.~Muller-Preussker and A.~Sternbeck,
  PoS {\bf LATTICE2009}, 237 (2009)
  [arXiv:0912.2249 [hep-lat]].



\bibitem{Maas:2009se}
  A.~Maas,
  arXiv:0907.5185 [hep-lat].

\bibitem{Maas:2009ph}
  A.~Maas, J.~M.~Pawlowski, D.~Spielmann, A.~Sternbeck and L.~von Smekal,
  arXiv:0912.4203 [hep-lat].


\bibitem{Singer:1978dk}
  I.~M.~Singer,
  Commun.\ Math.\ Phys.\  {\bf 60} (1978) 7.


\bibitem{Zwanziger:2007zz}
  D.~Zwanziger,
  Braz.\ J.\ Phys.\  {\bf 37} (2007) 127.


\bibitem{Sobreiro:2005vn}
  R.~F.~Sobreiro and S.~P.~Sorella,
  JHEP {\bf 0506}, 054 (2005)
  [arXiv:hep-th/0506165].


\bibitem{Bruckmann:2000xd}
  F.~Bruckmann, T.~Heinzl, A.~Wipf and T.~Tok,
  Nucl.\ Phys.\  B {\bf 584}, 589 (2000)
  [arXiv:hep-th/0001175].

\bibitem{Capri:2005tj}
  M.~A.~L.~Capri, V.~E.~R.~Lemes, R.~F.~Sobreiro, S.~P.~Sorella and R.~Thibes,
  Phys.\ Rev.\  D {\bf 72}, 085021 (2005)
  [arXiv:hep-th/0507052].



\bibitem{Ezawa:1982bf}
  Z.~F.~Ezawa and A.~Iwazaki,
  Phys.\ Rev.\  D {\bf 25}, 2681 (1982).

\bibitem{Suzuki:1989gp}
  T.~Suzuki and I.~Yotsuyanagi,
  Phys.\ Rev.\  D {\bf 42}, 4257 (1990).

\bibitem{Hioki:1991ai}
  S.~Hioki, S.~Kitahara, S.~Kiura, Y.~Matsubara, O.~Miyamura, S.~Ohno and T.~Suzuki,
  Phys.\ Lett.\  B {\bf 272}, 326 (1991)
  [Erratum-ibid.\  B {\bf 281}, 416 (1992)].



\bibitem{Nambu:1975ba}
  Y.~Nambu,
  Phys.\ Rept.\  {\bf 23}, 250 (1976).

\bibitem{Mandelstam:1974vf}
  S.~Mandelstam,
  Phys.\ Lett.\  B {\bf 53}, 476 (1975).

\bibitem{'tHooft:1982ns}
  G.~'t Hooft,
  Phys.\ Scripta {\bf 25} (1982) 133.




\bibitem{Amemiya:1998jz}
  K.~Amemiya and H.~Suganuma,
  Phys.\ Rev.\  D {\bf 60}, 114509 (1999)
  [arXiv:hep-lat/9811035].

\bibitem{Bornyakov:2003ee}
  V.~G.~Bornyakov, M.~N.~Chernodub, F.~V.~Gubarev, S.~M.~Morozov and M.~I.~Polikarpov,
  Phys.\ Lett.\  B {\bf 559}, 214 (2003)
  [arXiv:hep-lat/0302002].


\bibitem{Mendes:2006kc}
  T.~Mendes, A.~Cucchieri and A.~Mihara,
  AIP Conf.\ Proc.\  {\bf 892}, 203 (2007)
  [arXiv:hep-lat/0611002].




\bibitem{Min:1985bx}
  H.~Min, T.~Lee and P.~Y.~Pac,
  Phys.\ Rev.\  D {\bf 32}, 440 (1985).



\bibitem{Fazio:2001rm}
  A.~R.~Fazio, V.~E.~R.~Lemes, M.~S.~Sarandy and S.~P.~Sorella,
  Phys.\ Rev.\  D {\bf 64}, 085003 (2001)
  [arXiv:hep-th/0105060].

\bibitem{Dudal:2004rx}
  D.~Dudal, J.~A.~Gracey, V.~E.~R.~Lemes, M.~S.~Sarandy, R.~F.~Sobreiro, S.~P.~Sorella and H.~Verschelde,
  Phys.\ Rev.\  D {\bf 70}, 114038 (2004)
  [arXiv:hep-th/0406132].


\bibitem{Kondo:2001tm}
  K.~I.~Kondo, T.~Murakami, T.~Shinohara and T.~Imai,
  Phys.\ Rev.\  D {\bf 65}, 085034 (2002)
  [arXiv:hep-th/0111256].

\bibitem{Huber:2009wh}
  M.~Q.~Huber, K.~Schwenzer and R.~Alkofer,
  arXiv:0904.1873 [hep-th].



\bibitem{Capri:2008vk}
  M.~A.~L.~Capri, A.~J.~Gomez, V.~E.~R.~Lemes, R.~F.~Sobreiro and S.~P.~Sorella,
  Phys.\ Rev.\  D {\bf 79}, 025019 (2009)
  [arXiv:0811.2760 [hep-th]].


\bibitem{Capri:2006cz}
  M.~A.~L.~Capri, V.~E.~R.~Lemes, R.~F.~Sobreiro, S.~P.~Sorella and R.~Thibes,
  Phys.\ Rev.\  D {\bf 74}, 105007 (2006)
  [arXiv:hep-th/0609212].


\bibitem{Capri:2008ak}
  M.~A.~L.~Capri, V.~E.~R.~Lemes, R.~F.~Sobreiro, S.~P.~Sorella and R.~Thibes,
  Phys.\ Rev.\  D {\bf 77}, 105023 (2008)
  [arXiv:0801.0566 [hep-th]].

\bibitem{Sobreiro:2005ec}
  R.~F.~Sobreiro and S.~P.~Sorella,
  arXiv:hep-th/0504095.





\bibitem{Gomez:2009tj}
  A.~J.~Gomez, M.~S.~Guimaraes, R.~F.~Sobreiro and S.~P.~Sorella,
  Phys.\ Lett.\  B {\bf 683}, 217 (2010)
  [arXiv:0910.3596 [hep-th]].

\bibitem{Baulieu:2009ha}
  L.~Baulieu, D.~Dudal, M.~S.~Guimaraes, M.~Q.~Huber, S.~P.~Sorella, N.~Vandersickel and D.~Zwanziger,
  arXiv:0912.5153 [hep-th].

\bibitem{Sorella:2009vt}
  S.~P.~Sorella,
  Phys.\ Rev.\  D {\bf 80}, 025013 (2009)
  [arXiv:0905.1010 [hep-th]].

\bibitem{Piguet:1995er}
  O.~Piguet and S.~P.~Sorella,
  Lect.\ Notes Phys.\  {\bf M28} (1995) 1.

\bibitem{Capri:2007hw}
  M.~A.~L.~Capri, D.~Dudal, J.~A.~Gracey, S.~P.~Sorella and H.~Verschelde,
  JHEP {\bf 0801}, 006 (2008)
  [arXiv:0708.4303 [hep-th]].







\end{thebibliography}
\end{document}